\documentclass[usenatbib,numberedappendix]{emulateapj}

\newcommand{\be}{\begin{equation}}
\newcommand{\ee}{\end{equation}}
\newcommand{\bfig}{\begin{figure}}
\newcommand{\efig}{\end{figure}}
\newcommand{\bfige}{\begin{figure*}}
\newcommand{\efige}{\end{figure*}}
\newcommand{\bea}{\begin{eqnarray}}
\newcommand{\eea}{\end{eqnarray}}

\newcommand{\chandra}{{\it Chandra}~}

\shorttitle{M33 X-ray variability}
\shortauthors{Grimm et al.}

\begin{document}

\title{The X-ray binary population in M33: II. X-ray spectra and
variability}
\author{H.-J. Grimm\altaffilmark{1}, J.~McDowell\altaffilmark{1},
A. Zezas\altaffilmark{1}, D.-W. Kim\altaffilmark{1},
G.~Fabbiano\altaffilmark{1}}
\affil{Harvard-Smithsonian Center for Astrophysics, 60 Garden Street,
Cambridge, MA 02138}

\begin{abstract}
In this paper we investigate the X-ray spectra and X-ray spectral
variability of compact X-ray sources for 3 \chandra observations of
the Local Group galaxy M33. The observations are centered on the
nucleus and the star forming region NGC 604. In the observations 261
sources have been detected. For a total of 43 sources the number of
net counts is above 100, sufficient for a more detailed spectral
fitting. Of these sources, 25 have been observed in more than one
observation, allowing the study of spectral variability on
$\sim$months timescales. A quarter of the sources are found to be
variable between observations. However, except for two foreground
sources, no source is variable within any observation above the 99\%
confidence level. Only six sources show significant spectral
variability between observations. A comparison of $N_H$ values with HI
observations shows that X-ray absorption values are consistent with
Galactic X-ray binaries and most sources in M33 are intrinsically
absorbed. The pattern of variability and the spectral parameters of
these sources are consistent with the M33 X-ray source population being
dominated by X-ray binaries: Two thirds of the 43 bright sources have
spectral and timing properties consistent with X-ray binaries; we also
find two candidates for super-soft sources and two candidates for
quasi-soft sources.
\end{abstract}


\section{Introduction}
\label{sec:intro}
M33 is a late-type spiral galaxy, Sc II-III, and the third largest
galaxy in the Local Group. It is a unique galaxy in the Local Group
since morphologically it is of intermediate type between the large
early-type spiral galaxies and the numerous dwarf galaxies. Other
galaxies of this type cannot be investigated with the same depth even
with \chandra. At a distance of 840 kpc \citep{freedman:91} from the
Milky Way (MW) M33 is the second nearest major galaxy. It spans
roughly 73'$\times$45' on the sky. The line-of-sight absorption
column density is small, $N_H \sim 6 \times 10^{20}$ cm$^{-2}$
\citep{stark:92}. M33 is more actively star forming than either the MW
or M31 \citep{hippelein:03}, particularly compared to its much smaller
mass.

M33 has been studied with every X-ray mission since Einstein
\citep{markert:83}. But only recently have high angular resolution and
high sensitivity instruments like \chandra and XMM allowed us to study
the X-ray source population in depth. \citet{grimm:05} and Pietsch et
al. (2005) have provided source lists and fluxes for the X-ray source
population in M33, from the \chandra and XMM observations
respectively. In this paper we follow up on the \citet{grimm:05}
\chandra survey and present an analysis of the X-ray spectra and
variability behavior of X-ray sources in M33. Apart from fluxes,
variability and spectral energy distributions are an important
diagnostic tool for understanding the emission mechanism(s) in X-ray
sources, and for classifying these sources. Moreover, spectral and
time variability are the main characteristics of X-ray binaries, as
shown by the detailed work done for X-ray binaries in the Milky Way,
see e.g. \citet{klis:05}.

The paper is organized as follows. In Sec. \ref{sec:data} we describe
the data processing, followed in Sec. \ref{sec:analysis} by the
analysis procedures. The results and their implications for the nature
of the X-ray source population in M33 are discussed in
Sec. \ref{sec:results}; we also discuss individual bright sources
in detail in Sec. \ref{sec:sources}. We conclude with a summary of the
results in Sec. \ref{sec:summary}. The figures showing the results for
the spectral analysis for sources with more than one observation are
shown in the Appendix in the electronic edition only.

\section{Data}
\label{sec:data}

M33 has been observed with the ACIS instrument on \chandra four
times, see \citet{grimm:05}. In this paper we use three observations
whose ObsIds and dates are given in Table \ref{tab:obs}. Due to its
angular extent the observed parts of M33 cover all active chips, the
standard ACIS-S configuration for \dataset[ADS/Sa.CXO\#obs/00786]{ObsId
786}, and the standard ACIS-I configuration for
\dataset[ADS/Sa.CXO\#obs/01730]{ObsId 1730} and
\dataset[ADS/Sa.CXO\#obs/02023]{ObsId 2023}. There is considerable overlap
between the different observations. However, due to the decreasing
resolution/sensitivity  with increasing off-axis angle only the inner
part of M33 ($\sim$ 8-10 arc minutes) has a significant number of
sources in two observations, \dataset[ADS/Sa.CXO\#obs/00786]{786} and
\dataset[ADS/Sa.CXO\#obs/01730]{1730}. A fourth observation, ObsId
787, was disregarded because it was aimed at studying the nucleus and
suffers from both high background and small FOV.

\begin{table}[h]
\begin{center}
\caption{List of ACIS observations of M33}
\begin{tabular}{l|c|r|c}
\hline
ObsId & Date & Aim point & Duration [ks]\\
\hline
\dataset[ADS/Sa.CXO\#obs/00786]{786}  & 2000-08-30 &  Nucleus & 45\\
\dataset[ADS/Sa.CXO\#obs/01730]{1730} & 2000-07-12 &  Nucleus & 45\\
\dataset[ADS/Sa.CXO\#obs/02023]{2023} & 2001-07-06 &  NGC 604 & 90\\
\hline
\end{tabular}
\label{tab:obs}
\end{center}
\end{table}

The data from the three observations were processed according to the
standard data processing procedure with CIAO versions 3.1, including
exposure correction. Source detection was performed with {\em
wavdetect} with scales of 1, 2, 4, 8, 10, 12, and 16 in the energy
range 0.3--8 keV. The signal detection threshold was set to
$10^{-6}$. Source regions correspond to the 95\% encircled energy
area. For more details of the data analysis see \citet{grimm:05}.

\section{Analysis}
\label{sec:analysis}

We separate the analysis in three parts, short term,
long term, and spectral variability. The different methods are
discussed in the following subsections. We assume that the variability
of individual sources is independent, so there is no correlation
between short term and long term variability. Because the number of
counts observed in a single observation is not very large even for
bright sources we restrict the spectral variability analysis to a
comparison between different observations, when applicable.

The nucleus was excluded from the following analysis because it
suffers from strong pile-up in two observations.

\subsection{Short and long term variability}

In order to establish short timescale variability we perform a
Bayesian block analysis of the lightcurves of individual sources. A
Bayesian block analysis computes the best approximation to the
lightcurve shape in terms of piecewise constant flux levels or
blocks. The discriminator between a single flux level for the whole
lightcurve or two flux levels is the ratio of likelihoods of
describing a data segment with one or two blocks. In this analysis the
algorithm used is iterative. It starts with the whole lightcurve and
subsequently divides it until the likelihood ratio for dividing a
lightcurve segment becomes smaller than a predefined prior, in  
this case corresponding to $\sim$99\% confidence level. The Bayesian
block analysis is particularly suited for burst-like variability. For
more details about the principle of Bayesian blocks and the algorithm
used in this analysis see \citet{scargle:98}. The implementation used
here is the same as used by the CHAMP project and described in
\citep{kim:04}.

Since a Bayesian block analysis uses only the photon arrival times
for computation of the blocks, no binning is necessary. Therefore
there is no intrinsic restriction to the number of photons the
algorithm is applicable to. Obviously establishing variability
with a certain confidence becomes less likely for fainter sources.

Because a Bayesian block analysis is particularly sensitive to
burst-like variability we also performed a search for periodic
variability with the XRONOS v5.21 tool {\em efsearch}. This analysis
did not yield any source with significant short term variability.

To establish long timescale variability we use simple Poisson
statistics. This is justified because, with the exception of two
foreground sources, no source exhibits strong short term
variability. We compare the difference in fluxes between two
observations to the quadratically added errors for the fluxes. If a
source is undetected in an observation in which it was in the field of
view of \chandra, we calculate an upper limit to the source flux using
an algorithm developed by \citet{kraft:91}. Since the upper limit
value already corresponds to the 99\% confidence level, for a
comparison with a detected flux we compare the 3 sigma error of the
detection directly with the upper limit.

\subsection{Spectra and spectral variability}

For 43 out of the total 261 sources the number of counts in at least
one observation was larger than 100 net counts, which we consider
sufficient to attempt spectral fitting. Of these 43 sources 25 have
been observed in at least two observations. For these sources we also
compare the spectral properties with time. The spectral fitting was
done with XSPEC v11.3.1. Because of the generally low number of counts
the fitting was done for all sources with Cash statistics
\citep{cash:79}. At 100 counts we expect only $\sim$3 background
sources based on the CDF-N LogN--LogS \citep{alexander:03}, so the
vast majority of these sources are likely to belong to M33.

We first fit the bright sources with two simple absorbed spectral
models, power law and bremsstrahlung, to check the validity of our
assumptions about the spectral shape of the faint sources that do not
have sufficient counts for detailed spectral modeling (see derivation
of the X-ray luminosity function in \citet{grimm:05}). The absorption
was in one case a free parameter in the other case it was fixed to the
Galactic value. After validating that the assumptions about the
general power law shape are correct, we proceeded to fit more complex
model to the bright sources.

The results of the simple spectral fitting are presented in
Sec. \ref{sec:results}. Spectra of sources with more than one
observation with best fit values and corresponding confidence contour
plots are shown in Appendix \ref{sec:spectra} in the electronic
edition, except for the spectra of M33 X-7 (\object[]{CXO
J013334.1+303210}) and M33 X-9 (\object[]{CXO J013358.8+305004}),
which are shown in Figs. \ref{fig:x7} and \ref{fig:x9}, and will be
discussed in Sec. \ref{sec:sources}.

Based on the confidence regions for the fit parameters, shown in
Appendix \ref{sec:spectra}, only one of the X-ray sources shows
significant spectral variability from observation to observation, M33
X-4 (\object[]{CXO J013315.1+305317}). For the other sources in different
observations none of the parameters of a source spectrum show changes
corresponding to more than 99\% confidence level. This is despite the
fact that some sources show significant time variability between
observations. However, the number of counts available for most sources
for fitting is insufficient to determine spectral parameters to an
accuracy good enough to compare two observations. Moreover,
degeneracies between model parameters complicate the establishing of
variability.

Another approach to spectral variability is the use of hardness
ratios. These are relatively crude estimators of spectral changes but
because of the smaller number of degrees of freedom compared  to a
spectral fit can be statistically preferable. We therefore take all
sources that were observed in at least two observations, divide each
observation in three equally long parts, and construct hardness ratios
from these intervals. Because only sources with more than 100 counts
are used in this part of the analysis the separation in three parts
still gives sufficient number of counts ($\ge 20-30$) in each time bin
for a hardness ratio analysis \citep{prestwich:03}. Note that
\dataset[ADS/Sa.CXO\#obs/02023]{ObsId 2023} is roughly twice as long
as the other observations. The energy bands used are the same as in
\citet{grimm:05}; 0.3--1.0 keV for the soft band, 1.0--2.1 keV for the
medium band, and 2.1--8.0 keV for the hard band. The hardness ratios
are defined as
\begin{equation}
HR1 = \frac{M-S}{T},\hspace{1cm}
HR2 = \frac{H-M}{T}
\end{equation}
where $S$, $M$, and $H$ are the background and exposure corrected
counts in the soft, medium, and hard band, and $T$ are the corrected
counts in the whole energy band. To compare two observations
it is important to take into account the different effective areas of
the source location in each observation, especially if the source is
located on a front-illuminated chip and in another observation on a
back-illuminated chip. We compute the effective area of the source in
each observation for each energy band. For comparison with the other
observations of the source we normalize the effective area in each
band by the effective area value of the aim point of
\dataset[ADS/Sa.CXO\#obs/01730]{ObsId 1730}. Note that the choice of
the normalization constant is arbitrary. The results are combined for
all available observations. The results for all sources that show
evidence of variability are shown in Appendix \ref{sec:spectra} in the
electronic edition.

\section{Results and Discussion}
\label{sec:results}

\subsection{Short timescale variability}

Except for two sources which are foreground Galactic stars, no source
presents variability above the 99\% confidence level. In
Fig. \ref{fig:b1} we show the results of the Bayesian block analysis
for the only two sources that are variable on short timescales. The
dotted histograms represent the count rate binned in 400 second
intervals. Note that the binning is for plotting purposes only. The
green line is the power spectrum of the lightcurve. The thick solid
line is the result of the Bayesian block analysis. The small square to
the right of the lightcurve is a thumbnail image of the source from
the observation.

Source {\bf \object[]{CXO J013327.7+304645}} shows a clear X-ray outburst in
\dataset[ADS/Sa.CXO\#obs/02023]{ObsId 2023}. This outburst, lasting
about 10,000 seconds completely accounts for the long term variability
of the source. The lightcurve of the outburst is well fit by an
exponential decay with a decay timescale of 3900 seconds. The
persistent luminosity in \dataset[ADS/Sa.CXO\#obs/02023]{ObsId 2023}
is consistent within the errors with the luminosities in the other
observations. Outburst and persistent counts are not large enough to
investigate changes in spectral shape. An X-ray color comparison of
the burst and persistent emission shows a softening of the emission in
HR2 and a slight hardening in HR1 that is significant only at the
2-sigma level. The optical counterpart to source \object[]{CXO
J013327.7+304645} is a star in the USNO catalog with a V magnitude of
16.8. We have analyzed archival HST WFPC2 data. The object was
detected in bands corresponding to U, B, V, and I. The colors are
consistent with an early M type star, either on the main sequence or a
giant. Using normal V magnitudes for M stars this puts the star at a
distance of 90--320 pc for a main sequence star, or 24--29 kpc for a
giant. Considering the relatively high galactic latitude of M33,
$b=-31$ deg, the lower distance value, and thus a main sequence star
is more likely. Assuming a main sequence star the peak luminosity is
between $\sim10^{29}-1.5\times 10^{30}$ ergs s$^{-1}$, and the total
energy between $\sim10^{32}-1.5\times 10^{33}$ ergs.

\bfige[h]
  \resizebox{0.5\hsize}{!}{\includegraphics[angle=-90]{f1a.eps}}
  \resizebox{0.5\hsize}{!}{\includegraphics[angle=-90]{f1b.eps}}
  \caption{{\em Left}: Lightcurve and Bayesian block division of the
  outburst of CXO J013327.7+304645, an X-ray active
  star. The panel shows the lightcurve binned in 400 second intervals
  (dashed histogram). Note that the lightcurve binning is for plotting
  purposes only. The Bayesian blocks are shown as the thick black
  histogram. The right part of the panel shows a thumbnail picture of
  the source.{\em Right}: Lightcurve and Bayesian block division of
  the outburst of CXO J013341.8+303848, another X-ray
  active star.}
  \label{fig:b1}
\efige

Source {\bf \object[]{CXO J013341.8+303848}}, a foreground star as
well, has two outbursts in \dataset[ADS/Sa.CXO\#obs/00786]{ObsId
786}. The first outburst lasts about 2600 seconds, the second outburst
18000 seconds later lasts about 800 seconds. The persistent level of
X-ray emission increases slightly after each outburst. As shown in the
long term lightcurve the source also has a long term trend to
increasing luminosity for the 3 observations. The decay of the
outbursts cannot be fit uniquely; an exponential decay with a decay
constant of $\sim$1000 seconds or a linear decay are both
possible. During both outbursts the source becomes softer, but only at
the $1-2\sigma$ level. This source has no HST coverage.

\subsection{Long term variability}

Of a total of 261 sources, 198 have been detected in at least
two observations, and 62 in all three observations. The luminosities
for comparison are taken from spectral fits for the brighter sources
(more than $\sim$100 counts), or from the assumed spectrum, absorbed
power law with a photon index of 2 and Galactic absorption. We find
that 49 of 198 sources show variability between observations above 3
sigma. An additional 29 sources show signs of variability between 2
and 3 sigma. Of the 49 variable sources only 16 are apparently
persistent.

The strength of the variability can in fact be used to confirm the
nature of most of these sources as X-ray binaries. AGN in general show
variability on weeks or months timescale only up to factors of 2--3
\citep{mushotzky:93,paolillo:04}. Most of the sources in M33 show
stronger variability, particularly since the flux ratios for most
sources are only lower limits. Moreover, the sources with flux ratios
below 2 are bright sources that are unlikely to be AGN based only on
brightness.

Taking into account that $\sim$80--90 of the 198 sources are likely
background AGN, somewhat less than half the sources in M33 are
variable. Of these two thirds (34/49) may be candidate
transients. This number is somewhat higher than the about 50\% in the
Milky Way and Magellanic Clouds \citep{liu:00,liu:01}. But the
detection limits (few $10^{34}$ erg s$^{-1}$) do not allow to
establish true transient behavior for undetected sources so the number
of two thirds is only an upper limit.

For the brighter sources ($L_{X} > 5\times10^{36}$ erg s$^{-1}$) past
X-ray satellites also provide data for long term variability.
Fig. \ref{fig:long} shows data from Einstein HRI and IPC
\citep{trinchieri:88}, ROSAT  HRI \citep{schulman:95} and PSPC
\citep{long:96}, BeppoSAX \citep{parmar:01} as well as from this work
for the bright sources in the field of view of \chandra. The quoted
luminosities are converted to the \chandra band of 0.3--8.0 keV and,
if they are not individual measurements of spectra, converted to the
spectral shape of a power law with photon index of 2 and the Galactic
absorption value towards M33, $6\times10^{20}$ cm$^{-2}$. The error
bars in the plot only contain errors due to counting statistics. Other
errors, e.g. due to the conversion of energy bands and different
assumed spectral shapes, add generally another 20--30\% uncertainty in
the luminosity. In addition different instrument responses,
cross-calibration issues and other systematic effects add another
source of systematic errors, which is hard to quantify. An exception
is the supernova remnant {\bf M33 X-14}, for which converting the
luminosity given by \citet{long:96} would result in variability at
over $6\sigma$ compared to other observations. However,
\citet{long:96} assume a power law spectrum with a photon index of 2
to compute the source luminosity. Although this is a reasonable choice
for X-ray binaries or AGN, the spectrum of M33 X-14 is quite soft. We
extracted a spectrum of X-14 from the ROSAT observation rp600023a00,
the longest of the PSPC observations with $\sim$29 ks. The spectrum is
well fit by an absorbed blackbody model with a column density of
$4\times 10^{21}$ cm$^{-2}$ and a temperature of 0.09 keV. Although
this model is rather unphysical for a SNR we are interested only in
the flux, which is sufficiently accurate for the purpose of
comparison. The ROSAT data also agree with the spectrum obtained from
\chandra. Extrapolating this spectrum to the \chandra energy range
gives a luminosity of $3.7\times 10^{36}$ ergs s$^{-1}$ which agrees
very well with other observations of the source, in particular the
\chandra observed value of $3.6\times 10^{36}$ ergs s$^{-1}$.

It is clear from Fig. \ref{fig:long} that four of the bright X-ray
sources are variable. The sources that do not show evidence of
variability at the $3\sigma$ level are M33 X-1, X-2, and the SNR X-14.

\bfige
  \resizebox{\hsize}{!}{\includegraphics[angle=-90]{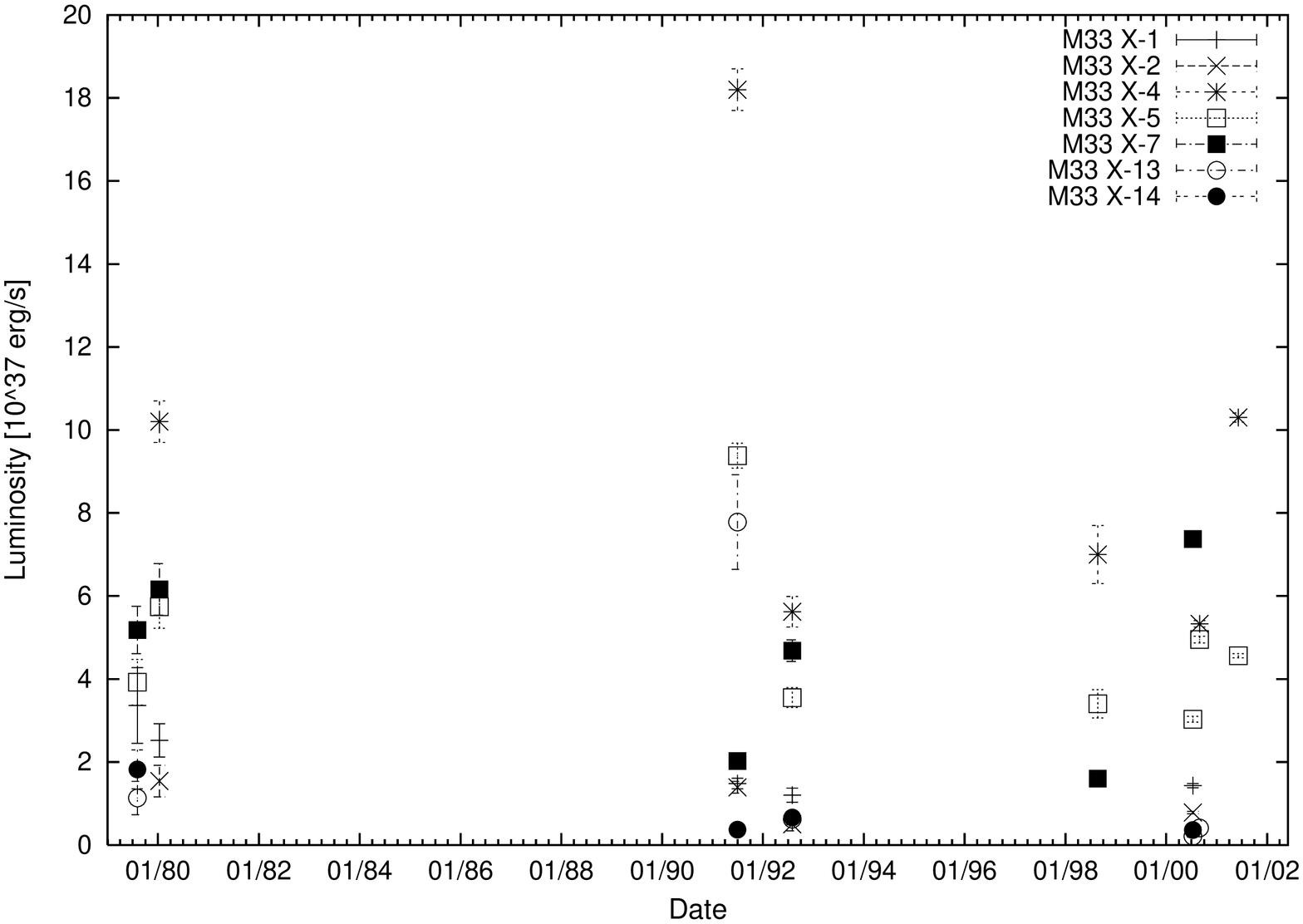}}
  \caption{Long term lightcurves of bright X-ray sources in M33
  observed with Einstein, ROSAT, BeppoSAX, and \chandra.}
  \label{fig:long}
\efige

The variability pattern is very similar to that observed in other
X-ray source populations, regardless of galaxy type, e.g. The
Antennae \citep{zezas:06}, M101 \citep{jenkins:05}, NGC 4697
\citep{sivakoff:05}. In particular that only very few, if any, sources
show short-term variability is very common, although this is most
likely due to limited photon statistics. On the other hand a
large fraction ($\sim$10-40\%) of sources exhibit long-term
variability. Moreover, the fraction of sources with strong long term
variability is quite confidently identified as X-ray binaries, and not
background AGN.

\subsection{Spectra}

Fitting all the 43 sources detected with more than 100 net counts with
a power law and a thermal bremsstrahlung spectrum, with column density
fixed to the Galactic value or as a free fit parameter, provides about
half of the sources with a good fit to the data. The other sources
require either different models, e.g. black body or other thermal
plasma models, or multiple components. The results of the simple power
law/bremsstrahlung fits support the validity of our assumptions of a
fixed spectral model for conversion from counts to fluxes for fainter
sources, assuming that the spectral properties of bright and faint
sources are not systematically different \citep{grimm:05}.

In Fig. \ref{fig:spec_prop} we show histograms for the best fit power
law and bremsstrahlung values for all sources.  The upper left panel
shows the comparison between photon indices $\Gamma$ of a power law
with column density being a free fit parameter versus column density
fixed to the Galactic value. The Galactic absorption towards M33 is
only $6 \times 10^{20}$ cm$^{-2}$. The upper right panel shows the
same for bremsstrahlung temperature $kT$. The lower left panel shows
the histogram of $\Gamma$ for the case of $N_H$ as a free fit
parameter. The power law slopes are concentrated around 2, the
canonical value for X-ray binaries and AGN. The peak at $\Gamma = 5$
comprises all sources with photon index larger than or equal to
5. Individually, all these sources are fit well with a blackbody or
plasma model with temperatures of 0.1--0.13 keV (blackbody) or
0.2--0.3 keV (plasma model). The lower right panel finally shows the
distribution of bremsstrahlung temperatures. Note that the influence
of $N_H$ is stronger in case of a power law than for a bremsstrahlung
spectrum as is evident in the generally smaller deviations from the
one-to-one correlation for the bremsstrahlung temperatures compared to
the photon indices in the upper panels.

\bfige
  \resizebox{0.5\hsize}{!}{\includegraphics{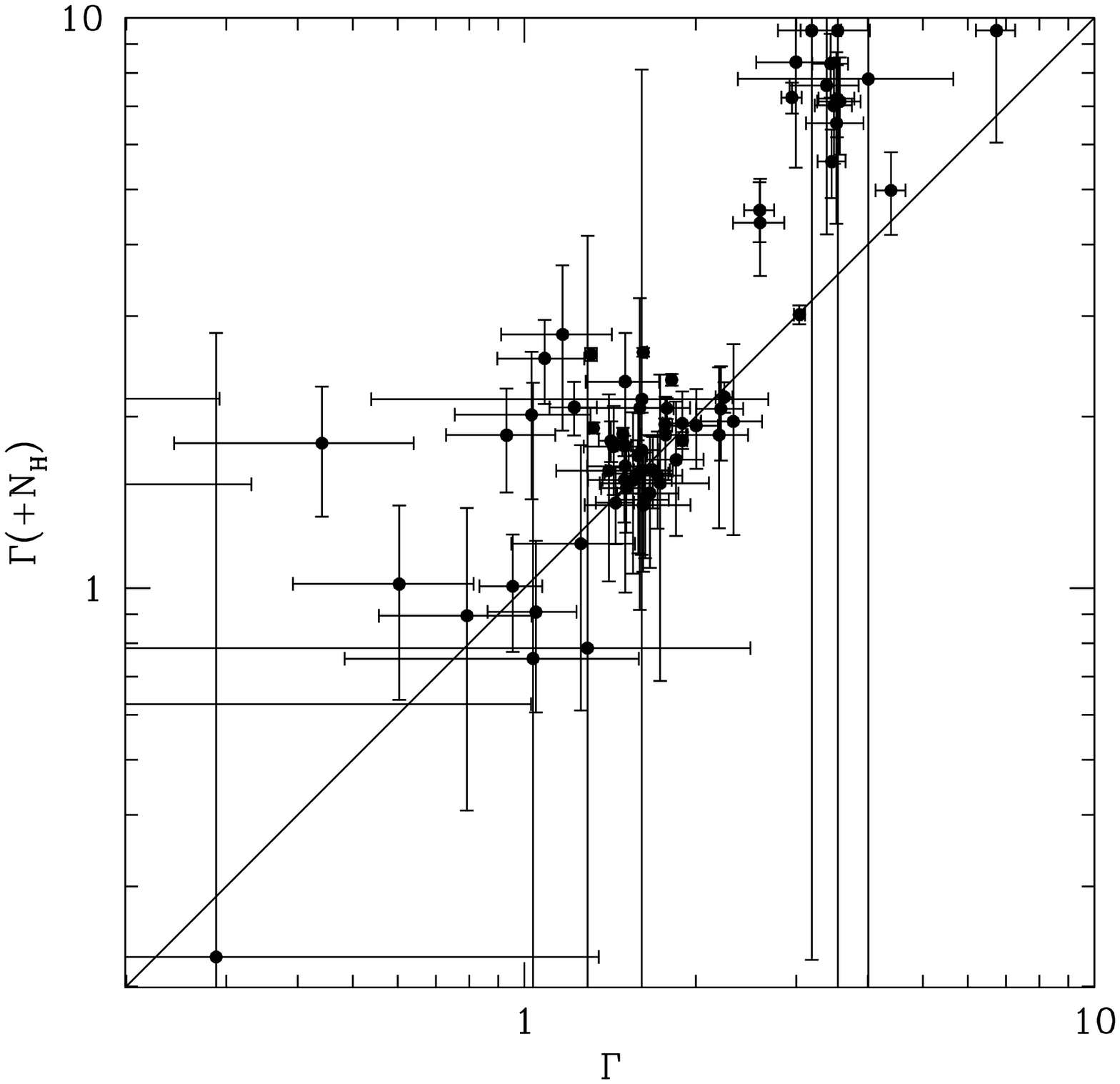}}
  \resizebox{0.5\hsize}{!}{\includegraphics{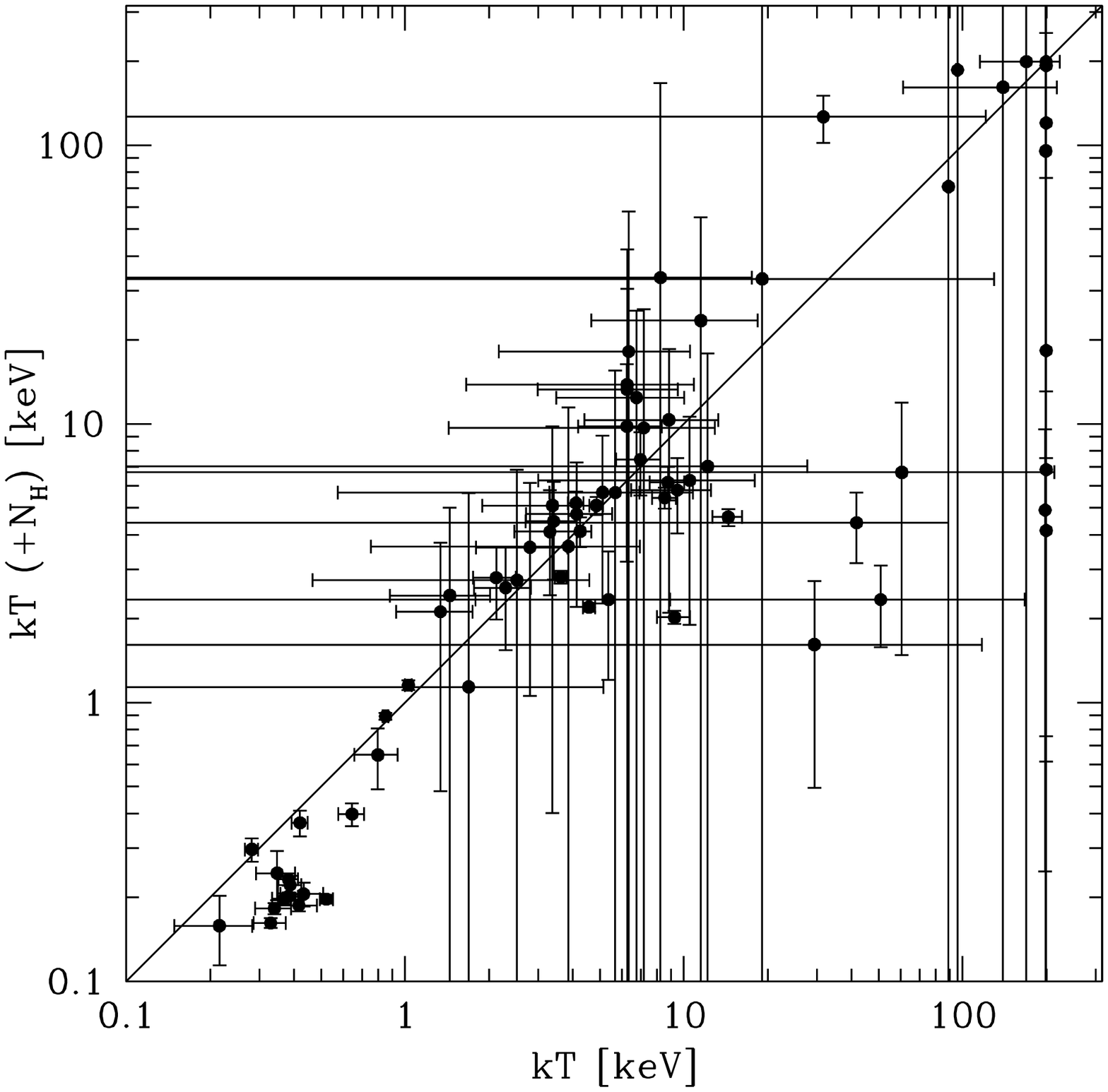}}
  \resizebox{0.5\hsize}{!}{\includegraphics{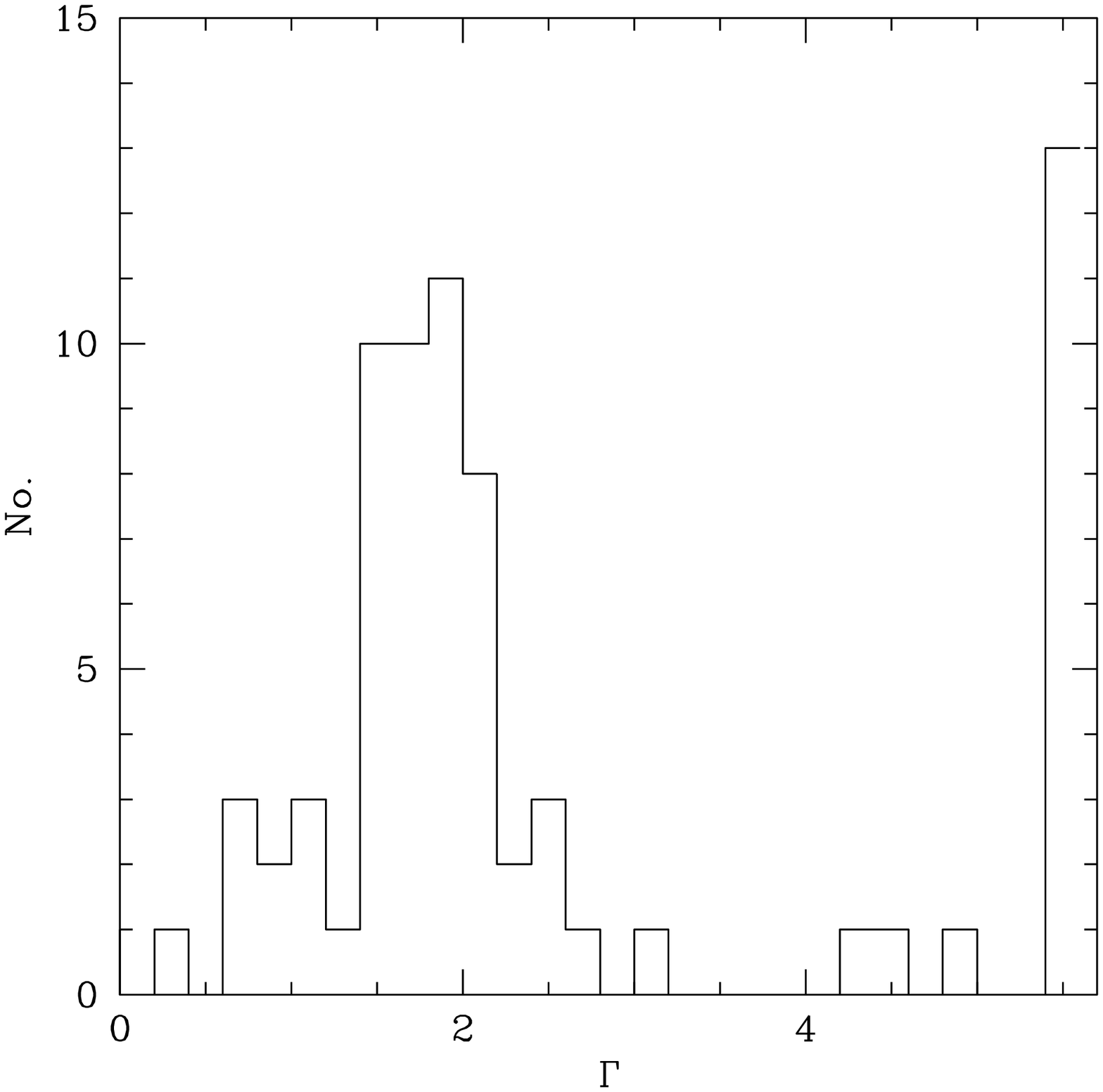}}
  \resizebox{0.5\hsize}{!}{\includegraphics{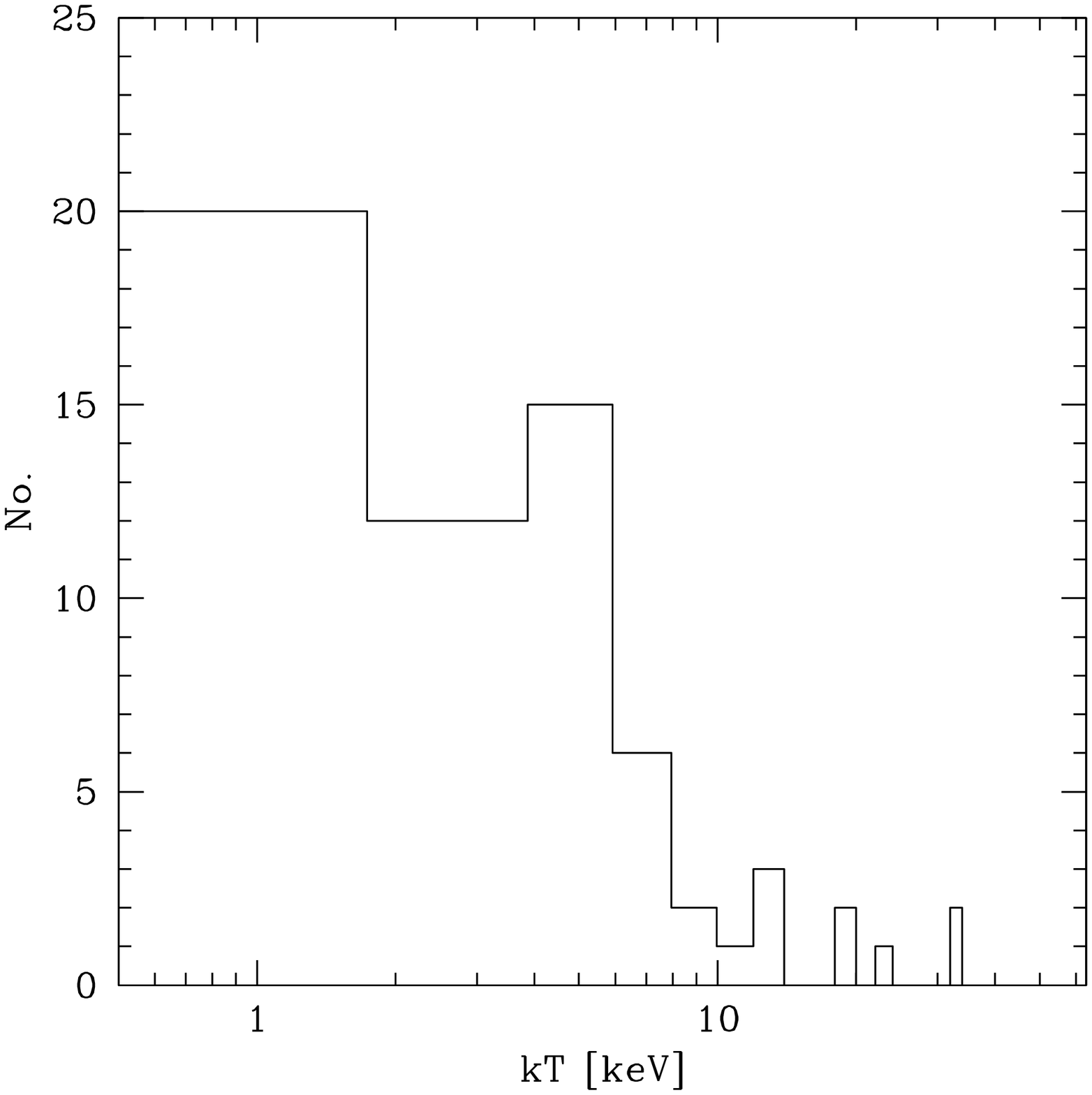}}
  \caption{Overall spectral properties of the sample of 43 bright
  X-ray sources in M33. The panels show the correlations of photon
  index $\Gamma$ with $N_H$ free versus fixed (upper left),
  bremsstrahlung temperature $kT$ with $N_H$ free versus fixed
  (upper right). Note that the influence of $N_H$ is stronger in case
  of a power law than for a bremsstrahlung spectrum as evident in the
  smaller deviations from the one-to-one correlation for the
  bremsstrahlung temperature. The lower panels show the distributions
  of $\Gamma$ (lower left) and $kT$ (lower right) for the case of
  $N_H$ as a free fit parameter. The power law slopes are mainly in
  the range from 1.4 to 2.5, the canonical values for X-ray binaries and
  AGN.}
  \label{fig:spec_prop}
\efige

As expected, fits with a fixed low column density produce smaller
photon indices or larger bremsstrahlung temperatures, respectively.
However, changes of temperature and photon index between fits with
fixed and free column density are not significantly larger than the
errors in the majority of cases. Comparing the difference between
photon indices divided by the square root of the errors shows that
$\sim$80\% of the sources have values of less than 3. For thermal
bremsstrahlung model the corresponding value is 73\%. Thus the
assumption of a general power law spectrum with $\Gamma = 2$, and
Galactic absorption is quite good \citep{grimm:05}.

\bfige[t]
  \resizebox{0.32\hsize}{!}{\includegraphics{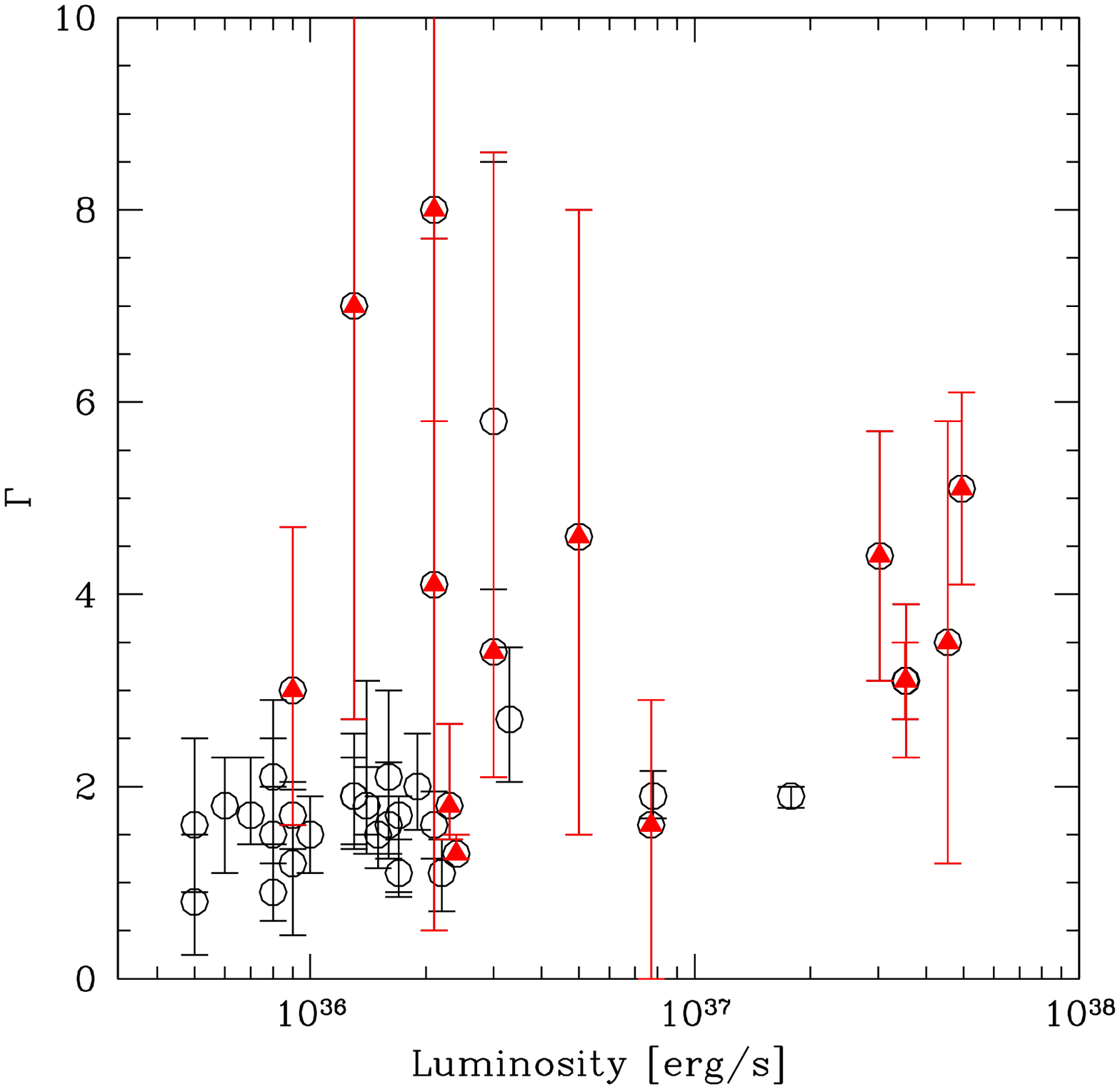}}
  \resizebox{0.32\hsize}{!}{\includegraphics{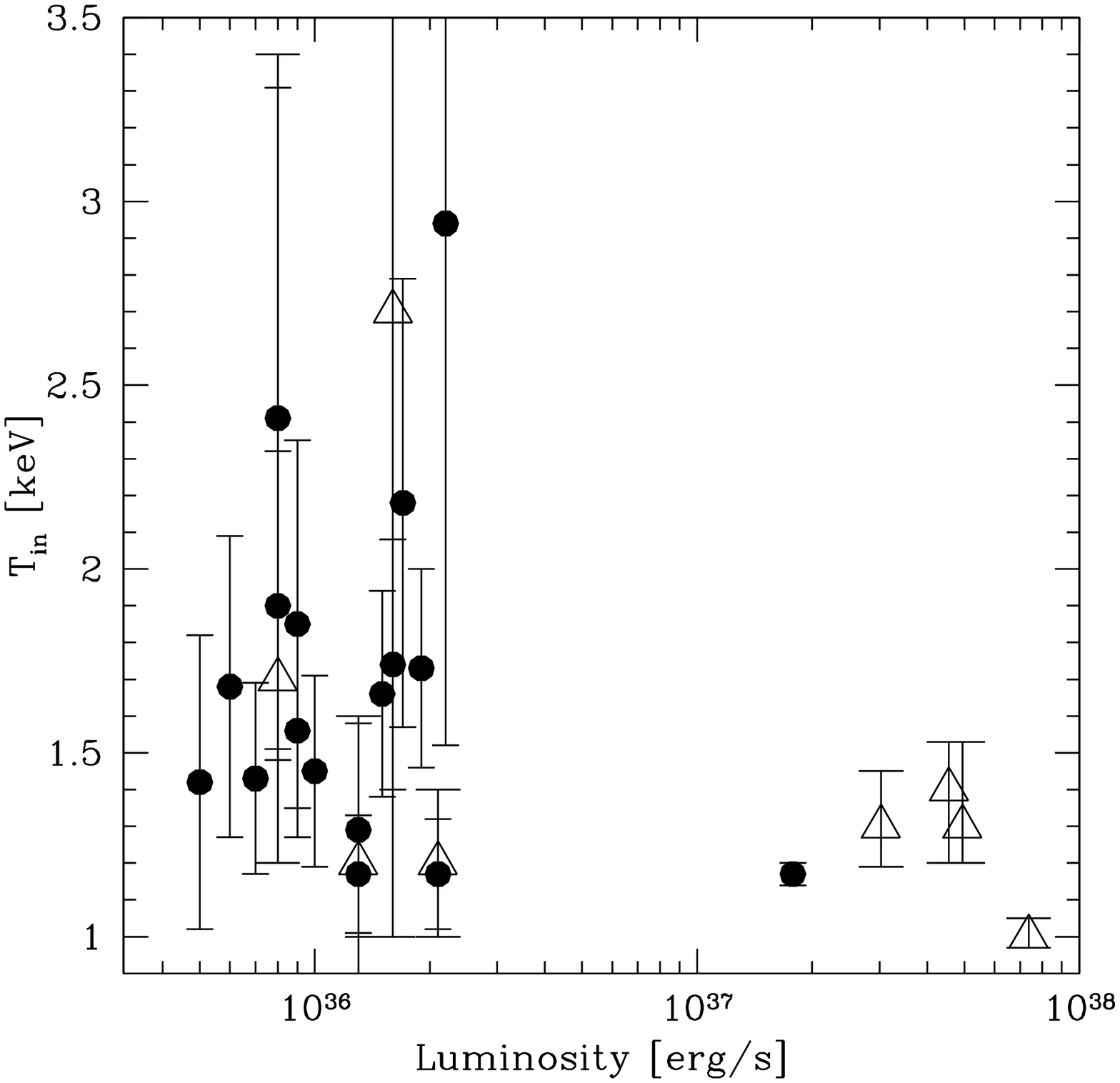}}
  \resizebox{0.32\hsize}{!}{\includegraphics{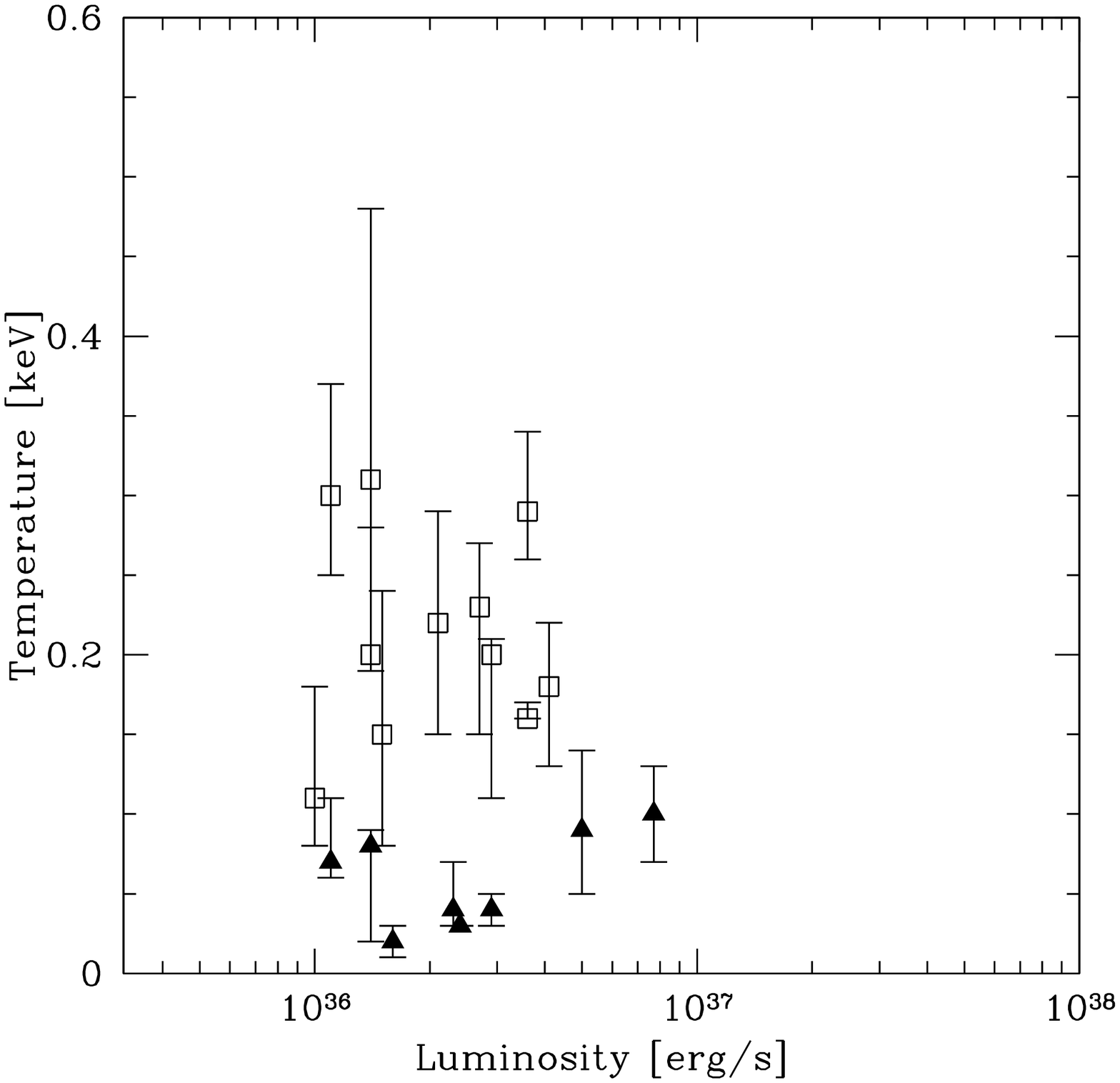}}
  \caption{The left panel shows the best fit photon indices for
  sources that are well fit by a single power law (open circles) or
  a multi-component spectrum containing a power law (open circles with
  filled triangles) versus luminosity. Two thirds of the single power
  law sources are also well fit by a disk black body model. The
  middle panel shows the disk black body temperatures versus
  luminosity for sources that are well fit by a single power law,
  but also by a simple disk black body model (filled circles). As
  expected for X-ray binaries the disk temperatures are in the range
  from 1--3 keV. The temperatures are also in the same range as the
  temperatures for sources that require a disk black body component in
  the spectrum (open triangles). The right panel shows temperatures of
  sources that are either well fit by a bremsstrahlung model (open
  squares) or a black body model (filled triangles) versus
  luminosity. With the exception of two spectra, the bremsstrahlung
  sources are also well fit by a thermal plasma model, XSPEC model
  {\it apec}, as discussed in the text. Errors are 90\% confidence
  level.}
  \label{fig:slopes}
\efige

Using the results of the more detailed spectral fitting of the 43
bright sources, we can confirm the results of the relatively blind
spectral fitting of all sources with a power law model. The left panel
of Fig. \ref{fig:slopes} shows the photon indices of spectra that are
well fit by either a single power law (open circles) or a combination
of a power law and other components (filled triangles) versus
luminosity. Errors are 90\% errors on the slope. There is a clear
trend that photon indices in multi-component fits are larger than for
single power law fits, and there is an indication that brighter
sources are more likely well fit by multi-component fits than by
single power laws. The reason is most likely that the single power law
sources have lower counts and multiple components are not
distinguishable. We also fit the single power law sources with a disk
black body model, XSPEC model {\it diskbb}, which has the same number
of degrees of freedom than the power law model. Two thirds of the
spectra are well fit by a disk black body model, the best fit
temperatures being in the range from 1 to 3 keV, as expected for X-ray
binaries \citep{tanaka:01}. The inner disk temperatures versus
luminosity are shown in the middle panel of Fig. \ref{fig:slopes}. The
values are the same as for the sources that require disk black bodies
for a good fit. This suggests that the single power law sources are
likely to be X-ray binaries and only the low number of counts allows a
good fit with a single power law.

There are 10 spectra (5 sources) that are best fit with a single
bremsstrahlung model with temperatures of $\sim$0.1--0.3 keV. The
bremsstrahlung temperatures versus luminosity are shown in the right
panel of Fig. \ref{fig:slopes} (open squares). Since these values are
unusually low for real bremsstrahlung sources, and the luminosities of
the sources relatively low, $1-5 \times 10^{36}$ erg s$^{-1}$, we
fit these spectra with a thermal plasma model (XSPEC model {\it apec})
as well. With one exception we obtain good fits with the plasma model
as well, the temperatures are in the range from 0.1 keV to 0.4 keV,
and metallicities from 0.05 to 0.3 solar with considerable
uncertainties. These values are at or below the expected metallicity
for M33 \citep{blair:85}. The good fit quality is not surprising
considering that the {\it apec} model has one degree of freedom more
than the simple bremsstrahlung model. However, given the low
temperatures and the consistent metallicities we consider the {\it
apec} model to be the more physical model for these sources. The
sources with good fits for the {\it apec} model are marked in Table
\ref{tab:spectra} (Appendix \ref{sec:spectra}) and the {\it apec}
model parameters are given, in Fig. B1 (Appendix
\ref{sec:spectra}) the contour plots for these fits are shown as
well. Also shown in the right panel of Fig. \ref{fig:slopes} are six
sources well fit by a black body model with very low temperatures at
or below 0.1 keV (filled triangles). For two sources the black body is
the only component in the spectrum. These sources are candidates for
super-soft sources and are discussed below in Sec. \ref{sec:sources}.
Three of the other four sources have an additional power law
component, that ranges from hard ($\Gamma\sim1.3$) to very soft
($\Gamma\sim4.6$). The black body temperatures are even lower than the
temperatures inferred from ULX intermediate mass black hole
candidates, see e.g. \citet{miller:04}. A truncated disk would be a
possibility to explain the low temperature, similar to the scenario
suggested by \citet{kubota:04}. However, given the luminosities the
sources would be in the hard state (e.g. \citet{maccarone:03}) but,
except for \object[]{CXO J013444.6+305535}, the photon indices are
larger than 2. \object[]{CXO J013444.6+305535} could indeed be in the
hard state with a photon index of 1.3 but that number is not well
constrained. This hard source could also be a candidate for a magnetic
CV that are known to have hard spectra and relatively high
luminosities \citep{kuulkers:06}. The soft power law in the other
sources might on the other hand be another thermal
component. Alternatively, the soft emission might be generated in an
outflow from the system, or heating of a surrounding medium.

Fig. \ref{fig:nh_comp} shows a comparison of measured values for
column density overlaid on a 1.49 GHz contour map of the central part
of M33 from VLA \citep{condon:87}. The resolution of the VLA map is
about 1 arcminute, and the confusion limit is given as 0.1 mJy. There
is no spatial correlation between the contour map and the magnitude of
the measure column densities. Thus the X-ray absorption value is in
parts due to location of sources in front of/behind HI gas, and
in parts due to intrinsic absorption around the X-ray source.

\bfige
  \resizebox{\hsize}{!}{\includegraphics[]{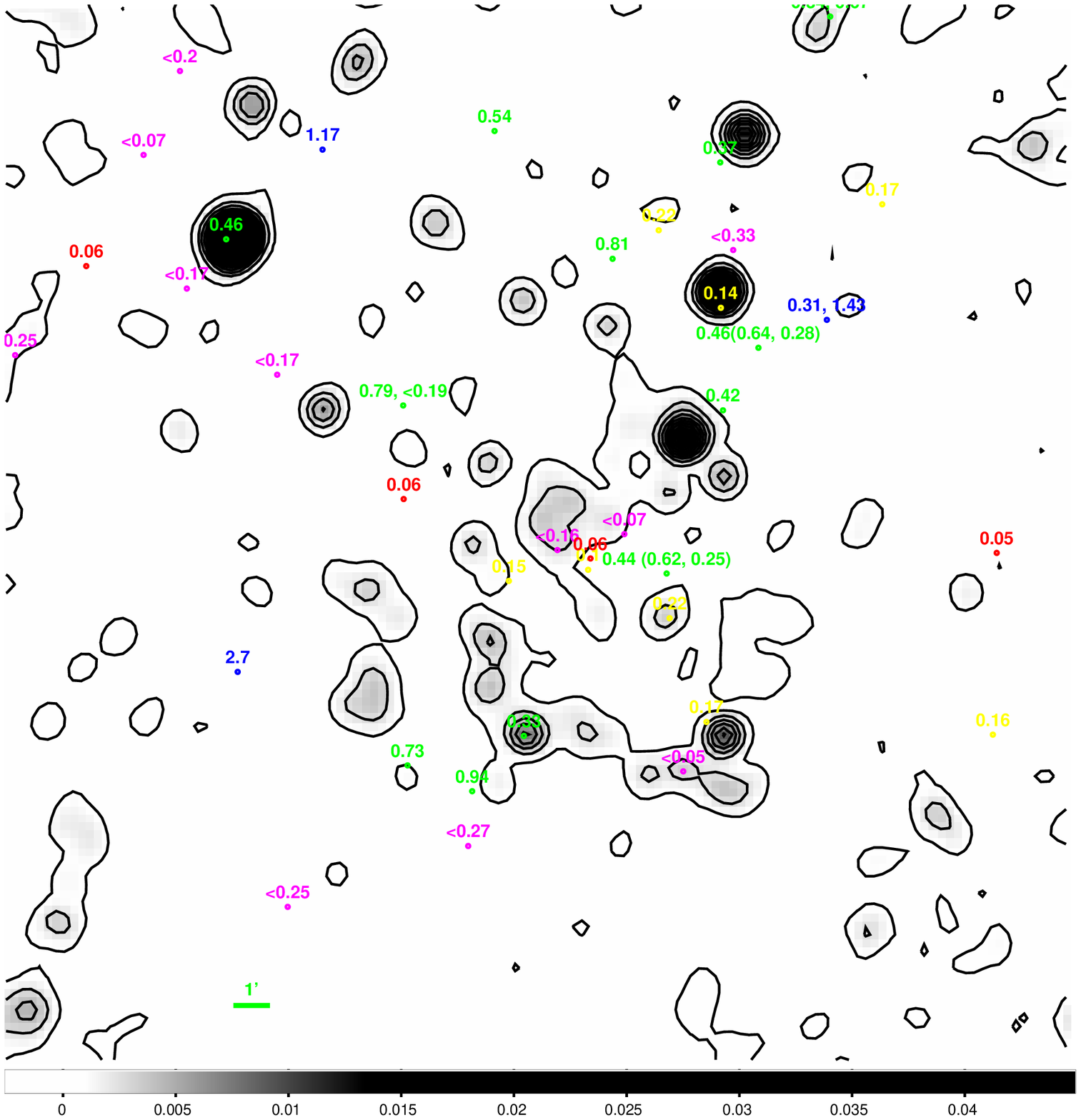}}
  \caption{Measured values for column density overlaid on a 1.49 GHz
   contour map of the central part of M33 from VLA \citep{condon:87}.
   The resolution of the VLA map is about 1 arcminute, and the RMS
   is about 0.1 mJy. Prominently visible is the HII region NGC 604 at
   the upper left. Also visible towards south of the center is the
   southern spiral arm. There is no correlation between the contour
   map and the magnitude of the measure column densities.}  
  \label{fig:nh_comp}
\efige

To compare actual values for the column density due to HI, we compute
the brightness temperature of the HI gas according to
\be
  S_{\nu} = \frac{2k\nu^{2}\Omega}{c^2}T_b,
\ee
with $S_{\nu}$ the radio flux in units of Jy/steradian at frequency
$\nu$ (1.49 GHz), $\Omega$ the opening angle of the beam (1'), and
$T_b$ the brightness temperature; $k$ is the Boltzmann constant, $c$
the speed of light. Assuming optically thin emission we compute the HI
column density according to
\be
  N_H = 1.83 \times 10^{18} T_b [\rm{\,K\, cm}^{-2}].
\ee
Within the size of a beam the radio flux has an RMS of $\sim$5 mJy
around the zero point. Therefore we take 5 mJy as the upper limit on
the sensitivity in a beam. This flux corresponds to $\sim$320 K or
$N_H$ of $5.7\times 10^{20}$ cm$^{-2}$, which is very close to the
Galactic absorption value. Fig. \ref{fig:nh_nh} shows the comparison
between the X-ray absorption values from spectral fitting and the
``expected'' absorption from HI.
\bfige
  \resizebox{\hsize}{!}{\includegraphics[]{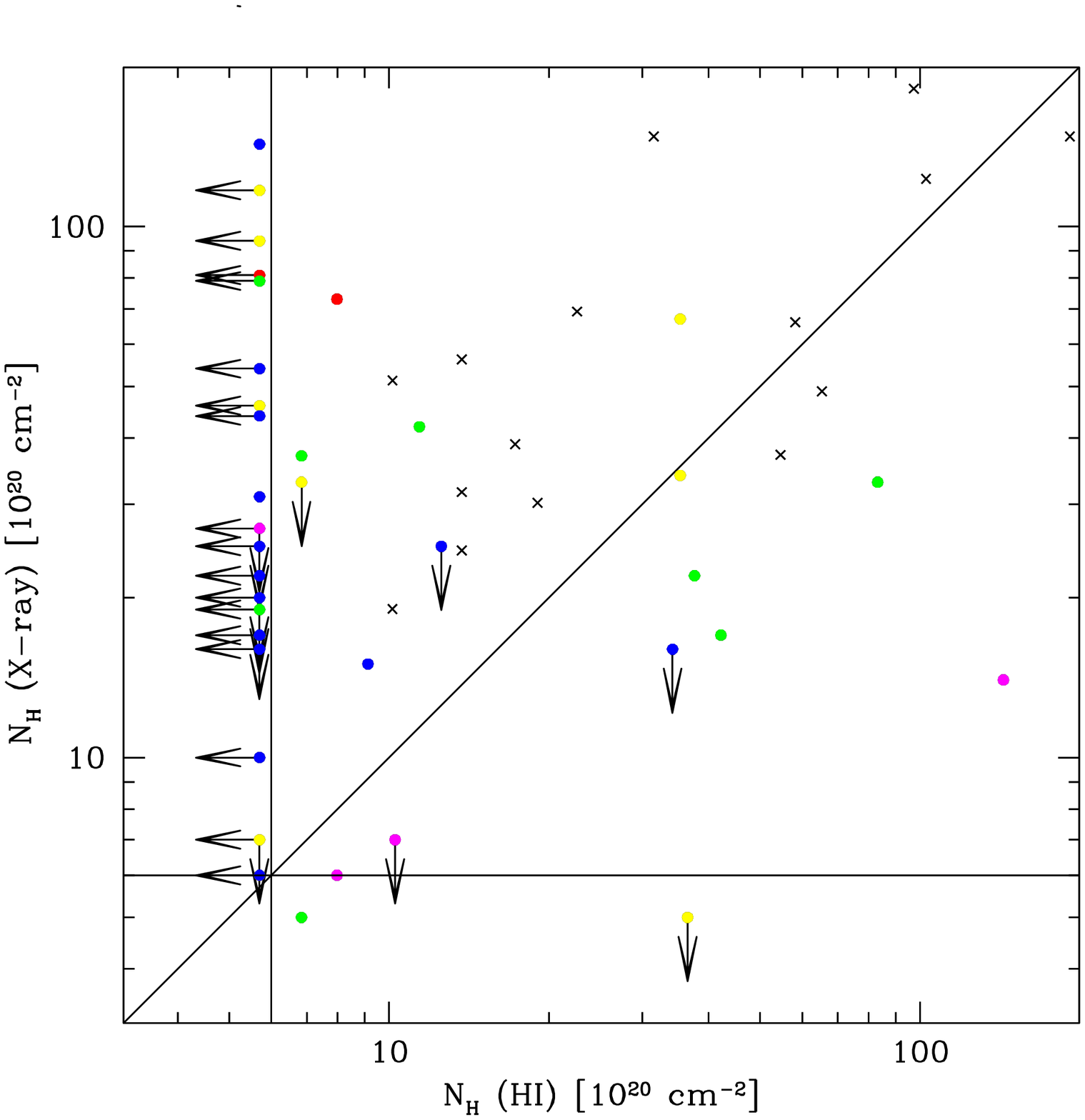}}
  \caption{Measured values for column density from X-ray spectral
   fitting versus absorption values inferred from HI radio
   flux. Upper limits are shown as arrows. Vertical and horizontal
   lines denote the Galactic absorption value. Filled circles are
   sources in M33, crosses are Galactic X-ray binaries, see text. The
   plot can be separated in three regions: 1. At/below the horizontal
   line, sources with apparently {\it less} absorption than the 
   Galactic value. 2. Below the diagonal line, sources are probably
   located in front of at least parts of the HI emission. 3. Above the
   diagonal line, sources are also intrinsically absorbed.}
  \label{fig:nh_nh}
\efige
Filled circles are sources in M33, crosses are a selection of Galactic
X-ray binaries from \citet{vrtilek:91}. Upper limits are shown as
arrows. The vertical and horizontal lines denote the Galactic
absorption value. The figure can be separated in roughly three
regions: 1. Sources at or below the horizontal line, i.e. sources with
apparently {\it less} absorption than the Galactic value, which
probably have an undetected thermal component in the spectrum.
2. Sources below the diagonal line which are probably located in front
of at least parts of the HI emission. These sources are most likely on
the Milky Way facing side of M33. 3. Sources above the diagonal line
are intrinsically absorbed. No conclusion as to their location can be
drawn.

Values for Galactic X-ray binaries are taken from \citet{vrtilek:91}
and \citet{liu:01} for $A_V$ values missing in the former. We convert
the optical extinction to neutral hydrogen column density according to
\citet{predehl:95} $N_H = 1.77 \times A_V$ in units of $10^{21}$
cm$^{-2}$, and use $A_V = 3\times E(B-V)$ in cases where only the
reddening in available. The comparison with Galactic X-ray binaries
shows that M33 X-ray sources also often show excess absorption at
levels comparable with Galactic X-ray binaries. The larger values for
$N_H$(HI) in the Milky Way can be explained simply by our location in
the Milky Way, compared to the nearly face-on orientation of M33.

Only six of the 25 sources observed in more than one observation with
more than 100 counts show significant variability in the hardness
ratio diagram, see Appendix \ref{sec:spectra}. Three of the sources
are the well known sources, M33 X-4, M33 X-5, and M33 X-13. The other
three are \object[]{CXO J013329.0+304216}, \object[]{CXO
J013329.3+304508}, and \object[]{CXO J013410.5+303946}. There is a
clear separation in that most sources have a very small significance
for moving in the hardness ratio diagram, whereas four of the ones
that show changes in hardness ratio are above 99.99\% confidence
level.

Based on spectral shape (photon indices, temperatures) and absorption
properties we conclude that M33 X-ray sources as a population are very
similar to Galactic X-ray binaries.

\subsection{Individual sources}
\label{sec:sources}

Here we discuss the properties of individual bright M33 sources. These
include previously known sources (M33 X-...) and sources newly
discovered in our \chandra observations. Wide band spectral studies
exist in the literature for some of these sources. In particular,
BeppoSAX has observed the brightest sources in M33 and measured
spectra of the sources M33 X-4 to M33 X-10 in the energy range from
2.0-10.0 keV \citep{parmar:01}. Of these sources X-6 and X-10 are not
in the field of view of the \chandra observations, and X-8 is the
nucleus which is not discussed here.

{\bf M33 X-4 (\object[]{CXO J013315.1+305317})}: The X-ray spectrum of
this soft source measured by BeppoSAX is well fit by either a power
law model with photon index of $\sim$3 or a bremsstrahlung model with
$kT\approx2$ keV. This source has been observed by \chandra twice, in
ObsIds \dataset[ADS/Sa.CXO\#obs/00786]{786} and
\dataset[ADS/Sa.CXO\#obs/02023]{2023}. Between the observations, the
flux increased by a factor of two from $5\times 10^{37}$ erg s$^{-1}$
to $10^{38}$ erg s$^{-1}$. The variability rules out the
interpretation of the source as a SNR \citep{okada:01}. Using the
overlapping energy range with the BeppoSAX observations, 2.0--8.0 keV,
we obtain the same values for absorption, photon index, and
bremsstrahlung temperature well within the errors and $\chi^2$-values
below 1 for observation \dataset[ADS/Sa.CXO\#obs/00786]{786}. For observation
\dataset[ADS/Sa.CXO\#obs/02023]{2023}, although the best fit values are
consistent with the BeppoSAX and \dataset[ADS/Sa.CXO\#obs/00786]{ObsId
786} values, the $\chi^2$-values are well above 2. Moreover, for the
whole \chandra range, 0.3--8.0 keV, none of these models is a good
fit. A statistically satisfactory fit for both ObsIds can be achieved
with a model consisting of a black body and a power law. The black
body temperatures ($\sim$0.65 keV) and photon indices (3-4.7) for the
two observations are consistent within the errors. However, the
absorption to the source changes. Taking into account the different
values for absorption there is a clear distinction at 99\% confidence
level in the contour plots, for $\Gamma$ versus $N_H$ and kT versus
$N_H$. The difference in $\Gamma$ versus kT is significant only at
90\% confidence level. The black body temperature is relatively low,
and does not change significantly, but similar values for disk black
body temperatures have been observed in Galactic BHs. These
temperatures occur at luminosities of $\sim$1--5\% of the Eddington
luminosity \citep{gierlinski:04}. With a luminosity increasing from
$\sim5\times 10^{37}$ erg s$^{-1}$ to $10^{38}$ erg s$^{-1}$, this
interpretation would make the compact object in X-4 a rather massive
but still stellar mass BH given the luminosity. The photon index of
the power law component is much softer than expected for a low-hard
state black hole \citep{tanaka:01} but is in agreement with high-soft
state Galactic black holes \citep{mcclintock:06}. The change in the
photon index between the observations (from $\sim$3 to 4.7) does not
indicate a state change from low-hard to high-soft. Even a photon
index of 3 is too large for the low-hard state ($\Gamma \sim$1.7). The
photon index is also in the range observed for the Galactic XRB Sco
X-1 which has a neutron star primary \citep{bradshaw:03}, but Sco X-1
does not exhibit a soft component like M33 X-4 \citep{kahn:84}. Thus
based on the X-ray data the source is likely to be a BH binary.

The hardness ratio analysis shows that the hardness of the source
varies significantly between the observations. The source becomes
harder in HR1 and softer in HR2, while the luminosity increases by a
factor of $\sim$2. This change is due to the increase in the medium
band (1.0--2.1 keV) with respect to the soft and hard band, which is
reflected in the spectral fits by a higher absorption column
and a lower power law photon index.

\begin{figure}
\caption{X-ray spectra and contour plots for best fit model. Models
and parameters are shown in Table \ref{tab:spectra}.\label{fig:spectra}}

\begin{minipage}[h]{\linewidth}
\resizebox{.3\linewidth}{!}{\includegraphics[angle=-90]{f9_1a.eps}}
\resizebox{.3\linewidth}{!}{\includegraphics[angle=-90]{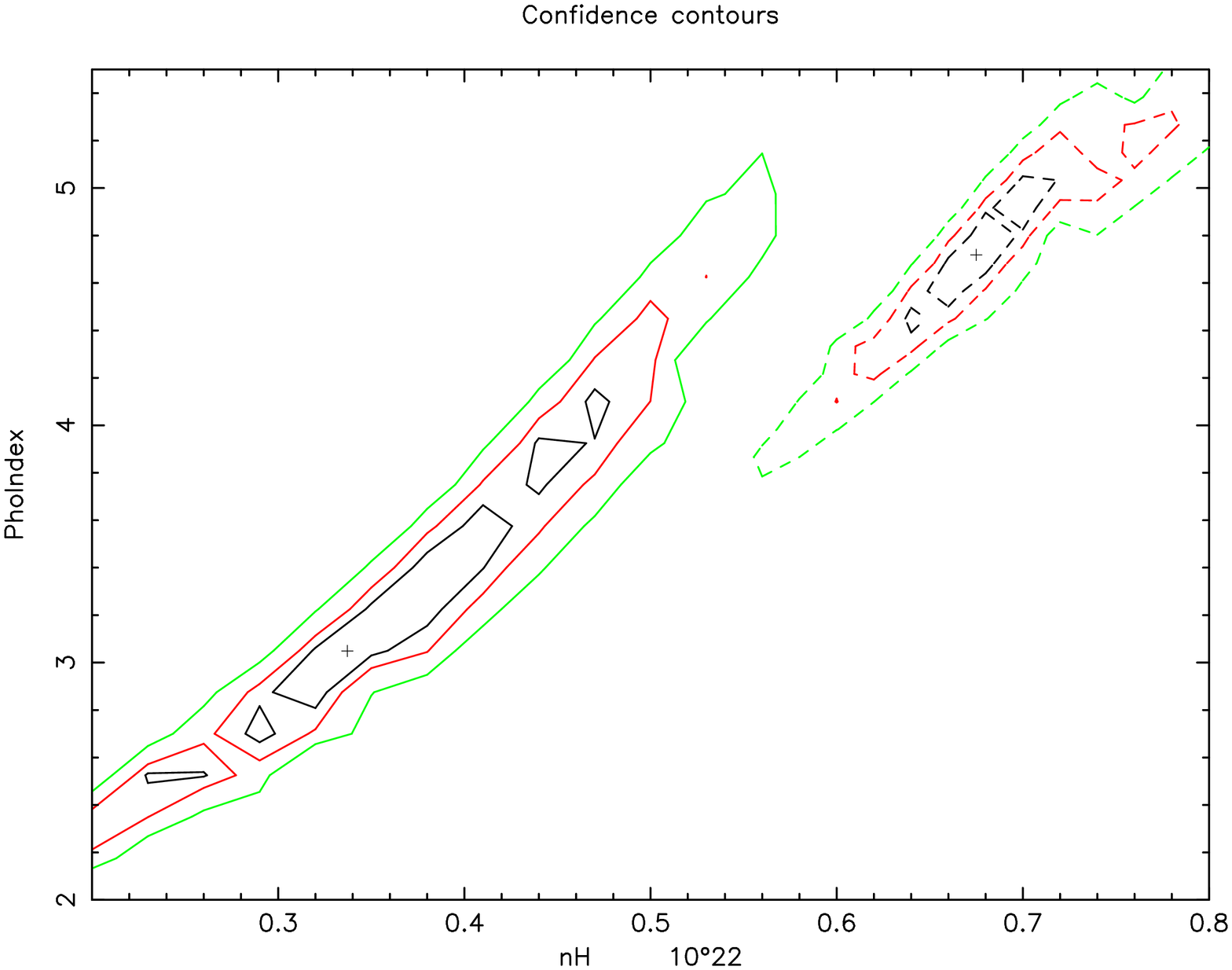}}
\resizebox{.3\linewidth}{!}{\includegraphics[angle=-90]{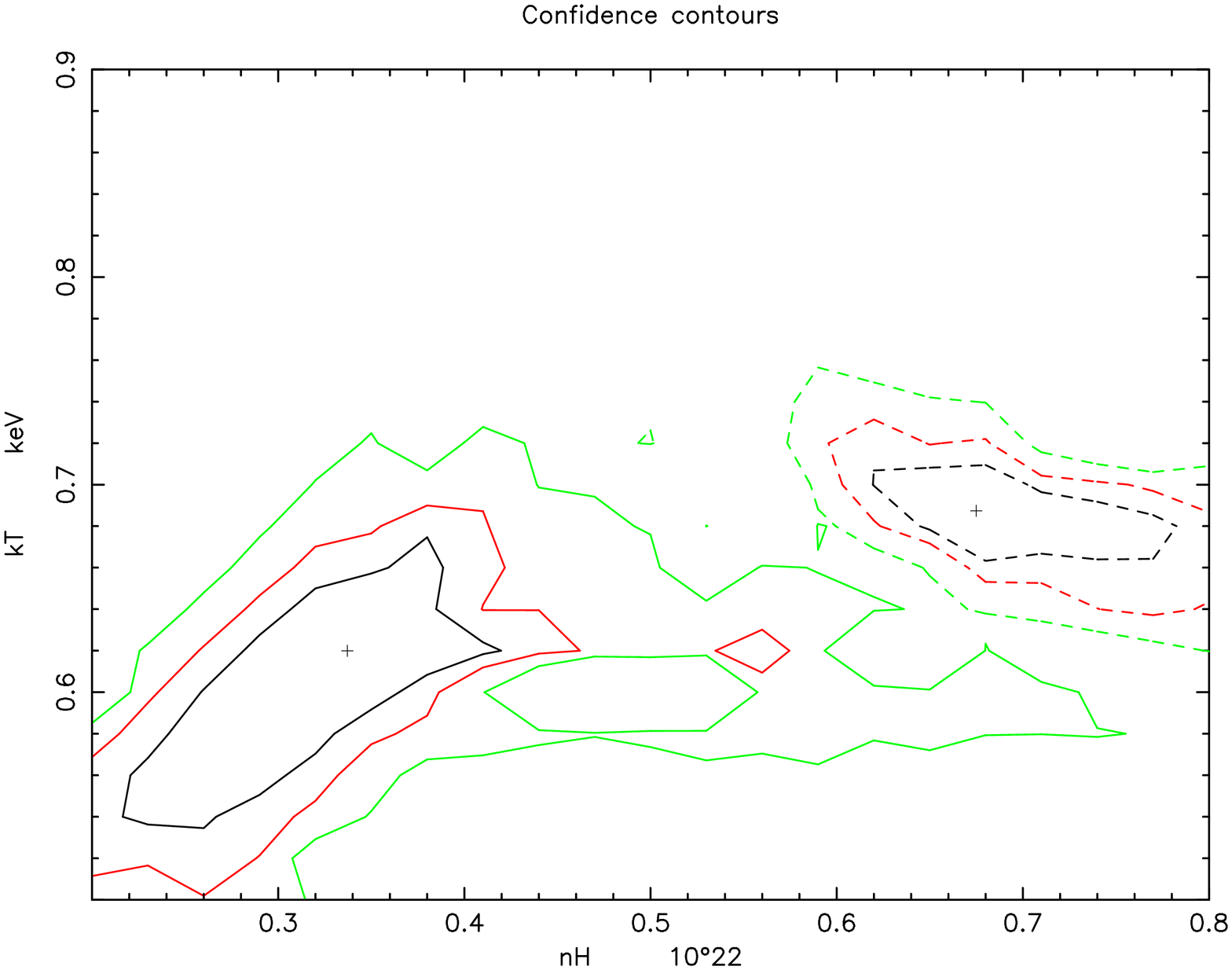}}
\end{minipage}

\begin{minipage}[h]{\linewidth}
\resizebox{.3\linewidth}{!}{\includegraphics[angle=-90]{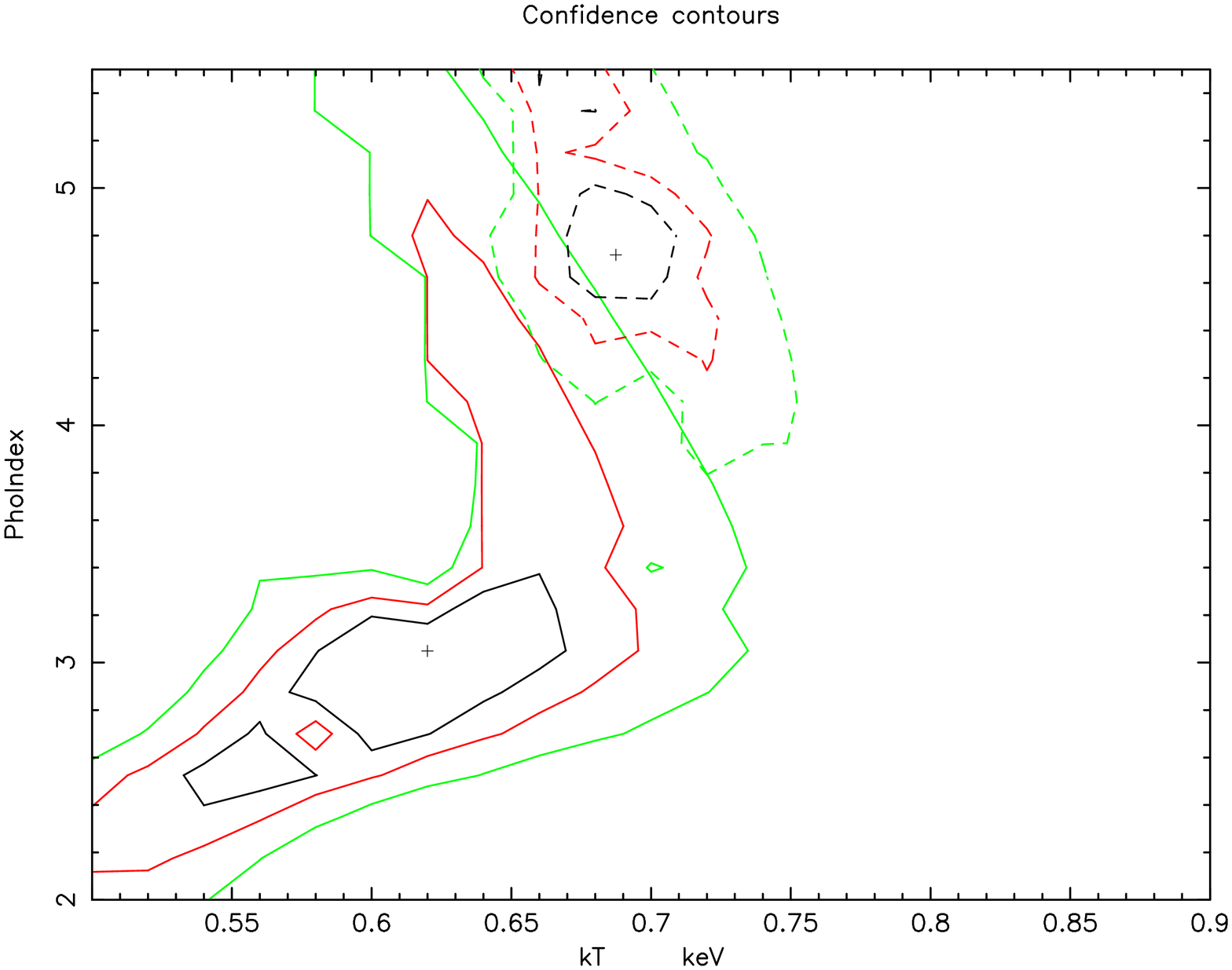}}
\end{minipage}

\begin{minipage}[h]{\linewidth}
\resizebox{.3\linewidth}{!}{\includegraphics{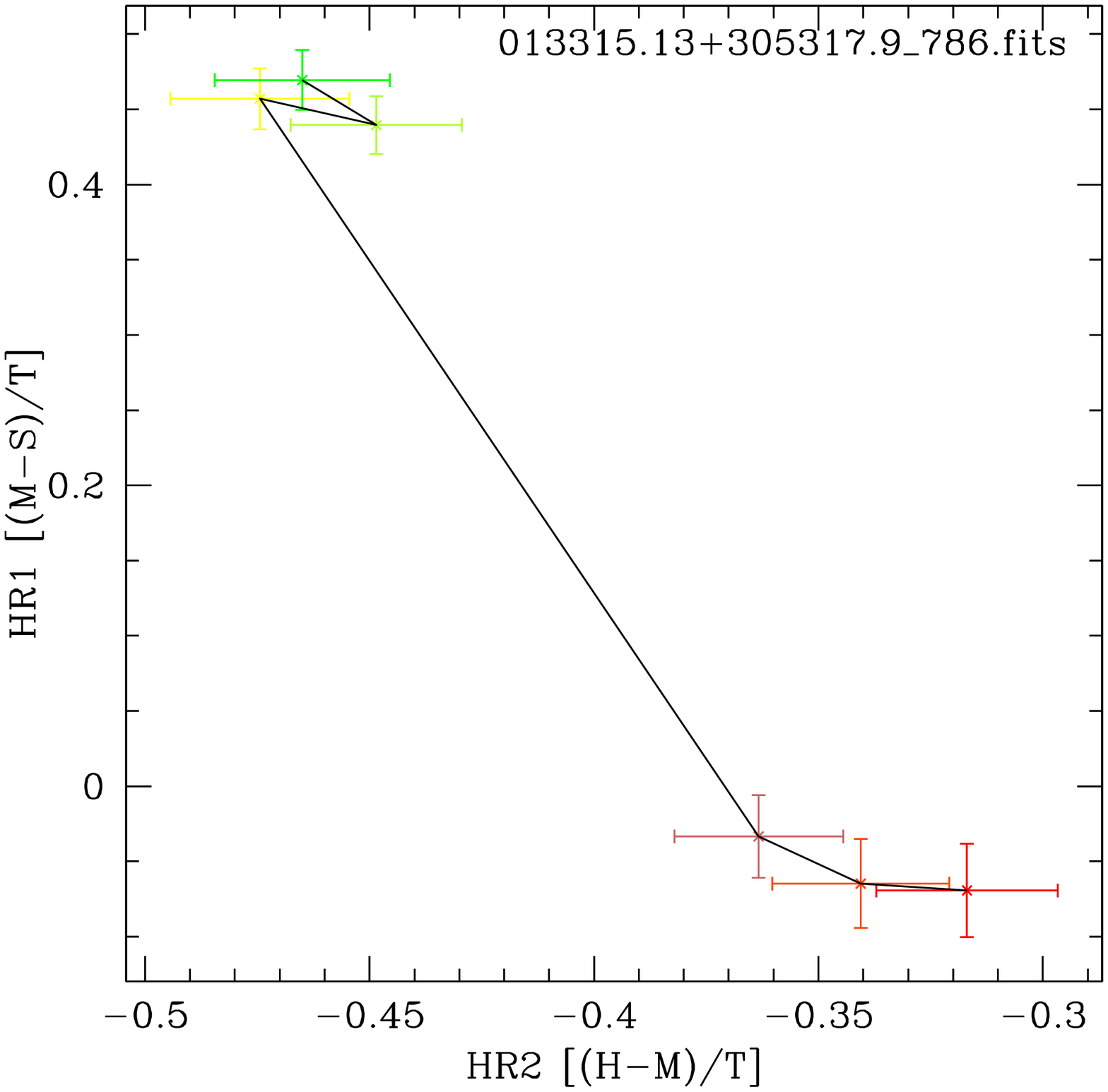}}
\resizebox{.3\linewidth}{!}{\includegraphics{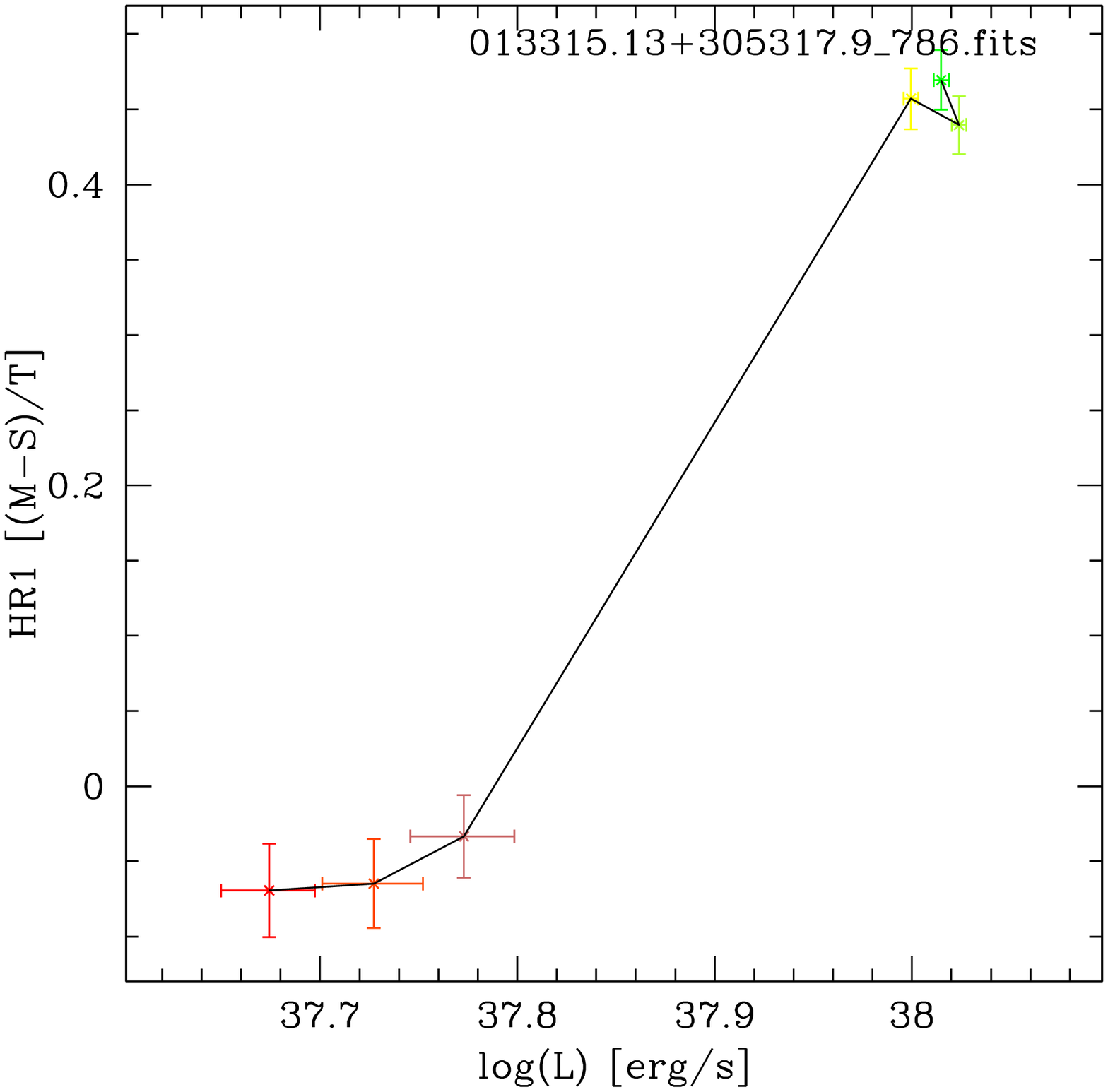}}
\resizebox{.3\linewidth}{!}{\includegraphics{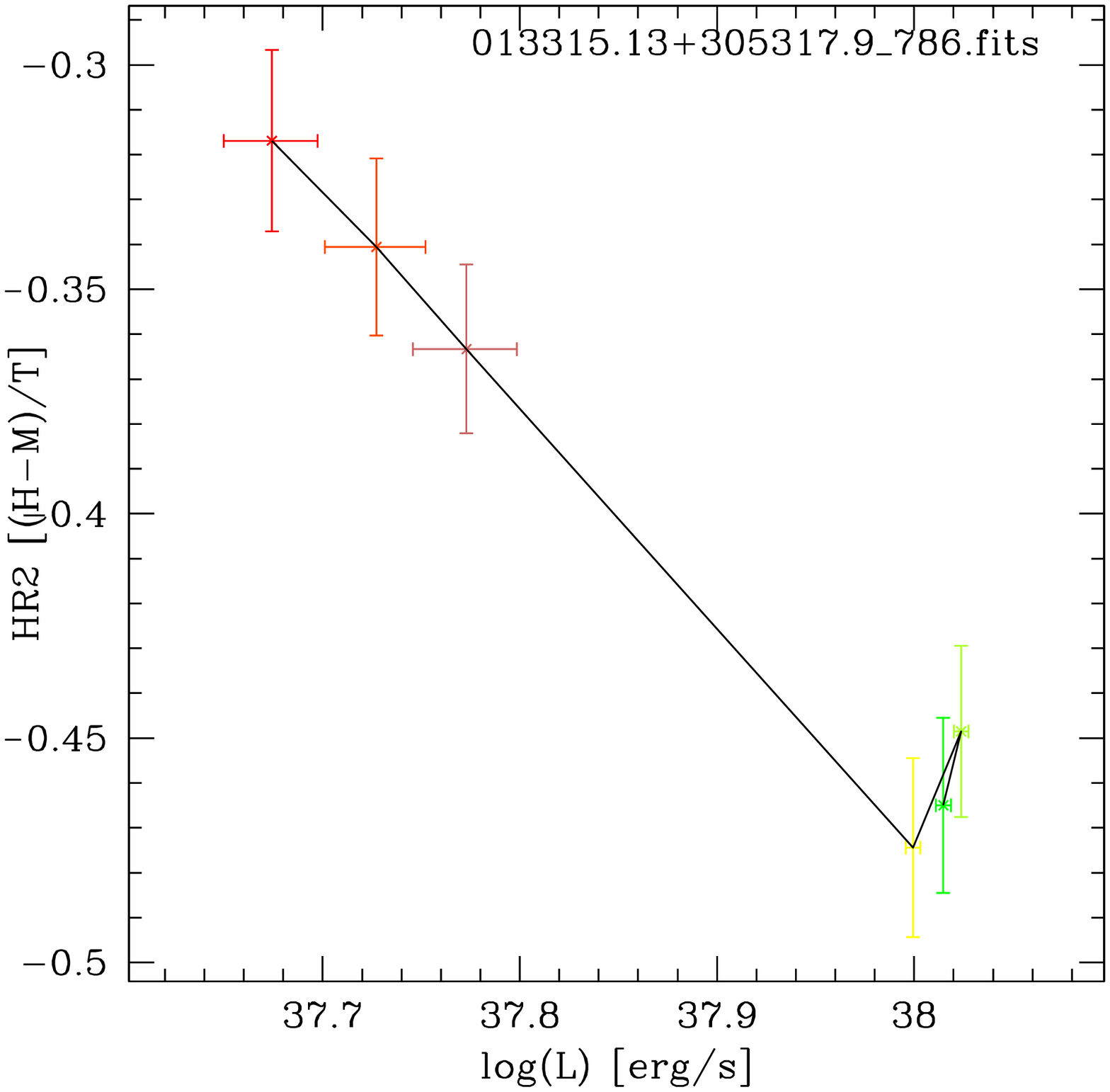}}
\end{minipage}

\begin{minipage}[h]{\linewidth}
\resizebox{.3\linewidth}{!}{\includegraphics[angle=-90]{f9_2a.eps}}
\resizebox{.3\linewidth}{!}{\includegraphics[angle=-90]{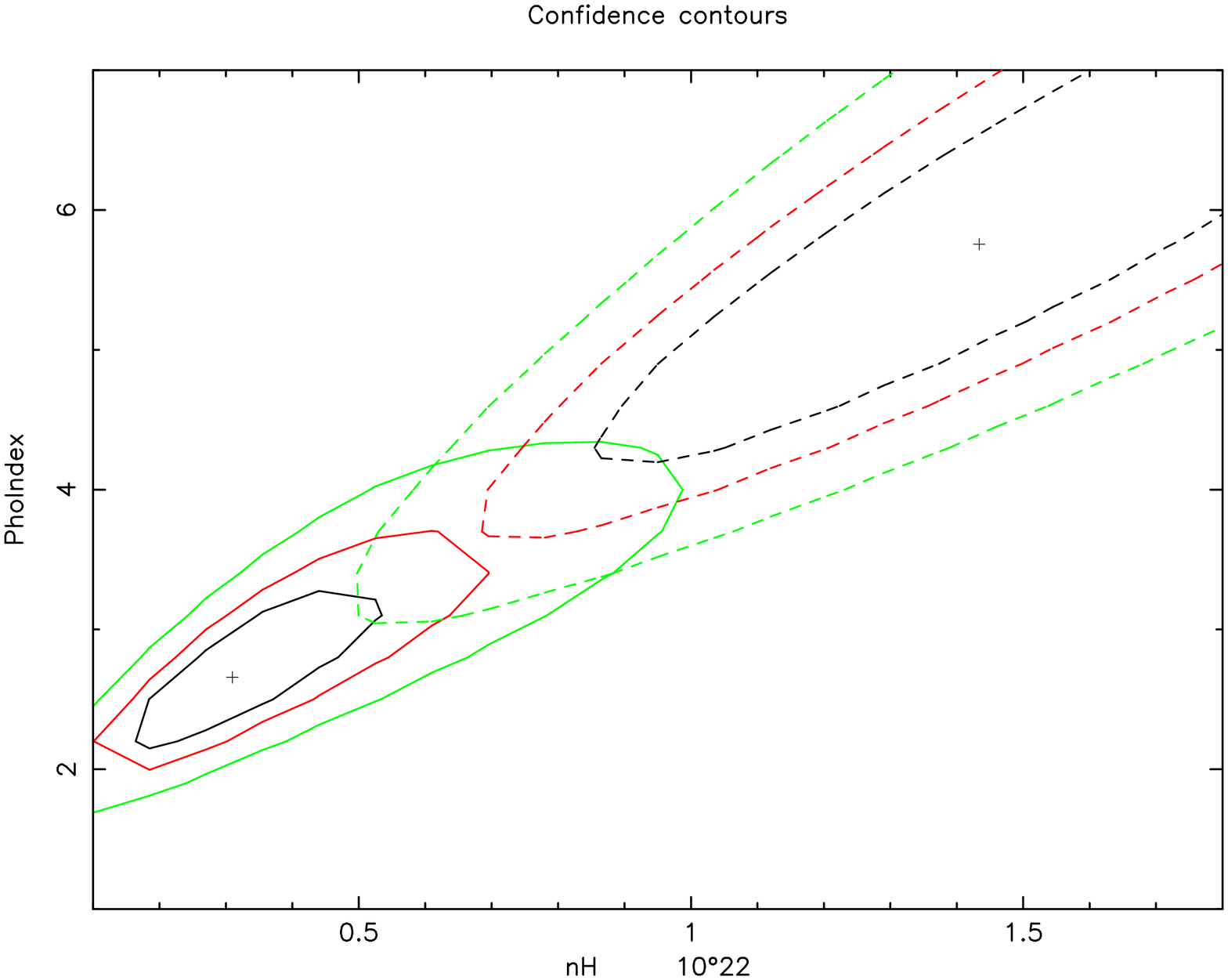}}
\end{minipage}

\begin{minipage}[h]{\linewidth}
\resizebox{.3\linewidth}{!}{\includegraphics[angle=-90]{f9_3a.eps}}
\resizebox{.3\linewidth}{!}{\includegraphics[angle=-90]{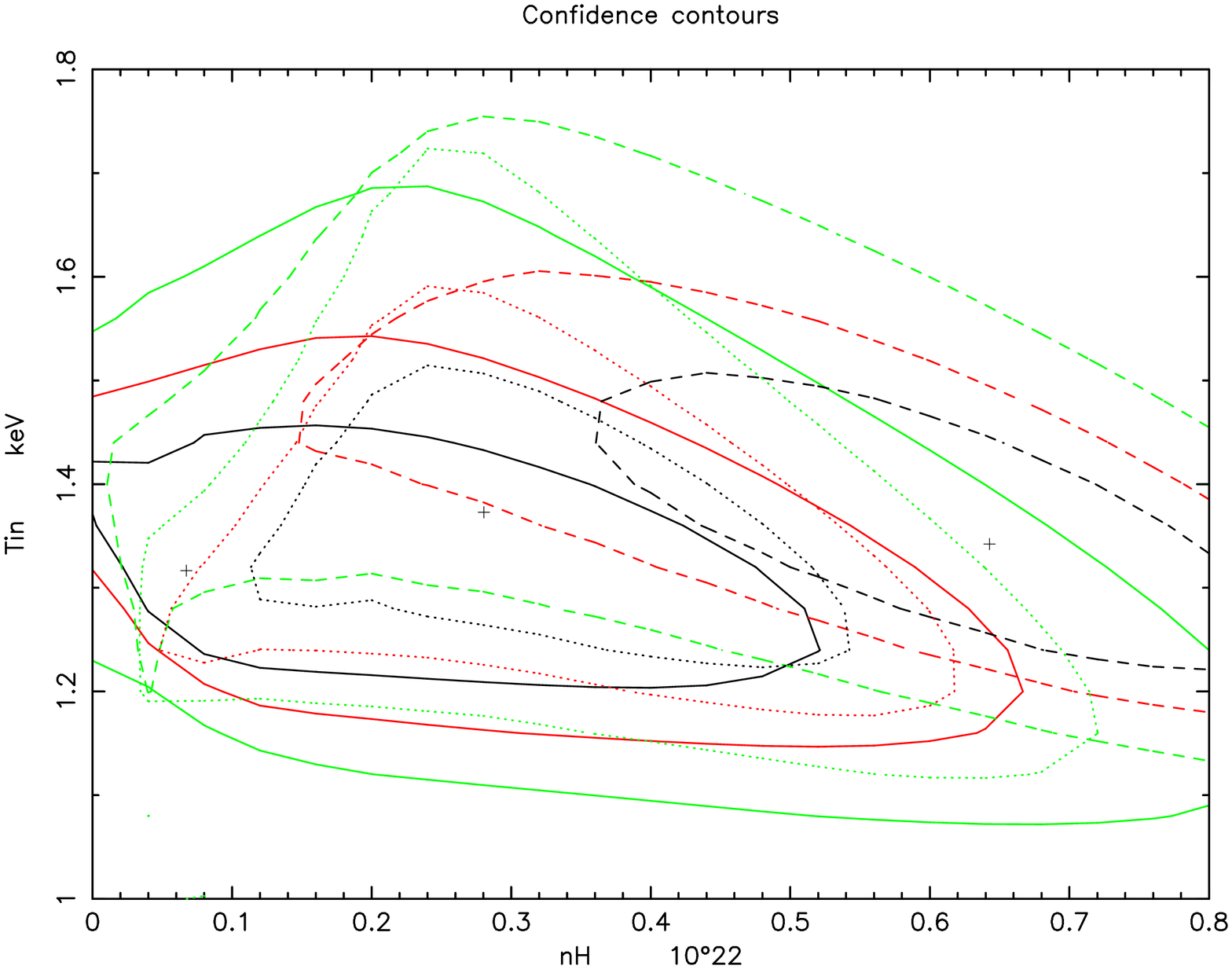}}
\end{minipage}

\end{figure}

{\bf M33 X-5 (\object[]{CXO J013324.4+304401})}: This source has been
observed in all three \chandra observations. For M33 X-5 (\object[]{CXO
J013324.4+304401}) BeppoSAX as well finds good fits with a soft power
law ($\Gamma\approx3$) or bremsstrahlung model ($kT\approx2.8$
keV). In all \chandra observations the overlapping energy range is fit
satisfactorily with these values. However, for the whole \chandra
energy range neither a single power law nor a bremsstrahlung spectrum
give good fits. A good fit is obtained with a disk black body
($\sim$1.3 keV) and power law ($\Gamma= 4-9$) combination. The contour
plots in Fig. \ref{fig:spectra} (Appendix \ref{sec:spectra}) show no
significant variation of the spectral parameters given the large
uncertainties. The errors on the photon index are large enough to not
make the changes between observations significant. However, a disk
black body alone does not yield a good fit. The disk black body
temperature is in good agreement with observations of Galactic BH
binaries at moderately sub-Eddington luminosities
\citep{gierlinski:04}. The luminosity of the sources changes between
$3\times 10^{37}$ erg s$^{-1}$ and $5\times 10^{37}$ erg s$^{-1}$;
thus it is significantly below the Eddington limit for a black hole
but in the range of transition luminosities from the low-hard to the
high-soft state of Galactic black holes \citep{maccarone:03}. Another
interpretation is that the source is a neutron star and the steep
power law corresponds to the harder thermal component from the neutron
star surface \citep{tanaka:01}. Due to the complexity of the combined
model only the disk temperature is well determined.

However, the hardness ratio analysis shows significant variation between
ObsIds \dataset[ADS/Sa.CXO\#obs/01730]{1730} and
\dataset[ADS/Sa.CXO\#obs/00786]{786} on the one hand and
\dataset[ADS/Sa.CXO\#obs/02023]{ObsId 2023} on the other. X-5 becomes
significantly softer in HR1 but does not change significantly in
HR2. There seems to be no correlation with luminosity as ObsIds
\dataset[ADS/Sa.CXO\#obs/00786]{786} and
\dataset[ADS/Sa.CXO\#obs/02023]{2023} have the same luminosity but
significantly different HR1 values.

{\bf M33 X-7 (\object[]{CXO J013334.1+303210})}: This source has been
observed only in \chandra observation
\dataset[ADS/Sa.CXO\#obs/01730]{1730}. For M33 X-7 (\object[]{CXO
013334.1+303210}) BeppoSAX as well finds good fits with a power law 
($\Gamma\approx2-3$) or bremsstrahlung model ($kT\approx3.7$ keV). In
the 2.0--8.0 keV energy range both models provide satisfactory (power
law) to good (bremsstrahlung) fits to the \chandra data. The photon
indices for fixed and variable absorption are larger than in the
BeppoSAX observations but still within the rather large errors. The
bremsstrahlung temperatures for fixed and variable absorption are
smaller, and in the case of fixed absorption inconsistent with the
lower limit of the BeppoSAX observation. Although the errors are quite
large this indicates that the source was softer in the \chandra
observations. Moreover, M33 X-7 was about five times brighter in the
\chandra observation ($\sim7\times 10^{37}$ erg s$^{-1}$) compared to
the time BeppoSAX observed the source ($\sim1.6\times 10^{37}$ erg
s$^{-1}$) (note that both luminosities are based on spectral
fits). The only satisfactory fit to the whole energy range of \chandra
is an absorbed disk blackbody with an inner disk temperature of $1\pm
0.02$ keV. This value is consistent with inner disk temperatures of
Galactic black holes \citep{mcclintock:06}. The column density is not
well constrained, but below $6.5\times 10^{20}$ cm$^{-2}$, putting the
upper limit of the absorption at the Galactic value. \citet{pietsch:04}
obtain the same values within the errors for an XMM observation, and
also new \chandra observations give the same spectral parameters
\citep{pietsch:06}. This might indicate the presence of an additional
soft component that due to insufficient counts presents itself as a
low absorption value. An additional soft component, however, does not
improve the fit, and its parameters are not well
determined. Fig. \ref{fig:x7} shows the spectrum and the contour plot
of absorption column density versus inner disk temperature. Given the
softer spectrum in the \chandra observation, a possible additional
soft component, and the five times higher luminosity, it is possible
that X-7 underwent a state transition. On the other hand the photon
index of the power law in the BeppoSAX spectrum is already softer than
expected for a low-hard state source \citep{tanaka:01}.

\bfige
  \resizebox{0.5\hsize}{!}{\includegraphics[angle=-90]{f7a.eps}}
  \resizebox{0.5\hsize}{!}{\includegraphics[angle=-90]{f7b.eps}}
  \caption{Energy spectrum and contour plot for M33 X-7. The model spectrum
  is an absorbed disk blackbody with an inner disk temperature of
  $1\pm 0.02$ keV and $N_H < 6.5\times 10^{20}$ cm$^{-2}$. The
  reduced $\chi^2$ is 1.2.}
  \label{fig:x7}
\efige

{\bf \object[]{CXO J013343.4+304630}} and {\bf \object[]{CXO
J013409.9+303219}}: These sources are the best candidates for
super-soft sources (SSS) in M33. \object[]{CXO J013343.4+304630} is
well fit by a black body spectrum with a temperature of about 75 eV
and moderate absorption (N$_H \sim 8\times 10^{21}$ cm$^{-2}$). The
source shows no significant spectral variability, but is variable on
timescales of months. \object[]{CXO J013409.9+303219} has been
observed only once. Its spectrum is also well fit by a black body with
a temperature of 37 eV and also moderate absorption (N$_H \sim
7\times10^{21}$ cm$^{-2}$). The temperatures are in the range for
SSSs, and the sources are not significantly detected above 1 keV, see
e.g. \citet{kahabka:97}. Given their luminosities of $\sim3-10\times
10^{35}$ erg s$^{-1}$ (\object[]{CXO J013343.4+304630}) and
$\sim3\times 10^{36}$ erg s$^{-1}$ (\object[]{CXO J013409.9+30321})
these sources are most likely nuclear burning white dwarfs.

{\bf M33 X-9 (\object[]{CXO J013358.8+305004})}: This source was observed in
\chandra observations \dataset[ADS/Sa.CXO\#obs/00786]{786} and
\dataset[ADS/Sa.CXO\#obs/02023]{2023}. It is well fit by a power law
with photon index of $\sim$1.2 in the BeppoSAX observation. For a
bremsstrahlung model only lower limits to the temperature are provided
($>2.6$ keV). It is important to note that in the BeppoSAX
observations X-9 is unresolved, but ROSAT observations have shown that
X-9 actually consists of 3 sources, at least two of which seem to be
variable and of comparable flux \citep{long:96}. Since BeppoSAX does
not resolve the sources, it is impossible to say if any of them
dominate the spectrum and if yes which one. The \chandra source is
associated with the source X-9a based on the ROSAT position. In the
whole \chandra energy range a power law is a good fit. The absorption
column is not well constrained in this fit, but the values for the
photon index are $\sim1.7$, with a reduced $\chi^2$ of
0.87. Fig. \ref{fig:x9} shows the spectrum and the contour plot of
absorption column density versus photon index. Assuming for the sake
of the argument that a significant fraction of the BeppoSAX flux
originated from X-9a, the source was thus significantly harder during
the BeppoSAX observation. Even if there is no spectral change in X-9
the source was then about a factor of ten brighter during the ROSAT
observations where it reached $\sim1.4\times 10^{37}$ erg
s$^{-1}$. Between the two \chandra observations the source luminosity
decreases by a factor of $\sim$2 from $1.7\times 10^{36}$ erg s$^{-1}$
to $7\times 10^{35}$ erg s$^{-1}$ without a significant change in
spectral parameters. Based on the variability, either spectral and/or
timing, and the luminosity the source is thus highly likely to be an
X-ray binary.

\bfige
  \resizebox{0.5\hsize}{!}{\includegraphics[angle=-90]{f8a.eps}}
  \resizebox{0.5\hsize}{!}{\includegraphics[angle=-90]{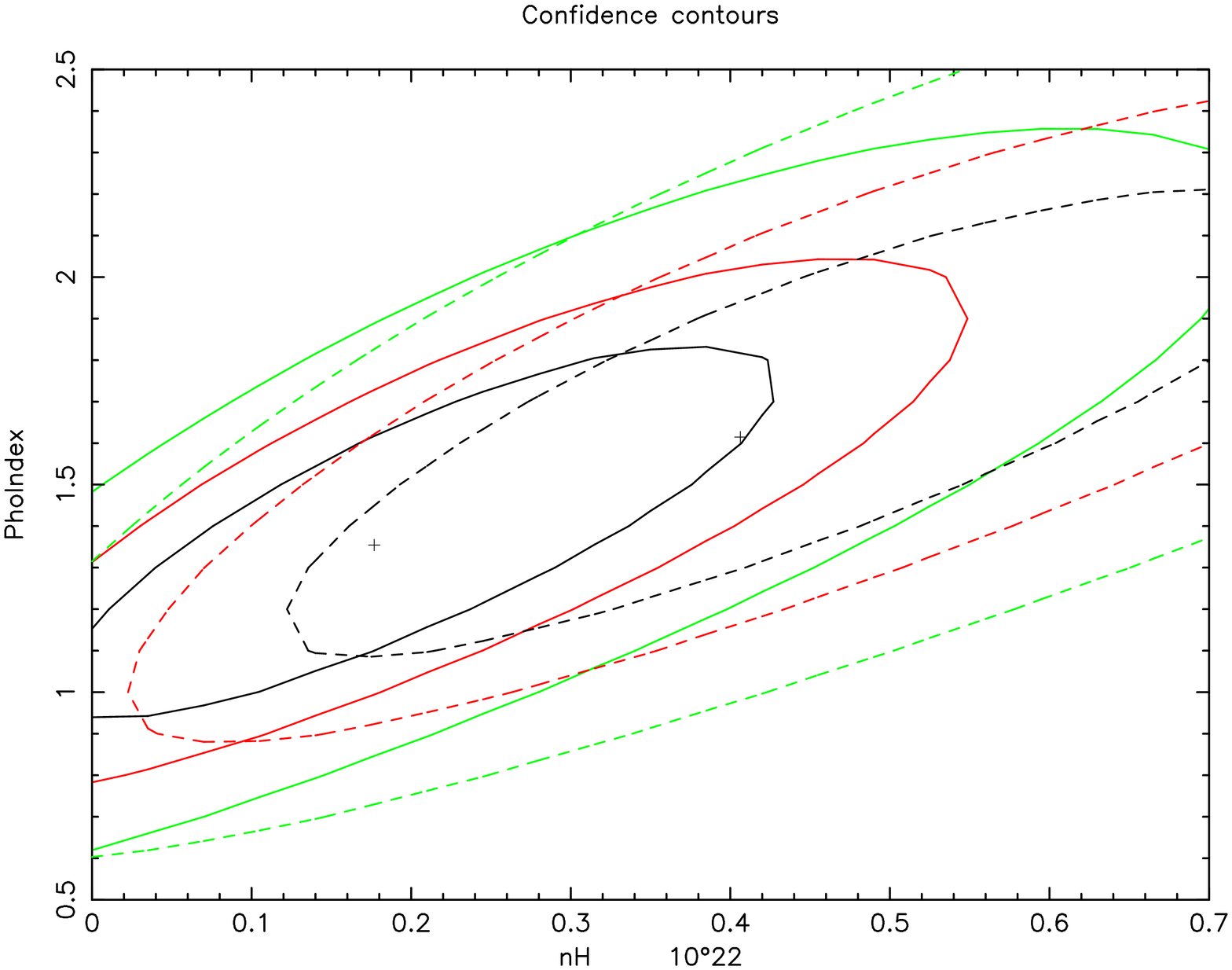}}
  \caption{Energy spectrum and contour plot for M33 X-9. The model spectrum
  is an absorbed power law with a photon index of $1.88 \pm 0.04$. The 
  reduced $\chi^2$ is 0.87.}
  \label{fig:x9}
\efige

{\bf M33 X-13 (\object[]{CXO J013354.8+303309})} and {\bf
\object[]{CXO J013329.0+304216}}: These two sources have very soft
bremsstrahlung temperatures ($\sim$ 0.2 keV) but are also variable on
timescales of months, thus excluding SNRs or HII regions as
counterparts.

M33 X-13 (\object[]{CXO J013354.8+303309}) is well fit with a
bremsstrahlung model with temperature $0.22$ keV and absorption of
$\sim3\times10^{21}$ cm$^{-2}$ in both observations. The source is
coincident with a SNR \citep{gordon:99}, but the flux varies by a
factor of two between two observations. It also shows significant
change in the hardness ratios. The source becomes softer in HR1 and
harder in HR2. The flattening of the spectrum is accompanied by an
increase in luminosity $2\times 10^{36}$ erg s$^{-1}$ to $4\times
10^{36}$ erg s$^{-1}$.

\object[]{CXO J013329.0+304216} shows a significant softening in HR1 and
hardening in HR2 from \dataset[ADS/Sa.CXO\#obs/01730]{ObsId 1730} to
\dataset[ADS/Sa.CXO\#obs/02023]{ObsId 2023}, the same behavior at
similar hardness ratios like M33 X-13 (\object[]{CXO
J013354.8+303309}). The spectrum of the source is well fit by a
bremsstrahlung spectrum with a temperature decreasing from
$0.24\pm0.01$ keV (\dataset[ADS/Sa.CXO\#obs/01730]{ObsId 1730}) to
$0.21\pm0.006$ keV (\dataset[ADS/Sa.CXO\#obs/00786]{ObsId 786}) and to
$0.19\pm0.006$ keV (\dataset[ADS/Sa.CXO\#obs/02023]{ObsId
2023}). However, the temperature is somewhat degenerate with the
column density which increases from $(0.29\pm0.1)\times10^{-22}$
cm$^{-2}$ (\dataset[ADS/Sa.CXO\#obs/01730]{ObsId 1730}) to
$(0.36\pm0.1)\times10^{-22}$ cm$^{-2}$
(\dataset[ADS/Sa.CXO\#obs/00786]{ObsId 786}) and to
$(0.42\pm0.17)\times10^{-22}$ cm$^{-2}$
(\dataset[ADS/Sa.CXO\#obs/02023]{ObsId 2023}). Although the 
column density changes systematically, the large errors make it
impossible to choose which, if any, value is correct. Thus although
the errors on the temperatures are small the significance of the
temperature change from the spectral fit is not large. Between the
\chandra observations the luminosity increases from $2.7\times
10^{36}$ erg s$^{-1}$ to $3.6\times 10^{36}$ erg s$^{-1}$.

Temperature and variability of these sources indicate that they could
be quasi-soft sources, although a simple black body does not provide a
good fit. However, more complicated models (e.g. black body plus power
law) do not improve the fits due to the small number of observed
counts. For quasi-soft sources the luminosities would be relatively
low (a few $10^{36}$ erg s$^{-1}$) but still in the range observed in
other galaxies \citep{distefano:04}. The sources are too bright ($\sim
2-4 \times 10^{36}$ erg s$^{-1}$) for normal CVs ($\sim10^{30-32}$ erg
s$^{-1}$), and magnetic CVs that can reach such luminosities have very
hard spectra with effective temperatures of several keV
\citep{kuulkers:06}. On the other hand the sources have too high
temperatures for super-soft sources \citep{kahabka:97}.

{\bf \object[]{CXO J013329.3+304508}} and {\bf \object[]{CXO
J013410.5+303946}}: The two sources show evidence for spectral
variability in their hardness ratio diagrams. Source \object[]{CXO
J013329.3+304508} shows a significant softening in HR1 but no change
in HR2 from \dataset[ADS/Sa.CXO\#obs/01730]{ObsId 1730} to
\dataset[ADS/Sa.CXO\#obs/02023]{ObsId 2023}. The best fit model for
the spectrum is a bremsstrahlung model, however the temperature is not
well determined in the fit. Moreover, the source is variable between
ObsIds \dataset[ADS/Sa.CXO\#obs/00786]{786} and
\dataset[ADS/Sa.CXO\#obs/02023]{2023}, between which the luminosity
decreases by $\sim$30\% from $3\times 10^{36}$ erg s$^{-1}$ to
$2\times 10^{36}$ erg s$^{-1}$. Source \object[]{CXO J013410.5+303946} shows a
softening in HR1 from \dataset[ADS/Sa.CXO\#obs/01730]{ObsId 1730} to
\dataset[ADS/Sa.CXO\#obs/00786]{ObsId 786} but no significant change
in HR2. From \dataset[ADS/Sa.CXO\#obs/00786]{ObsId 786} to
\dataset[ADS/Sa.CXO\#obs/02023]{ObsId 2023} the HR1 hardens again so
that the hardness ratio for \dataset[ADS/Sa.CXO\#obs/02023]{ObsId 2023}
is consistent with the hardness ratio for
\dataset[ADS/Sa.CXO\#obs/01730]{ObsId 1730}. The spectrum of ObsIds
\dataset[ADS/Sa.CXO\#obs/01730]{1730} and
\dataset[ADS/Sa.CXO\#obs/02023]{2023} is well fit by a disk black body
with an inner disk temperature 1.2 keV. The spectrum of
\dataset[ADS/Sa.CXO\#obs/00786]{ObsId 786} on the other hand is well
fit by a power law with a photon index of 1.7. Both photon index and
disk temperatures are consistent with Galactic X-ray binary spectra
\citep{tanaka:01}. It is interesting to note that the luminosity
increases between the first two observations and stays high in the
third, although the spectral shape in the first and third observation
are basically identical. The source increased its luminosity between
ObsIds \dataset[ADS/Sa.CXO\#obs/01730]{1730} and
\dataset[ADS/Sa.CXO\#obs/00786]{786} from $\sim1.3\times 10^{36}$ erg
s$^{-1}$ to $\sim2.1\times 10^{36}$ erg s$^{-1}$ and had the same high
luminosity during \dataset[ADS/Sa.CXO\#obs/02023]{ObsId 2023}.
Although the power law spectrum is relatively hard, the softening of
the hardness ratio and the increasing luminosity do not indicate a
state change in the system.

The spectral and timing behavior of the bright X-ray sources in M33 is
varied but can be well described within the framework of Galactic
X-ray sources and X-ray sources in other galaxies observed with
\chandra.

\section{Summary and Conclusion}
\label{sec:summary}

In the three \chandra observations of M33, performed between the years
2000 and 2001, 261 sources have been detected in the range of X-ray
luminosities from $L_{X}\sim 10^{34-38}$ erg s$^{-1}$ (0.3--8.0
keV). Of this total, 198 sources have been detected in at least two
observations, and 62 sources have been observed in all three
observations. We find that 49 sources show variability between
observations above 3 sigma. For a total of 43 sources, 25 of which in
more than one observation, the number of counts is sufficient for a
more detailed spectral fitting. Given the angular extent of our survey
we expect $\sim$3--4 AGN interlopers for sources with luminosities
above $10^{36}$ erg s$^{-1}$, and about one for sources with
luminosities above $10^{37}$ erg s$^{-1}$ (see \citet{grimm:05}).
All sources in M33, except for the nucleus, have luminosities below
$\sim10^{38}$ erg s$^{-1}$.

{\bf Time variability:} Roughly a quarter (49/198) of all \chandra
sources exhibits long timescale variability ($\sim$weeks--years),
therefore we can exclude SNRs and HII regions in M33 as their
counterparts. Except for two Galactic star interlopers none of the M33
sources show any short timescale X-ray variability
($\sim$seconds--hours). Detection of such variability would separate
stellar mass objects from AGNs, since in AGNs the  light travel time
is too large to produce variability on time scales shorter than
hours. However, the amplitude of the long-term variability (larger
than a factor of 2--3) makes it unlikely for the variable sources to
be AGN (generally less than a factor of 2--3).

A significant number of sources showing long-term variability but
little to no short-term variability is very similar to other
extragalactic X-ray source populations. The comparison of the
amplitude of long term variability with AGN and known X-ray binaries
shows that a large majority of these sources are X-ray binaries.

{\bf X-ray spectra:} The large majority of source spectra is well fit
by models containing either a power law or a thermal plasma
(bremsstrahlung or {\it apec}), but in the most luminous sources more
complex models are required. Of the 43 sources with enough counts for
detailed spectral modeling,
\begin{itemize}

\item ten are well fit with spectral models consistent with accreting
X-ray binaries, namely a disk black body or a black body plus power
law. Three of the sources well described with a black body plus power law
combination have black body temperatures at or below 0.1 keV. These
are even lower than temperatures inferred from ULX intermediate mass
black hole candidates, see e.g. \citet{miller:04}. The luminosities of
these sources also do not indicate the presence of an intermediate
mass black hole.

\item In addition, 16 sources are well fit by a single power law. With
one exception all the photon indices are in the range $\sim$1--2,
consistent with Galactic X-ray binaries \citep{tanaka:01}. The photon
indices are not inconsistent with AGN but we expect only $\sim$3
background AGN among the bright sources. Two thirds of the sources
well fit by a single power law are also well fit by a disk black body
model with inner disk temperatures in the range 1.0--3.0 keV, which
would be consistent with Galactic BH and NS binaries
\citep{tanaka:01}.

\item There are also 4 sources with bremsstrahlung spectra and
temperatures between $\sim$2--7 keV.
\end{itemize}

Moreover, 12 of 30 sources also show long term variability, confirming
their nature as accreting objects. Therefore more than half (30) of
the sources are consistent with being X-ray binaries based on the
X-ray spectrum.

The majority (12) of the remaining 13 sources are well fit by thermal
plasma models (bremsstrahlung or {\it apec} models as discussed
above), 7 of which with relatively low temperatures around 0.2
keV. Five of these  12 sources are variable. Three sources with
unusually low bremsstrahlung temperatures from 0.1--0.3 keV are also
well fit by an {\it apec} model as well; one is variable. The {\it
apec} temperatures of these sources are around 0.3 keV and abundances
are consistent with the overall metallicity of M33 \citep{blair:85}
(see Fig. \ref{fig:slopes}). A black body model does not provide a
good fit to the data. Five non-variable of the 12 sources are
coincident with SNRs \citep{gordon:99}. Variable sources in this
category are candidates for quasi-soft sources.

Two sources are well fit by a single black body with temperatures
of $\sim$75 eV and $\sim$37 eV respectively. One of the sources is a
long term variable with no significant spectral variability. The other
source has been observed only once. Both sources are not significantly
detected above 1 keV. These sources are the best candidates for
super-soft sources in M33.

Given the spectra, and variability, the majority of these sources are
likely X-ray binaries, with a sizable fraction of SNRs/HII regions
confirming a young X-ray source population.

{\bf Spectral variability:} Six of the bright sources show evidence
for spectral variability. The type of variability is varied, as has
been seen in other galaxies as well \citep{zezas:06}. The spectral
variability is sometimes correlated with luminosity (M33 X-4, X-13),
and sometimes not (M33 X-5). In the source X-4 and X-5 the spectral
change seems to originate in the column density--photon index
degeneracy. In both cases, the thermal component (black body or disk
black body) remains constant, whereas the photon index and column
density change their values. With the current data it is not possible
to decide, if there is a change in column density within the system or
a genuine change in the hard component, assuming that the model is a
correct description. M33 X-13 and \object[]{CXO J013329.0+304216} both
show a continuous softening of the spectrum that is correlated in both
cases with an increase in luminosity, and they could be identified
with quasi-soft sources. The other two sources show significant change
only in the soft band with only moderate change in luminosity.

Given the brightness, spectra, and variability (temporal and spectral)
these sources are highly likely X-ray binaries.

A comparison of $N_H$ values measured from the X-ray spectra and HI
radio observations of M33 shows that the X-ray absorption is larger
for most sources than inferred from HI column density. A few sources
have lower X-ray absorption than expected and may be located in front
of the HI gas. However, the majority of sources is intrinsically
absorbed. The absorption in X-rays can be compared with values observed
in Galactic X-ray binaries, and is generally moderate at a few
$10^{21}$ cm$^{-2}$.

This comparison allows the localization of a few sources in M33, and
indicates for most sources intrinsic absorption at the same level as
observed in Galactic X-ray binaries.

The luminosity function of M33 sources and optical counterparts from
\citet{grimm:05} have shown that the X-ray source population in M33 is
dominated by young objects. The long term variability, spectra, and
spectral variability presented in this paper provide evidence that a
large majority of sources are X-ray binaries. The properties of the
X-ray binaries (variability, spectral parameters including absorption)
are very similar to HMXBs in the Milky Way or the Magellanic
Clouds. Thus X-ray binary populations do not show strong variations
over the parameters, e.g. metallicity, covered by the Milky Way, M33,
and the Magellanic Clouds. Thus the M33 X-ray source population
confirms our knowledge about HMXB populations in the Milky Way and
other galaxies. Therefore M33 can also be used as a comparison template
for similar galaxies at larger distances, which is important as
galaxies of this type are numerous.

Comparisons with previous X-ray missions show that repeated
observations of a galaxy are important to study the X-ray source
population in detail. This analysis also shows that especially with
repeat observations it is possible to identify the major parts of an
X-ray population on X-ray data alone. The ongoing deep \chandra survey
of M33 will add significantly more data to such an analysis, and
provide a more detailed picture of the M33 X-ray source population.

\section{Acknowledgments}
This work has been supported by NASA grant GO2-3135X and by grant
AR6-7007X from Chandra X-ray Center operated by the SAO for NASA. This
research has made use of the NASA/IPAC Extragalactic Database (NED)
which is operated by the Jet Propulsion Laboratory, California
Institute of Technology, under contract with the National Aeronautics
and Space Administration. We thank the referee for constructive
comments on the paper.

\bibliographystyle{apj}
\bibliography{ms}

\begin{appendix}

\section{Time Variability}
\label{sec:timevar}
Table A1 presents the analysis results of variability
between observations. We use Poissonian statistics for the
background-subtracted source counts. The table shows only sources with
variability above 3$\sigma$ between any two observation. The first
column gives the name of the source, the next three columns give the
fluxes, with errors in parentheses, for each observation if the source
was detected, or 3 $\sigma$ upper limits on the flux. The last three
columns give the significance of variability between any two
observations. Comparisons between upper limits are set to zero.

\begin{table}[h]
\begin{center}
\caption{List of sources variable at more than $3\sigma$ between
any observation. Comparisons between upper limits are set to
zero.\label{tab:var}}
\begin{tabular}{lcccccc}
\tableline
\tableline
Source & \multicolumn{3}{c}{Flux [$10^{-7}$ cts/s/cm$^2$]} & \multicolumn{3}{c}{Significance [$\sigma$]}\\
 & 1730& 786 & 2023 & $1730$& $786$ & $1730$\\
 & & & & $\rightarrow786$& $\rightarrow2023$ & $\rightarrow2023$\\
\tableline
CXO J013308.3+304802 &    ---        &     66 (8.4) &  $<$4.5       &   ---  &    7.3 &   ---  \\
CXO J013315.1+305317 &    ---        &   3000 (46)  &    5800 (62)  &   ---  &   36.3 &   ---  \\
CXO J013315.5+304448 &     180 (18)  &    170 (19)  &  $<$4.3       &    0.4 &    8.7 &    9.8 \\
CXO J013321.7+303858 &      24 (5.2) &   ---        &  $<$6.2       &   ---  &   ---  &    3.4 \\
CXO J013321.9+303921 &      22 (5.0) &   ---        &  $<$4.3       &   ---  &   ---  &    3.5 \\
CXO J013323.9+304821 &    ---        &     46 (7.4) &  $<$9.2       &   ---  &    5.0 &   ---  \\
CXO J013324.4+304401 &    1700 (39)  &   2800 (45)  &    2500 (31)  &   18.5 &    5.5 &   16.1 \\
CXO J013327.7+304645 &      14 (4.3) &     20 (5.4) &      52 (6.6) &    0.9 &    3.8 &    4.8 \\
CXO J013328.6+304321 &      20 (4.7) &     19 (4.9) &  $<$3.5       &    0.1 &    3.2 &    3.5 \\
CXO J013329.0+304216 &     150 (12)  &    160 (11)  &     200 (10)  &    0.6 &    2.7 &    3.2 \\
CXO J013329.2+304537 &      34 (6.0) &     39 (6.5) &  $<$6.2       &    0.6 &    5.0 &    4.6 \\
CXO J013329.2+304508 &     140 (11)  &    170 (12)  &     120 (8.2) &    1.8 &    3.4 &    1.5 \\
CXO J013333.0+304920 &    ---        &     28       &  $<$6.8       &   ---  &    3.5 &   ---  \\
CXO J013334.1+303714 &      43 (6.4) &   ---        &  $<$4.3       &   ---  &   ---  &    6.0 \\
CXO J013335.5+303728 &      12 (4.0) &     19 (3.7) &  $<$4.3       &    1.3 &    4.0 &    1.9 \\
CXO J013336.3+303742 &      90 (9.1) &     48 (5.5) &  $<$7.3       &    3.9 &    7.4 &    9.1 \\
CXO J013337.4+304718 &      71 (8.4) &    100 (9.7) &      29 (5.2) &    2.3 &    6.5 &    4.3 \\
CXO J013339.2+304049 &      50 (6.7) &     47 (8.4) &  $<$6.2       &    0.3 &    4.9 &    6.5 \\
CXO J013340.0+304323 &      32 (5.6) &     23 (4.7) &  $<$9.5       &    1.2 &    2.9 &    4.0 \\
CXO J013341.5+304136 &      26 (5.0) &     37 (5.6) & $<$10.7       &    1.5 &    4.7 &    3.1 \\
CXO J013341.8+303848 &      70       &   150        &     200       &    6.8 &    3.8 &   10.4 \\
CXO J013342.5+304253 &      15 (4.2) &     56 (6.8) & $<$11.8       &    5.1 &    6.5 &    0.8 \\
CXO J013343.4+304630 &      20 (8.8) &     61 (7.7) &      79 (7.5) &    3.5 &    1.7 &    5.1 \\
CXO J013344.2+304026 &      12       &   30.8       &    $<$5       &    2.3 &    3.9 &    1.5 \\
CXO J013346.2+303807 &  $<$3.5       & $<$4.3       &      90 (9.0) &    0.0 &    9.5 &    9.6 \\
CXO J013350.5+303821 &      27 (5.1) &     57 (5.7) & $<$11.8       &    3.9 &    7.9 &    3.0 \\
CXO J013353.6+303605 & $<$10.4       &    9.9 (2.9) &    ---        &    0.2 &   ---  &   ---  \\
CXO J013354.8+303309 &     120 (10)  &    230 (12)  &    ---        &    7.0 &   ---  &   ---  \\
CXO J013356.8+303729 &      37 (8.4) &     83 (6.7) &    ---        &    4.3 &   ---  &   ---  \\
CXO J013358.8+305004 &    ---        &     93 (9.8) &      36 (4.5) &   ---  &    5.3 &   ---  \\
CXO J013409.8+305044 &    ---        &    3.8       &      33 (4.5) &   ---  &    5.3 &   ---  \\
CXO J013410.3+305346 &    ---        &   $<$5       &      23 (3.8) &   ---  &    4.7 &   ---  \\
CXO J013410.5+303946 &      72 (7.9) &    120 (8.0) &     120 (7.7) &    4.3 &    0.0 &    4.4 \\
CXO J013416.7+305101 &    ---        & $<$7.7       &      27 (3.7) &   ---  &    5.2 &   ---  \\
CXO J013419.2+304942 &    ---        &   $<$5       &      17 (3.0) &   ---  &    4.0 &   ---  \\
CXO J013424.6+304428 &  $<$5.1       &     26 (6.0) &  $<$6.8       &    3.5 &    3.2 &    0.0 \\
CXO J013425.3+304157 & $<$15.2       &     33 (6.2) &      11 (2.9) &    2.9 &    3.2 &    1.4 \\
CXO J013426.7+304811 &    ---        & $<$3.5       &      14 (2.8) &   ---  &    3.8 &   ---  \\
CXO J013429.1+304212 &      13 (4.5) &$<$14.7       &     7.5 (2.2) &    0.4 &    3.3 &    1.1 \\
CXO J013429.7+305026 &    ---        & $<$7.7       &      34 (3.9) &   ---  &    6.7 &   ---  \\
CXO J013432.1+305158 &    ---        &   $<$5       &      93 (13)  &   ---  &    6.8 &   ---  \\
CXO J013432.5+305035 &    ---        &   $<$5       &      15 (3.0) &   ---  &    3.3 &   ---  \\
CXO J013432.7+303436 &      51 (8.3) & $<$3.6       &    ---        &    5.7 &   ---  &   ---  \\
CXO J013433.7+304701 &    ---        &    430 (27)  &     280 (14)  &   ---  &    4.9 &   ---  \\
CXO J013435.0+304439 &    ---        & $<$7.7       &      17 (3.0) &   ---  &    3.1 &   ---  \\
CXO J013436.4+304713 &    ---        &   50.3       &      23 (3.3) &   ---  &    3.1 &   ---  \\
CXO J013446.7+304449 &    ---        & $<$6.4       &      21 (3.3) &   ---  &    4.4 &   ---  \\
CXO J013449.0+304446 &    ---        & $<$6.4       &      34 (4.0) &   ---  &    6.9 &   ---  \\
CXO J013451.1+304356 &    ---        & $<$3.5       &      30 (3.8) &   ---  &    7.0 &   ---  \\
CXO J013451.9+304615 &    ---        &    120 (14)  &      48 (4.8) &   ---  &    4.9 &   ---  \\
\tableline
\end{tabular}
\end{center}
\end{table}

\section{Spectra}
\label{sec:spectra}
The spectra are combined in one plot. Black corresponds to the first
observation, red to a second observation, and green to a third
obervation. For easier comparison contour plots are also overlaid for
all observations. Solid contours correspond to the first observation,
dashed contours to the second observation, and dotted to the third
observation. Note that not all sources are observed in all
observations.

Some sources are well-fit by a bremsstrahlung model with low
temperatures. We also fit these spectra with an APEC model which also
provides a good fit. In case the APEC model provides a good fit as
well we give the parameters and contour plots in the table and
figures.

\LongTables
\def\arraystretch{1.5}
\begin{deluxetable}{lrrc@{\,}c@{\,}cc@{\,}c@{\,}c}
\tabletypesize{\small}
\tablecaption{Spectral parameters.\label{tab:spectra}}
\tablewidth{0pt}
\tablehead{\\[0.2cm]
\colhead{Source} & \colhead{ObsId} & \colhead{XSPEC model} & \colhead{$N_H$} & \multicolumn{5}{c}{parameters}\\
\colhead{} & \colhead{} & \colhead{} & \colhead{[$10^{22}$ cm$^{-2}$]} & \multicolumn{5}{c}{kT, T0, Tin, E, $\sigma$: [keV],}\\
\colhead{} & \colhead{} & \colhead{} & \colhead{} & \multicolumn{5}{c}{Z: [solar abund.]}
}
\startdata
\vspace{0.3cm}
                &      &            &             &           & & & & \\
                &      &            &             &     kT      & & & & \vspace{0.15cm}\\

\object[]{CXO J013253.4+303817} & 1730 & bremss     & 0.05$_{-0.03}^{+0.04}$ & 7.1$_{-2.0}^{+3.55}$   & & & &  \\
(M33 X-1)       &      &            &             &          & & & & \vspace{0.3cm}\\

                &      &            &             &  $\Gamma$   & & & &\vspace{0.15cm} \\

\object[]{CXO J013253.9+303312} & 1730 & pow        & 0.16$_{-0.05}^{+0.07}$ & 1.9$_{-0.23}^{+0.26}$   & & & & \\
(M33 X-2)       &      &            &             &          & & & & \\

                &      &            &             &     kT      & & & &\vspace{0.15cm} \\
\object[]{CXO J013308.3+304802} &  786 & bremss     & 0.17$_{-0.09}^{+0.12}$ & 2.3$_{-1.2}^{+4.3}$   & & & &\vspace{0.3cm}\\

                &      &            &             &    kT       &   $\Gamma$   & & &\vspace{0.15cm} \\
\object[]{CXO J013315.1+305317} &  786 & bbody+pow  & 0.34$_{-0.13}^{+0.05}$ & 0.62$_{-0.11}^{+0.07}$ & 3.0$_{-0.7}^{+0.3}$ & & & \\
(M33 X-4)       & 2023 & bbody+pow  & 0.67$_{-0.09}^{+0.12}$ & 0.69$\pm0.04$ & 4.7$_{-0.25}^{+0.75}$ & & & \vspace{0.3cm}\\

                &      &            &             &  $\Gamma$   & & & &\vspace{0.15cm} \\
\object[]{CXO J013315.5+304448} & 1730 & pow        & 0.31$_{-0.18}^{+0.26}$ & 2.7$_{-0.65}^{+0.75}$  & & & & \\
                &  786 & pow        & 1.43$_{-0.63}^{+1.02}$ & 5.8$_{-1.75}^{+2.7}$   & & & & \vspace{0.3cm}\\

                &      &            &             &    Tin      &  $\Gamma$   & & &\vspace{0.15cm} \\
\object[]{CXO J013324.4+304401} & 1730 & diskbb+pow & $<0.56$     & 1.3$_{-0.11}^{+0.15}$ & 9.5\tablenotemark{u} & & & \\
(M33 X-5)       &  786 & diskbb+pow & 0.64$_{-0.32}^{+0.24}$ & 1.3$_{-0.1}^{+0.23}$ & 5.7$_{-1.9}^{+1.3}$ & & & \\
                & 2023 & diskbb+pow & 0.28$_{-0.2}^{+0.28}$ & 1.4$_{-0.2}^{+0.13}$ & 3.5$_{-0.9}^{+1.9}$ & & & \vspace{0.3cm}\\

                &      &            &             &    kT       &  $\Gamma$   & \vspace{0.15cm}\\
\object[]{CXO J013327.7+304645} & 2023 & bbody+pow  & $<0.33$     &  1.2\tablenotemark{u} & 3.0$_{-1.4}^{+1.7}$ \vspace{0.3cm}\\

                &      &            &             &     kT       & & & & \vspace{0.15cm}\\
\object[]{CXO J013329.0+304216} & 1730 & bremss     & 0.34$_{-0.1}^{+0.18}$  &  0.23$_{-0.08}^{+0.04}$ & & & & \\
                &  786 & bremss     & 0.37$_{-0.01}^{+0.44}$ &  0.20$_{-0.09}^{+0.01}$ & & & & \\
                & 2023 & bremss     & 0.55$_{-0.01}^{+0.04}$ &  0.16$_{-0.0}^{+0.01}$ & & & & \\

                &      &            &             &    kT        &      Z     & & & \vspace{0.15cm}\\
                & 1730 & apec       & $<0.31$     &  0.36$_{-0.1}^{+0.3}$ & 0.11$_{-0.07}^{+0.19}$ & & & \\
                &  786 & apec       & $<0.16$     &  0.4$_{-0.06}^{+0.14}$ & 0.22$_{-0.12}^{+0.34}$ & & & \\
                & 2023 & apec       & 0.25$_{-0.1}^{+0.15}$  &  0.28$\pm0.05$ & 0.22$_{-0.13}^{+2.2}$ & & & \vspace{0.3cm}\\

                &      &            &             &    kT        &      Z     &  & & \vspace{0.15cm}\\
\object[]{CXO J013329.3+304508} & 1730 & apec       & 0.14$_{-0.09}^{+0.1}$ & 4.7$_{-1.4}^{+5.9}$ & $<4.5$ & & & \\
                &  786 & apec       & $<0.13$               & 8.2$_{-4.3}^{+16.8}$ & $<1.9$ & & & \\
                & 2023 & apec       & $<0.09$               & 17.2$_{-10.2}^{+\infty}$ & 2.0\tablenotemark{u} & & & \vspace{0.3cm}\\

                &      &            &             &     kT       & & & & \vspace{0.15cm}\\
\object[]{CXO J013329.4+304912} &  786 & bremss     & 0.49$_{-0.34}^{+0.51}$ & 0.15$_{-0.07}^{+0.09}$ & & & & \\
                & 2023 & bremss     & 0.37$_{-0.02}^{+0.08}$ & 0.2$_{-0.01}^{+0.08}$ & & & & \\

                &      &            &             &    kT        &      Z     & & & \vspace{0.15cm}\\
                &  786 & apec       & $<0.31$     & 0.31$_{-0.1}^{+0.18}$ & 0.3$_{-0.26}^{+2.2}$ & & & \\
                & 2023 & apec       & $<0.6$      & 0.26$_{-0.13}^{+0.37}$ & 0.05$_{-0.04}^{+0.23}$ & & & \vspace{0.3cm}\\

                &      &            &             &   kT &   E$_{1/2}$ & $\sigma_{1/2}$ &  &  \vspace{0.15cm}\\
\object[]{CXO J013331.1+303333} & 1730 & bremss & 0.17$_{-0.05}^{+0.06}$ & 0.29$_{-0.03}^{+0.05}$ & 0.89$_{-0.03}^{+0.03}$ & $<0.045$ & &\\
(M33 X-14)      &      &   +2*gauss &             &          & 1.35$_{-0.02}^{+0.02}$ & $<0.035$ & & \vspace{0.3cm}\\
\newpage
\vspace{0.3cm}
                &      &            &             &           & & & & \\
                &      &            &             &    Tin       & & & & \vspace{0.15cm}\\
\object[]{CXO J013334.1+303210} & 1730 & diskbb     & $<0.05$     & 1.0$_{-0.03}^{+0.05}$ & & & & \\
(M33 X-7)       &      &            &             &          & & & & \vspace{0.3cm}\\

                &      &            &             &     kT       & & & & \vspace{0.15cm}\\
\object[]{CXO J013335.9+303627} & 1730 & bremss     & 0.22$_{-0.16}^{+0.29}$ & 0.31$_{-0.12}^{+0.17}$  & & & & \\
                &      &            &             &    kT        &      Z     & & & \vspace{0.15cm}\\
                & 1730 & apec       & $<0.38$     &  0.40$_{-0.12}^{+0.28}$ & 0.12$_{-0.09}^{+0.55}$ & & & \vspace{0.3cm}\\

                &      &            &             &  $\Gamma$    & & & & \vspace{0.15cm}\\
\object[]{CXO J013336.3+303742} & 1730 & pow        & 0.62$_{-0.33}^{+0.4}$ & 2.1$_{-0.8}^{+0.9}$   & & & & \\
                &  786 & pow        & 0.25$_{-0.22}^{+0.3}$ & 1.2$_{-0.75}^{+0.85}$ & & & & \vspace{0.3cm}\\

                &      &            &             &   $\Gamma$   & & & &\vspace{0.15cm} \\
\object[]{CXO J013337.4+304718} & 1730 & pow        & $<0.33$     & 1.9$_{-0.55}^{+0.65}$ & & & & \\
                &  786 & pow        & 0.22$_{-0.15}^{+0.17}$ & 2.0$_{-0.45}^{+0.55}$ & & & & \\
                & 2023 & pow        & $<0.49$     & 0.8$_{-0.55}^{+0.7}$  & & & & \vspace{0.3cm}\\

                &      &            &             &     kT       &    Z     & & &\vspace{0.15cm} \\
\object[]{CXO J013341.8+303848} & 1730 & apec       & $<0.07$     &  0.70$\pm0.15$ & 0.17$_{-0.12}^{+0.19}$ & & & \\
                &  786 & apec       & $<0.05$     &  0.70$_{-0.12}^{+0.09}$ & 0.12$_{-0.06}^{+0.07}$ & & & \\
                & 2023 & apec       & $<0.07$     &  0.72$_{-0.15}^{+0.17}$ & $<0.04$ & & & \vspace{0.3cm}\\

                &      &            &             &     kT       & & & & \vspace{0.15cm}\\
\object[]{CXO J013343.4+304630} & 1730 & bbody      & 0.79\tablenotemark{f}   & 0.07\tablenotemark{f}    & & & & \\
                &  786 & bbody      & 0.79$_{-0.44}^{+0.32}$  & 0.07$_{-0.005}^{+0.035}$ & & & & \\
                & 2023 & bbody      & 0.83$_{-0.08}^{+0.52}$  & 0.08$_{-0.055}^{+0.005}$ & & & & \vspace{0.3cm}\\

                &      &            &             &     kT       &    kT       &     Z       & & \vspace{0.15cm}\\
\object[]{CXO J013346.2+303807} & 2023 & bbody+apec & 0.06\tablenotemark{f} & 0.02$\pm0.01$& 3.1$_{-1.5}^{+7.3}$ & 0.3\tablenotemark{f} & & \vspace{0.3cm}\\

                &      &            &             &  $\Gamma$    & & & & \vspace{0.15cm}\\
\object[]{CXO J013346.5+303748} & 1730 & pow        & $<0.14$     & 1.6$\pm0.35$ & & & & \\
                &  786 & pow        & 0.1$_{-0.08}^{+0.1}$ & 1.1$_{-0.25}^{+0.35}$ & & & & \vspace{0.3cm}\\

                &      &            &             &  $\Gamma$    & & & & \vspace{0.15cm}\\
\object[]{CXO J013350.5+303821} & 1730 & pow        & $<0.39$     & 1.6$_{-0.7}^{+0.9}$ & & & & \\
                &  786 & pow        & $<0.16$     & 1.5$_{-0.4}^{+0.4}$ & & & & \vspace{0.3cm}\\

                &      &            &             &     kT       & & & & \vspace{0.15cm}\\
\object[]{CXO J013354.8+303309} & 1730 & bremss     & 0.30$_{-0.12}^{+0.27}$ & 0.22$_{-0.07}^{+0.07}$ & & & & \\
(M33 X-13)      &  786 & bremss     & 0.36$_{-0.11}^{+0.15}$ & 0.18$_{-0.05}^{+0.04}$  & & & & \vspace{0.3cm}\\

                &      &            &             &  $\Gamma$    & & & & \vspace{0.15cm}\\
\object[]{CXO J013356.8+303729} & 1730 & pow        & 0.15\tablenotemark{f} & 1.5\tablenotemark{f} & & & & \\
                &  786 & pow        & 0.15$_{-0.09}^{+0.1}$ & 1.5$_{-0.35}^{+0.4}$ & & & & \vspace{0.3cm}\\

                &      &            &             &  $\Gamma$    & & & & \vspace{0.15cm}\\
\object[]{CXO J013358.8+305004} &  786 & pow        & $<0.46$     & 1.7$_{-0.8}^{+0.2}$  & & & & \\
(M33 X-9a)      & 2023 & pow        & 0.54$_{-0.45}^{+0.22}$ & 1.8$_{-0.7}^{+0.5}$  & & & & \vspace{0.3cm}\\

                &      &            &             &    kT        &      Z     &  $\Gamma$   & \vspace{0.15cm}\\
\object[]{CXO J013401.5+303136} &  786 & apec+pow   & 0.94$_{-0.54}^{+0.5}$ & 6.3\tablenotemark{u} & 0.3\tablenotemark{f} & 3.4$_{-1.3}^{+5.2}$ & \vspace{0.3cm}\\

                &      &            &             &    kT        &  $\Gamma$  & & &\vspace{0.15cm} \\
\object[]{CXO J013402.0+303004} &  786 & bbody+pow  & $<0.27$     & 0.04$_{-0.01}^{+0.025}$ & 1.8$_{-0.35}^{+0.85}$ & & & \vspace{0.3cm}\\

                &      &            &             &    kT        &  & & &\vspace{0.15cm} \\
\object[]{CXO J013409.9+303219} &  786 & bbody    &0.73$_{-0.25}^{+0.34}$ & 0.037$_{-0.006}^{+0.007}$ & & & & \\
\newpage
\vspace{0.3cm}
                &      &            &             &           & & & & \\
                &      &            &             &    Tin     & & & \\
\object[]{CXO J013410.5+303946} & 1730 & diskbb     & 0.06\tablenotemark{f} & 1.2$_{-0.3}^{+0.4}$ & & & \\
                &      &            &             &  $\Gamma$    & & & & \\
                &  786 & pow        & 0.06\tablenotemark{f} & 1.7$\pm 0.1$ & & & \\
                &      &            &             &    Tin     & & & \\
                & 2023 & diskbb     & 0.06\tablenotemark{f} & 1.2$\pm 0.2$ & & & \vspace{0.3cm}\\

                &      &            &             &    kT     &  &  &\vspace{0.15cm} \\
\object[]{CXO J013410.6+304223} & 1730 & bremss     & 0.79$_{-0.43}^{+0.51}$ & 0.11$_{-0.03}^{+0.07}$ & & & \\
                & 2023 & bremss     & $<0.19$     & 0.3$_{-0.05}^{+0.07}$ & & & \vspace{0.3cm}\\

                &      &            &             &    Tin      & & & &\vspace{0.15cm} \\
\object[]{CXO J013421.2+304932} & 2023 & diskbb     & 1.17$_{-0.48}^{+0.55}$  & 1.7$_{-0.5}^{+1.7}$ & & & & \vspace{0.3cm}\\

                &      &            &             & $\Gamma_1$  & E$_{break}$ & $\Gamma_2$  & &\vspace{0.15cm}\\
\object[]{CXO J013425.4+302821} &  786 & bknpow     & $<0.25$     & 2.9$_{-1.1}^{+1.6}$ & $>1.2$ & 0.16$_{-1.0}^{+2.1}$ & &\vspace{0.3cm}\\

                &      &            &             & $\Gamma$    & & & & \vspace{0.15cm}\\
\object[]{CXO J013425.5+305514} &  786 & pow        & 0.16$_{-0.06}^{+0.07}$ & 2.1$_{-0.15}^{+0.25}$ & & & & \\
                & 2023 & pow        & 0.17$_{-0.04}^{+0.03}$ & 2.0$_{-0.04}^{+0.14}$ & & & & \vspace{0.3cm}\\

                &      &            &             &  $\Gamma$ & & & & \vspace{0.15cm}\\
\object[]{CXO J013427.0+304314} &  786 & pow        & $<0.27$     & 0.9$_{-0.3}^{+1.6}$ & & & & \\
                & 2023 & pow        & $<0.17$     & 1.7$_{-0.3}^{+0.6}$ & & & & \vspace{0.3cm}\\

                &      &            &             &     kT    & & & & \vspace{0.15cm}\\
\object[]{CXO J013432.0+303455} & 1730 & bremss     & 2.7$_{-1.1}^{+0.8}$  & 2.7$_{-0.9}^{+11.3}$ & & & & \vspace{0.3cm}\\

                &      &            &             &    kT       &  $\Gamma$ & & & \vspace{0.15cm}\\
\object[]{CXO J013433.7+304701} &  786 & bbody+pow  & $<0.9$      & 0.1$\pm0.03$ & $<2.9$   & & & \\
                & 2023 & bbody+pow  & 0.46$_{-0.33}^{+0.54}$ & 0.09$_{-0.04}^{+0.05}$ & 4.6$_{-3.1}^{+3.4}$ & & & \vspace{0.3cm}\\

                &      &            &             &     kT    & & & & \vspace{0.15cm}\\
\object[]{CXO J013435.1+305646} & 2023 & bremss     & 1.1$_{-0.38}^{+0.36}$  & 4.7$_{-2.4}^{+5.5}$ & & & & \vspace{0.3cm}\\

                &      &            &             &  $\Gamma$  &  & & &\vspace{0.15cm} \\
\object[]{CXO J013438.8+304538} &  786 & pow        & $<0.18$     &  1.6$_{-0.35}^{+0.65}$ &  & & & \\
                & 2023 & pow        & $<0.17$     &  1.8$_{-0.3}^{+0.4}$ &  & & & \vspace{0.3cm}\\

                &      &            &             &  $\Gamma$ & & & & \vspace{0.15cm}\\
\object[]{CXO J013438.8+305504} & 2023 & pow        & 0.06$\pm0.04$ & 1.9$_{-0.12}^{+0.1}$ & & & & \vspace{0.3cm}\\

                &      &            &             &  $\Gamma$ & & & & \vspace{0.15cm}\\
\object[]{CXO J013439.8+305143} & 2023 & pow        & $<0.2$      & 1.5$_{-0.3}^{+0.5}$  & & & & \vspace{0.3cm}\\

                &      &            &             &   Tin     & & & &\vspace{0.15cm}\\
\object[]{CXO J013444.5+304922} &  786 & diskbb     & $<0.07$     & 2.7$_{-1.7}^{+25}$ & & & &\\
                & 2023 & diskbb     & $<0.08$     & 0.46$_{-0.07}^{+0.09}$ & & & &\vspace{0.3cm}\\

                &      &            &             &    kT         &  $\Gamma$ & & & \vspace{0.15cm}\\
\object[]{CXO J013444.6+305535} & 2023 & bbody+pow  & 0.15$_{-0.04}^{+0.1}$  & 0.03\tablenotemark{u} & 1.3$_{-0.05}^{+0.2}$ & & & \vspace{0.3cm}\\

                &      &            &             &   $\Gamma$  & & & &\vspace{0.15cm} \\
\object[]{CXO J013451.9+304615} &  786 & pow        & 0.06\tablenotemark{f} & 1.1$_{-0.4}^{+0.35}$ & & & & \\
                & 2023 & pow        & 0.06\tablenotemark{f} & 1.7$_{-0.35}^{+0.27}$ & & & & \vspace{0.3cm}\\

                &      &            &             &   $\Gamma$ & & & &\vspace{0.15cm} \\
\object[]{CXO J013453.2+305718} & 2023 & pow        & 0.27$_{-0.23}^{+0.29}$ & 2.1$_{-0.7}^{+0.8}$ & & & & \vspace{0.3cm}\\

                &      &            &             &  $\Gamma$  & & & &\vspace{0.15cm} \\
\object[]{CXO J013501.1+304345} &  786 & pow        & $<0.54$     & 1.8$_{-0.5}^{+1.3}$ & & & & \\
                & 2023 & pow        & $<0.25$     & 1.9$_{-0.5}^{+0.4}$ & & & &
\enddata
\tablenotetext{f}{Parameter fixed}
\tablenotetext{u}{Parameter unconstrained at 90\% CL}
\end{deluxetable}

\newpage
\begin{figure*}

\begin{minipage}[h]{\linewidth}
\resizebox{.3\linewidth}{!}{\includegraphics{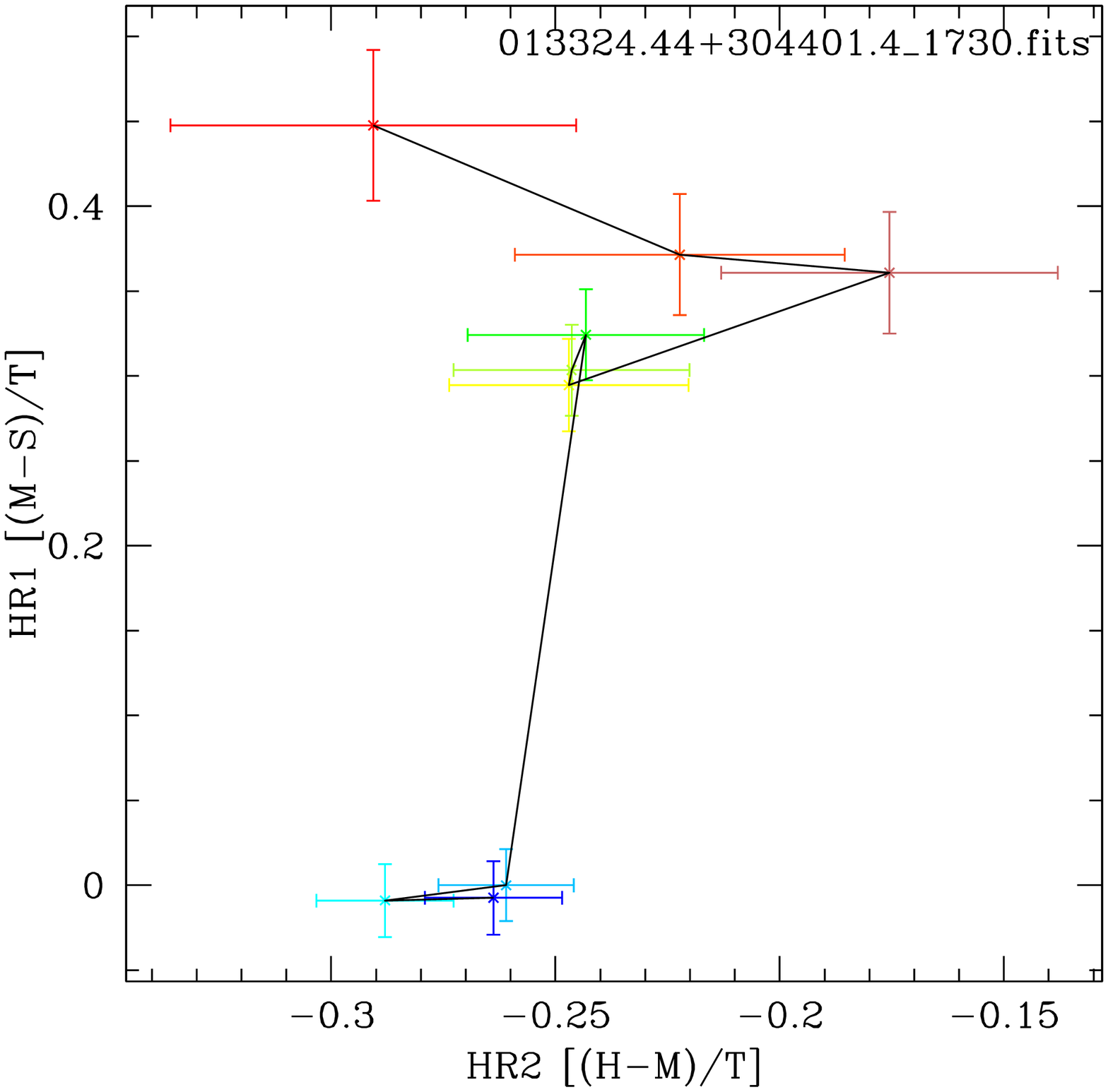}}
\resizebox{.3\linewidth}{!}{\includegraphics{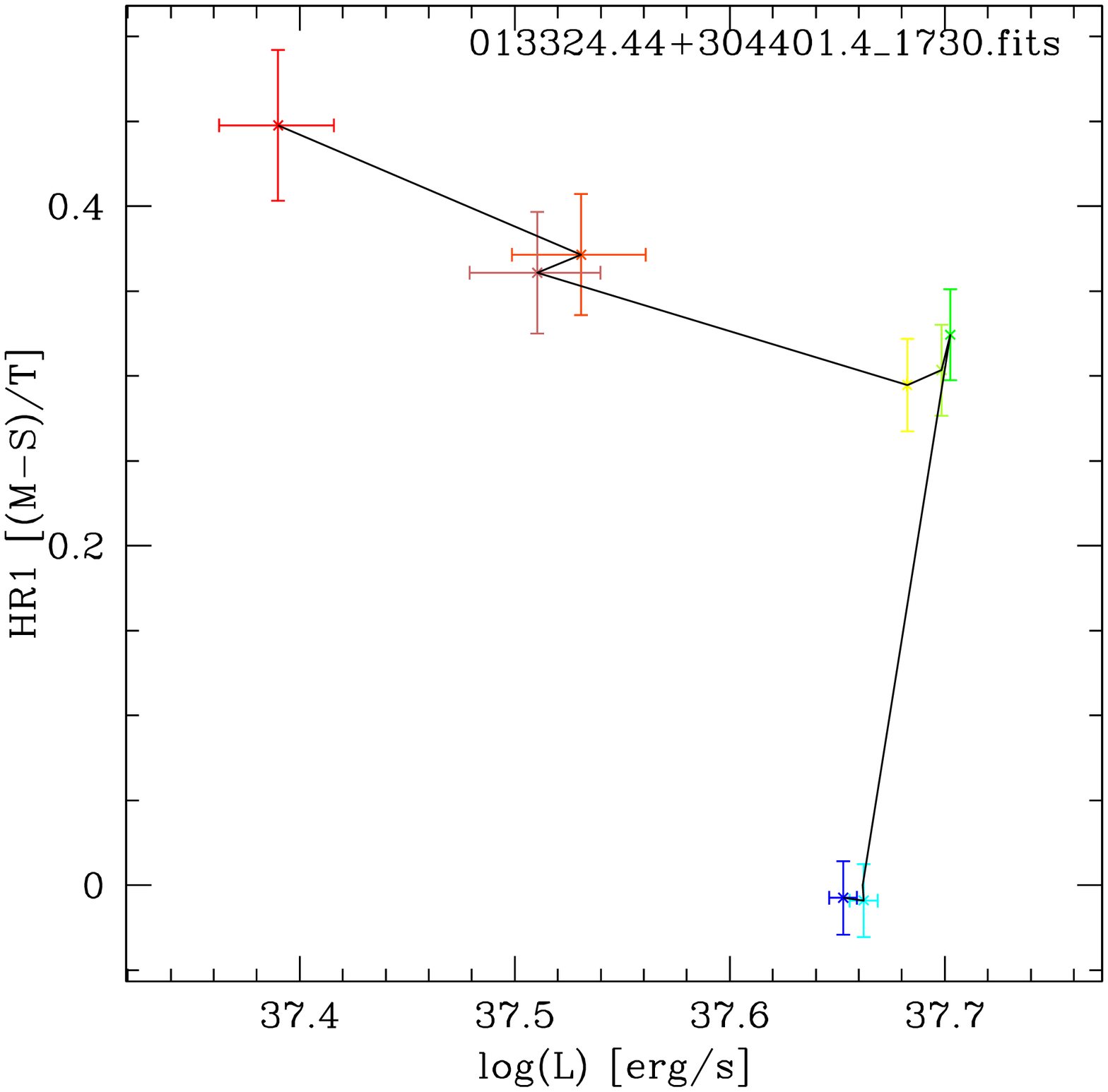}}
\resizebox{.3\linewidth}{!}{\includegraphics{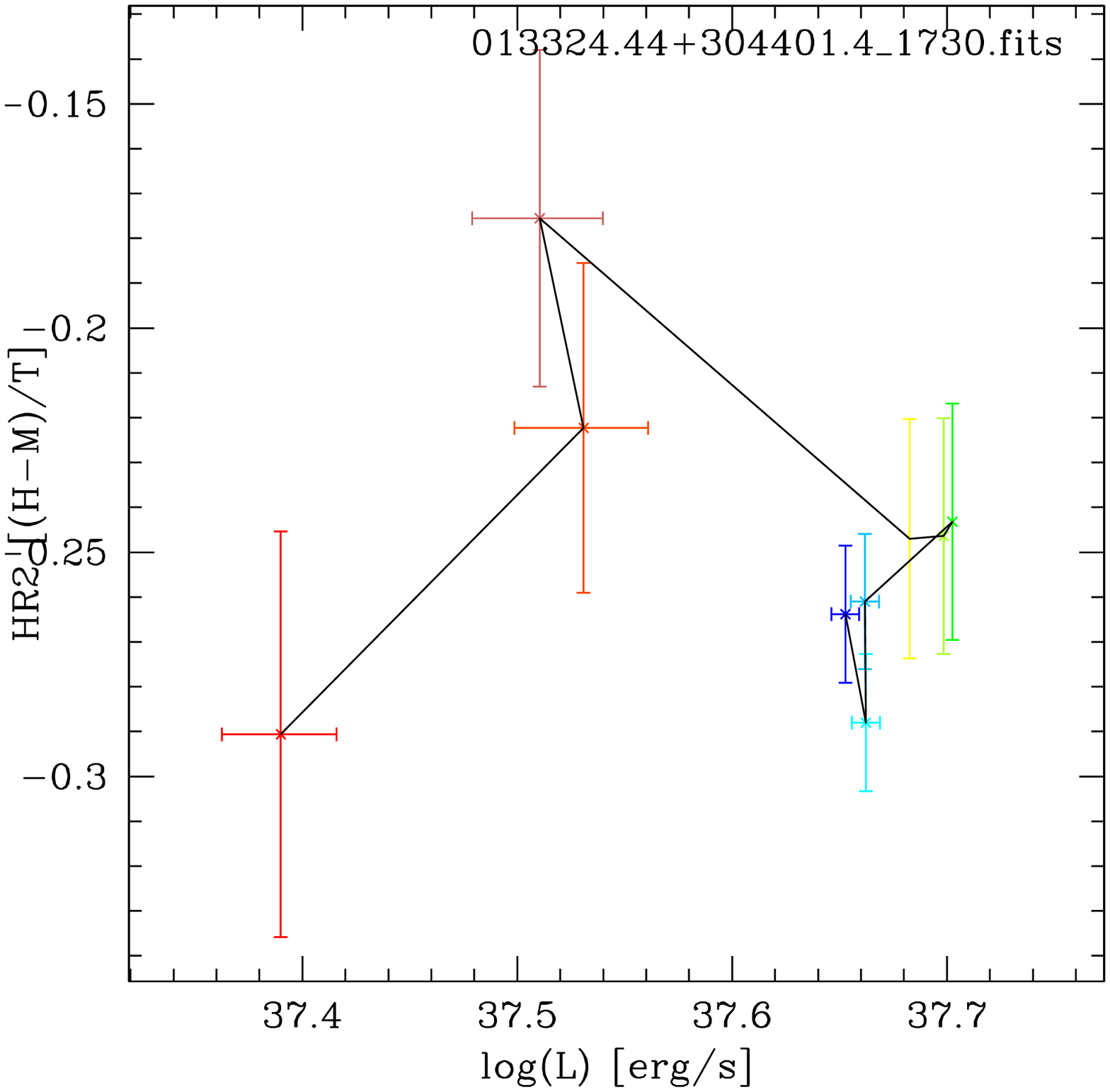}}
\end{minipage}

\begin{minipage}[h]{\linewidth}
\resizebox{.3\linewidth}{!}{\includegraphics[angle=-90]{f9_4a.eps}}
\resizebox{.3\linewidth}{!}{\includegraphics[angle=-90]{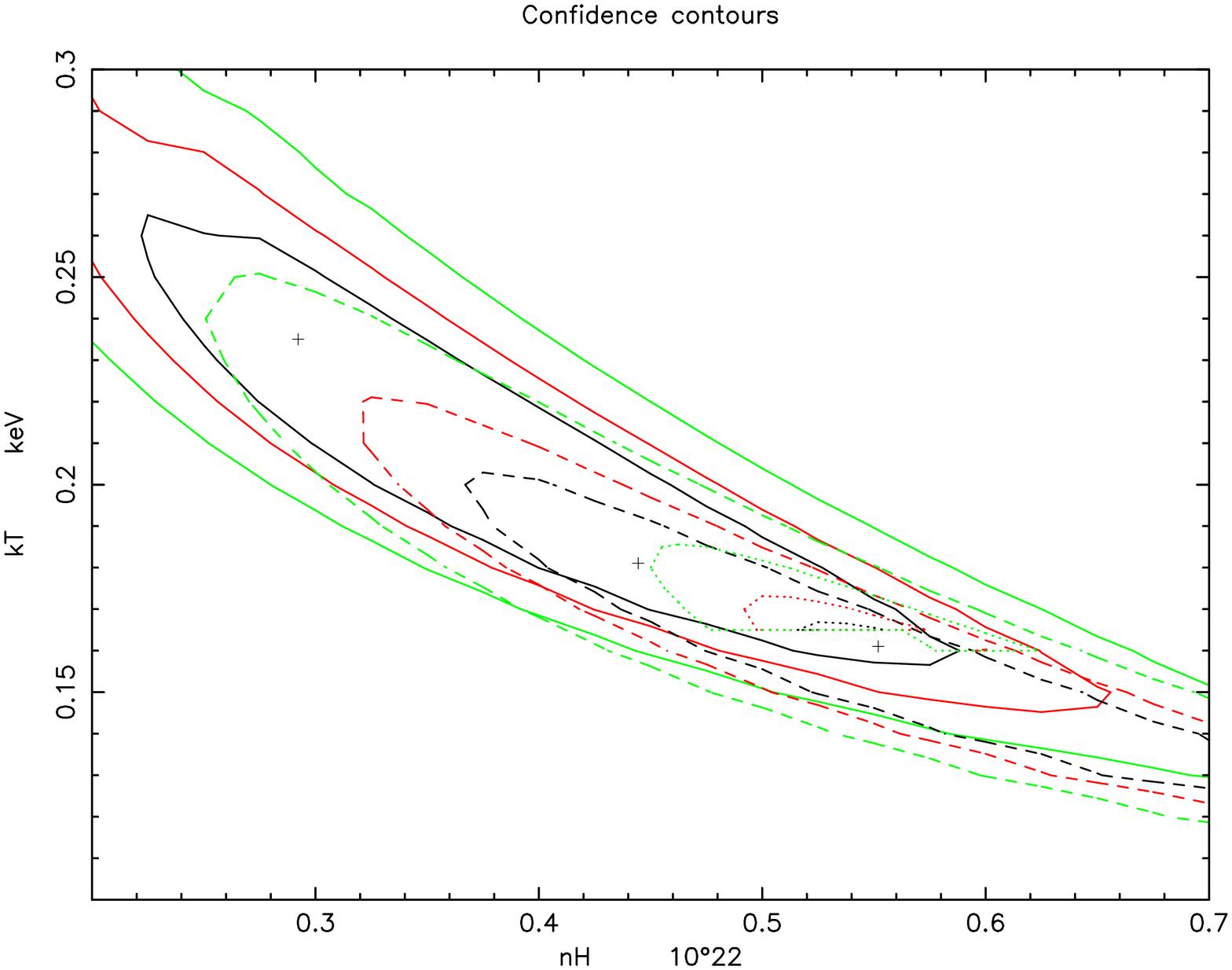}}
\end{minipage}

\begin{minipage}[h]{\linewidth}
\resizebox{.3\linewidth}{!}{\includegraphics{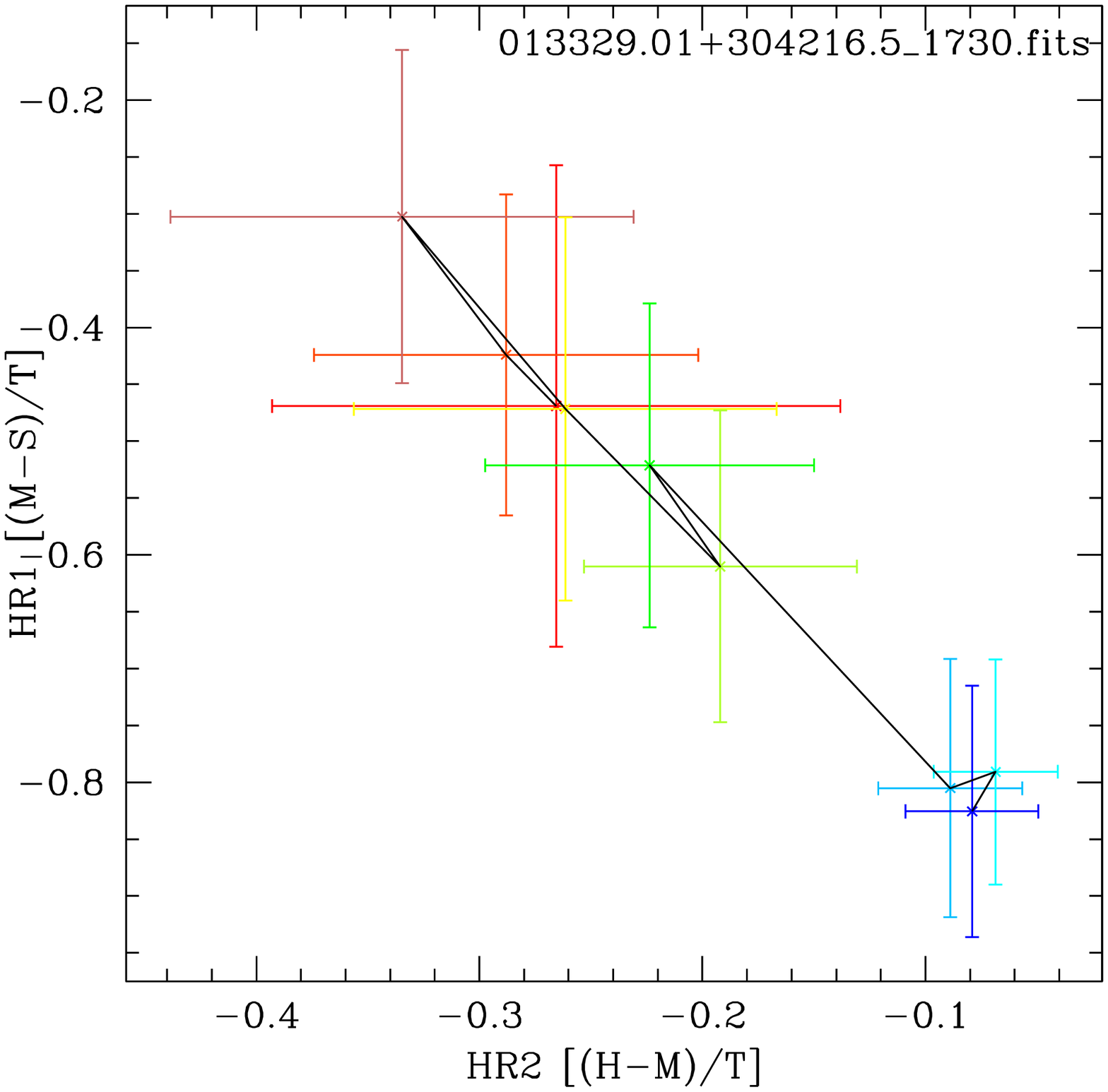}}
\resizebox{.3\linewidth}{!}{\includegraphics{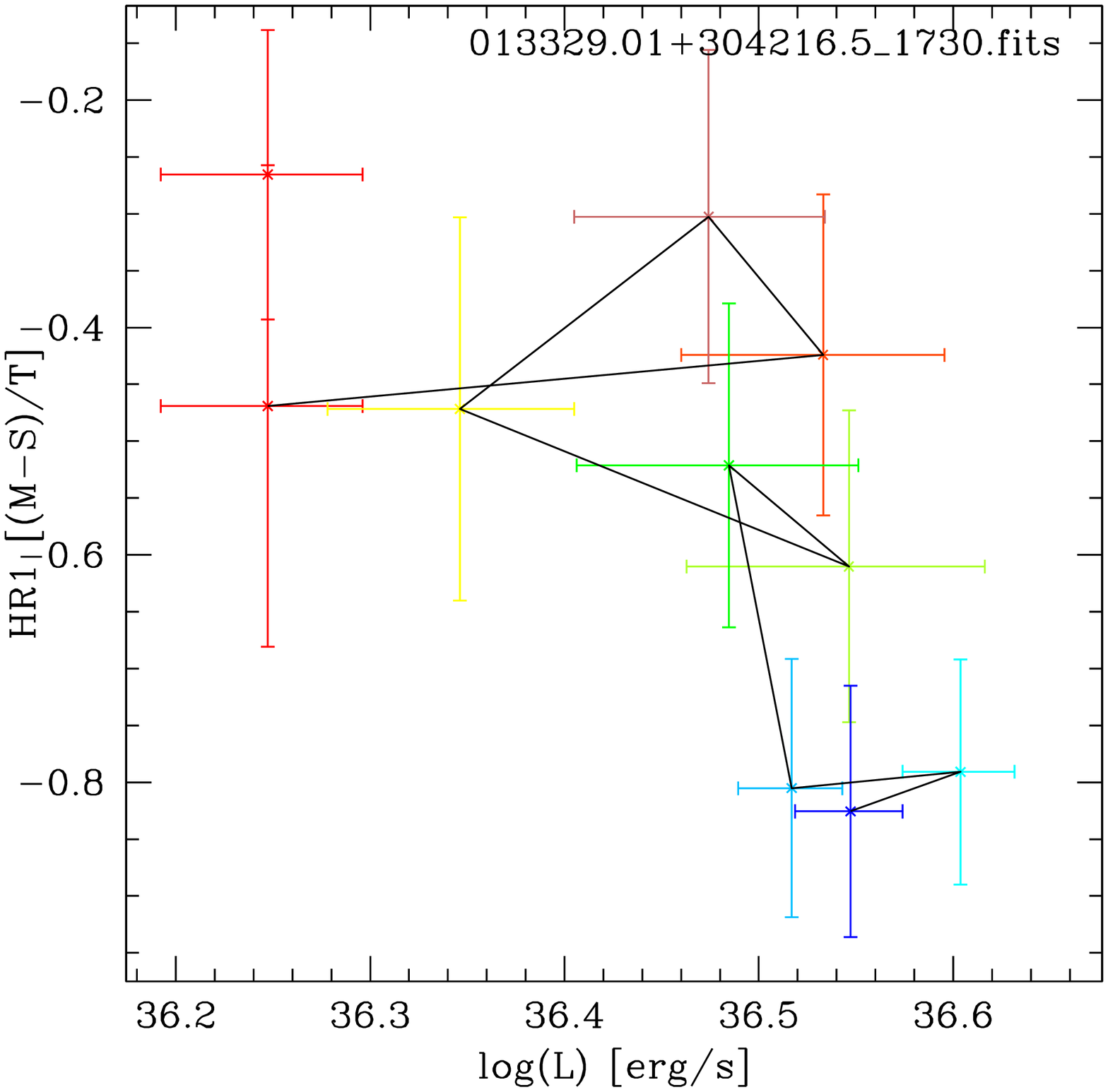}}
\resizebox{.3\linewidth}{!}{\includegraphics{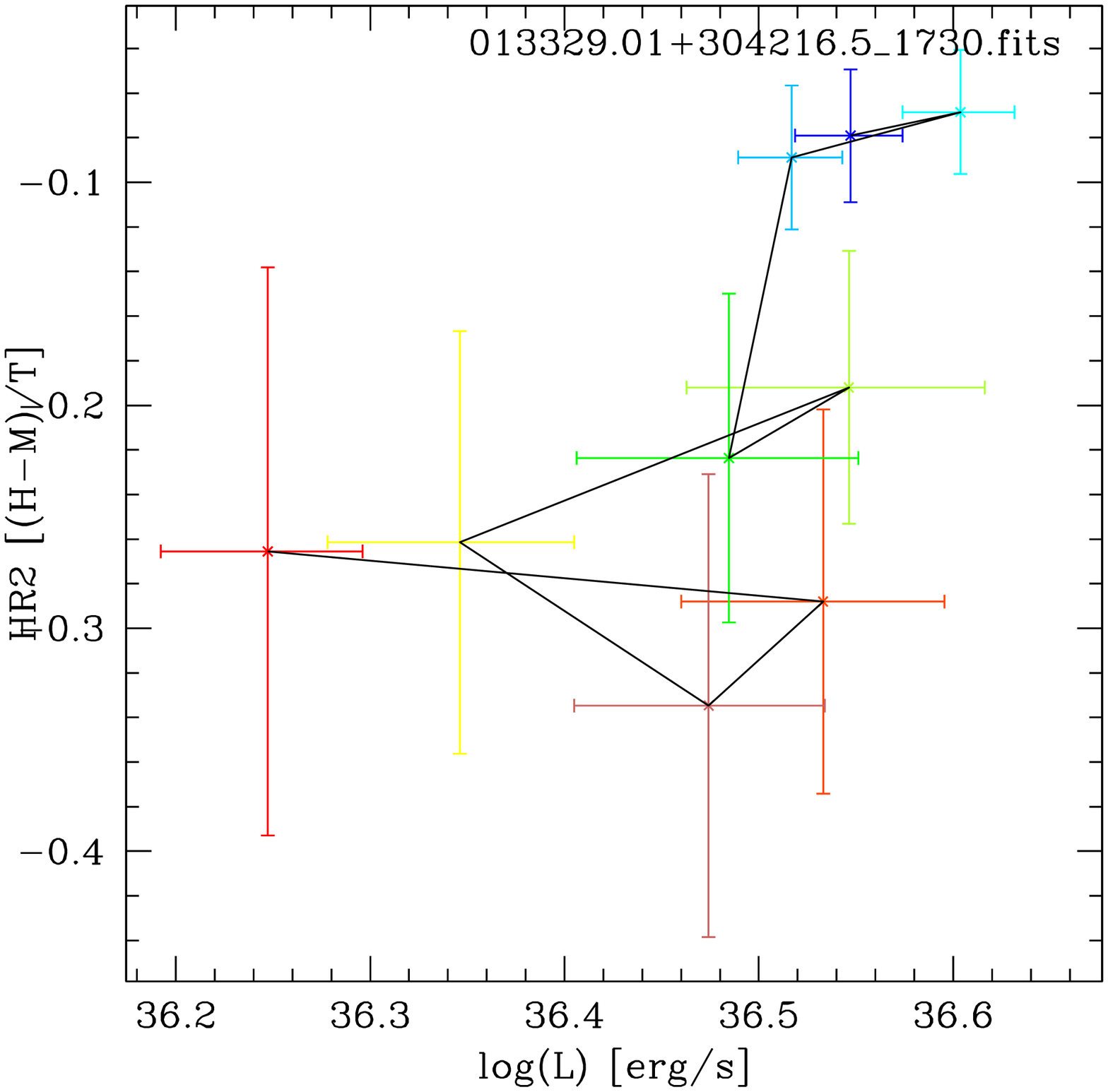}}
\end{minipage}

\begin{minipage}[h]{\linewidth}
\resizebox{.3\linewidth}{!}{\includegraphics[angle=-90]{f9_5a.eps}}
\end{minipage}

\begin{minipage}[h]{\linewidth}
\resizebox{.3\linewidth}{!}{\includegraphics{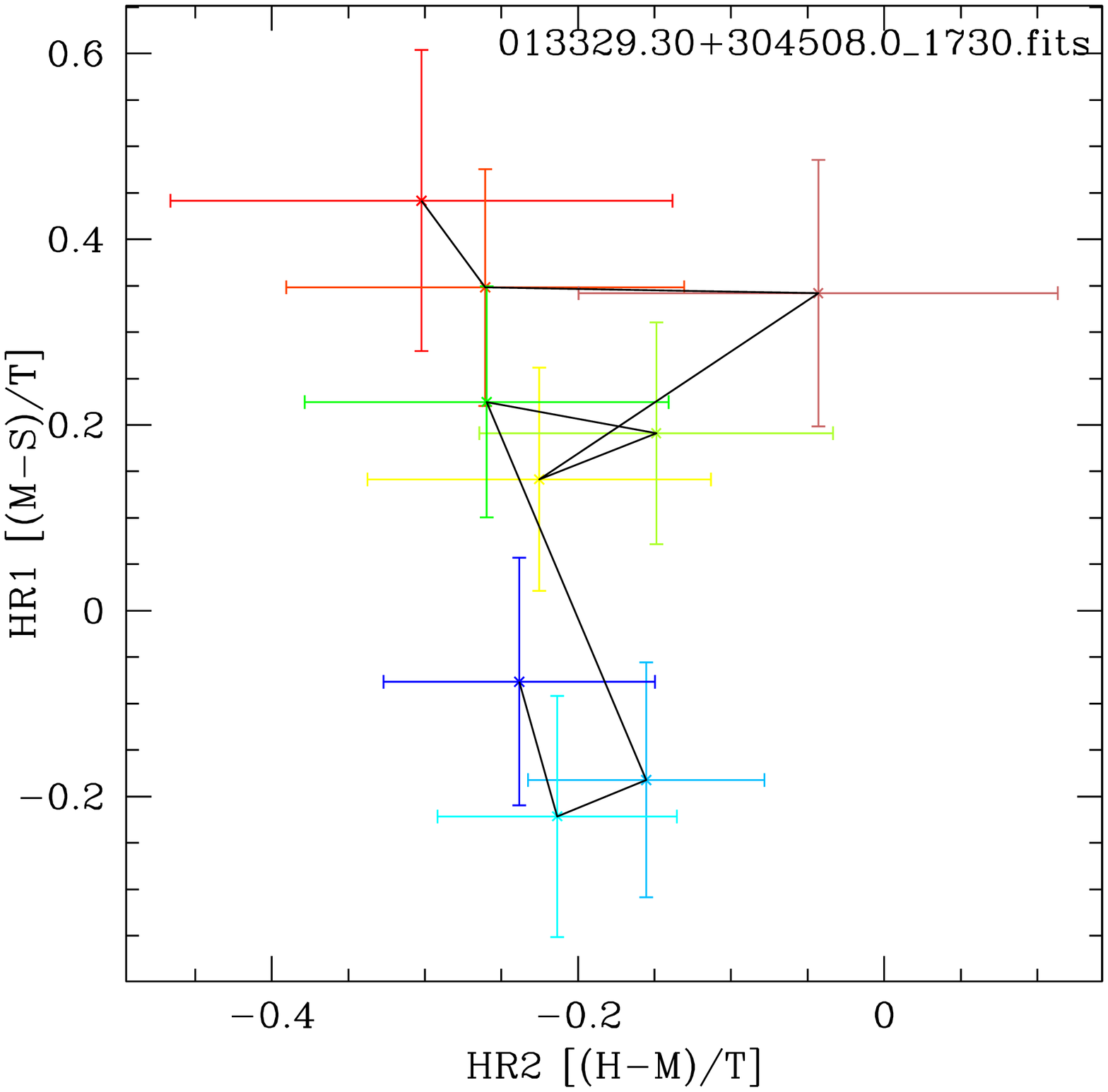}}
\resizebox{.3\linewidth}{!}{\includegraphics{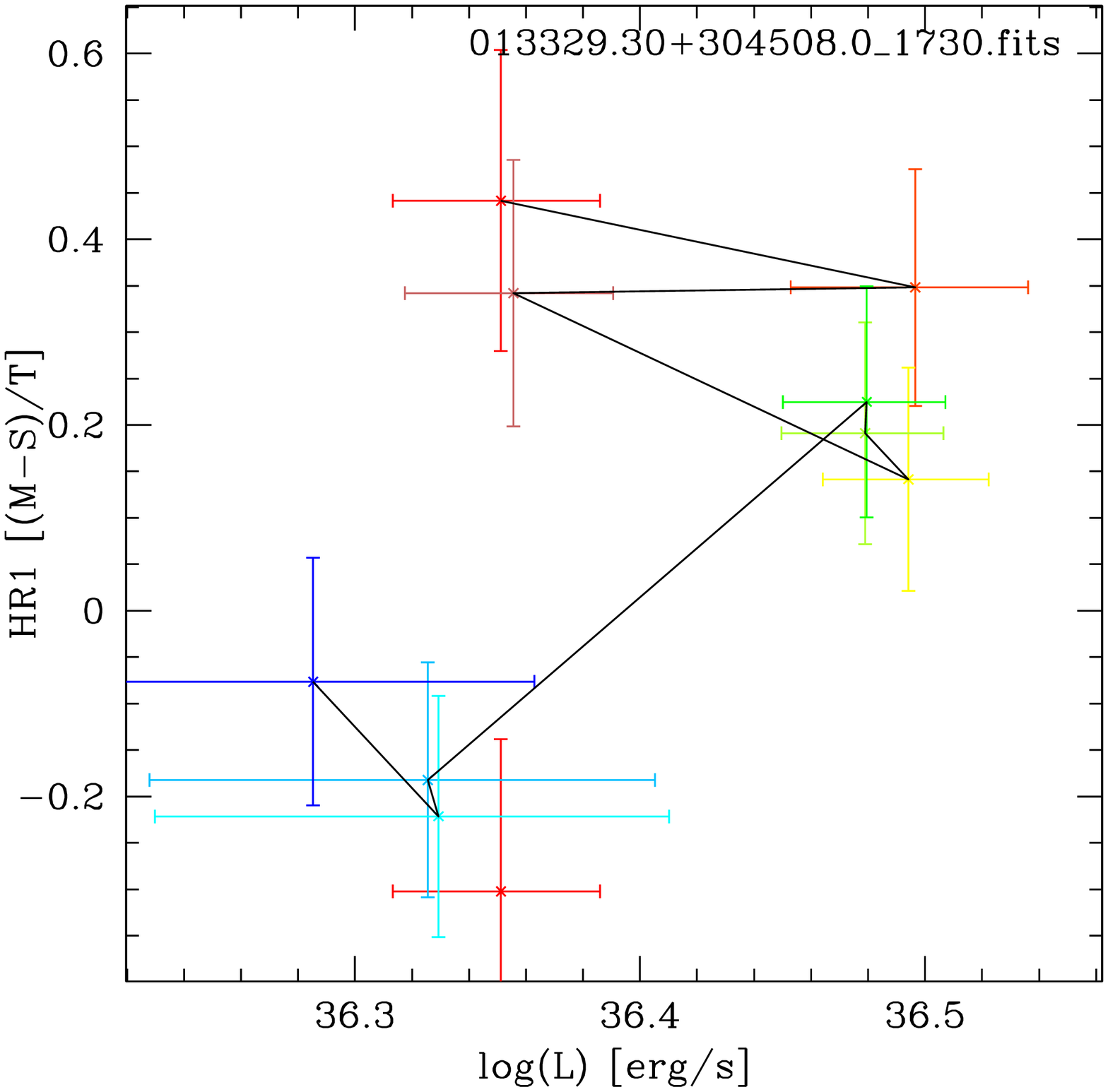}}
\resizebox{.3\linewidth}{!}{\includegraphics{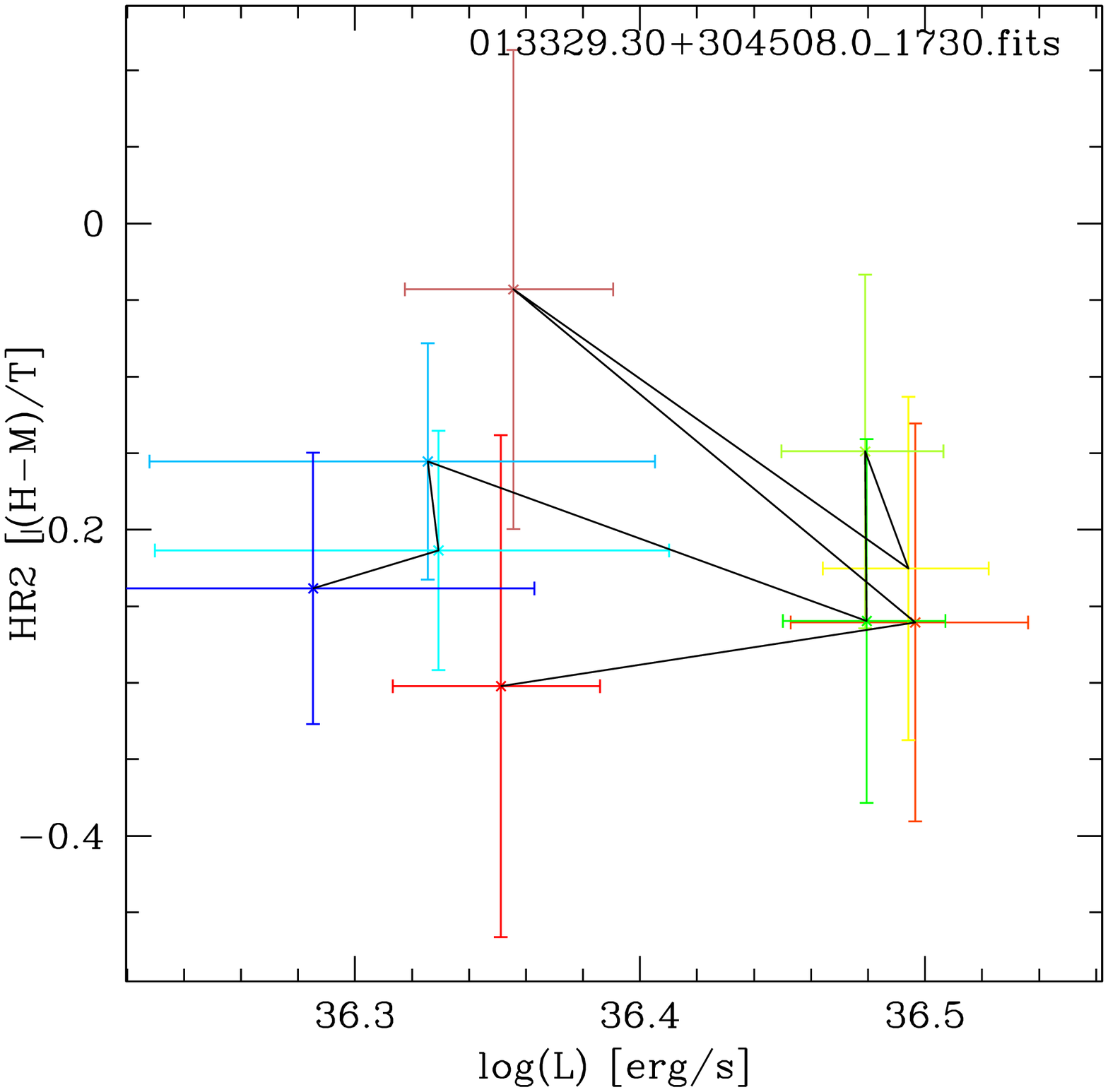}}
\end{minipage}

\end{figure*}
\newpage
\begin{figure*}

\begin{minipage}[h]{\linewidth}
\resizebox{.3\linewidth}{!}{\includegraphics[angle=-90]{f9_6a.eps}}
\resizebox{.3\linewidth}{!}{\includegraphics[angle=-90]{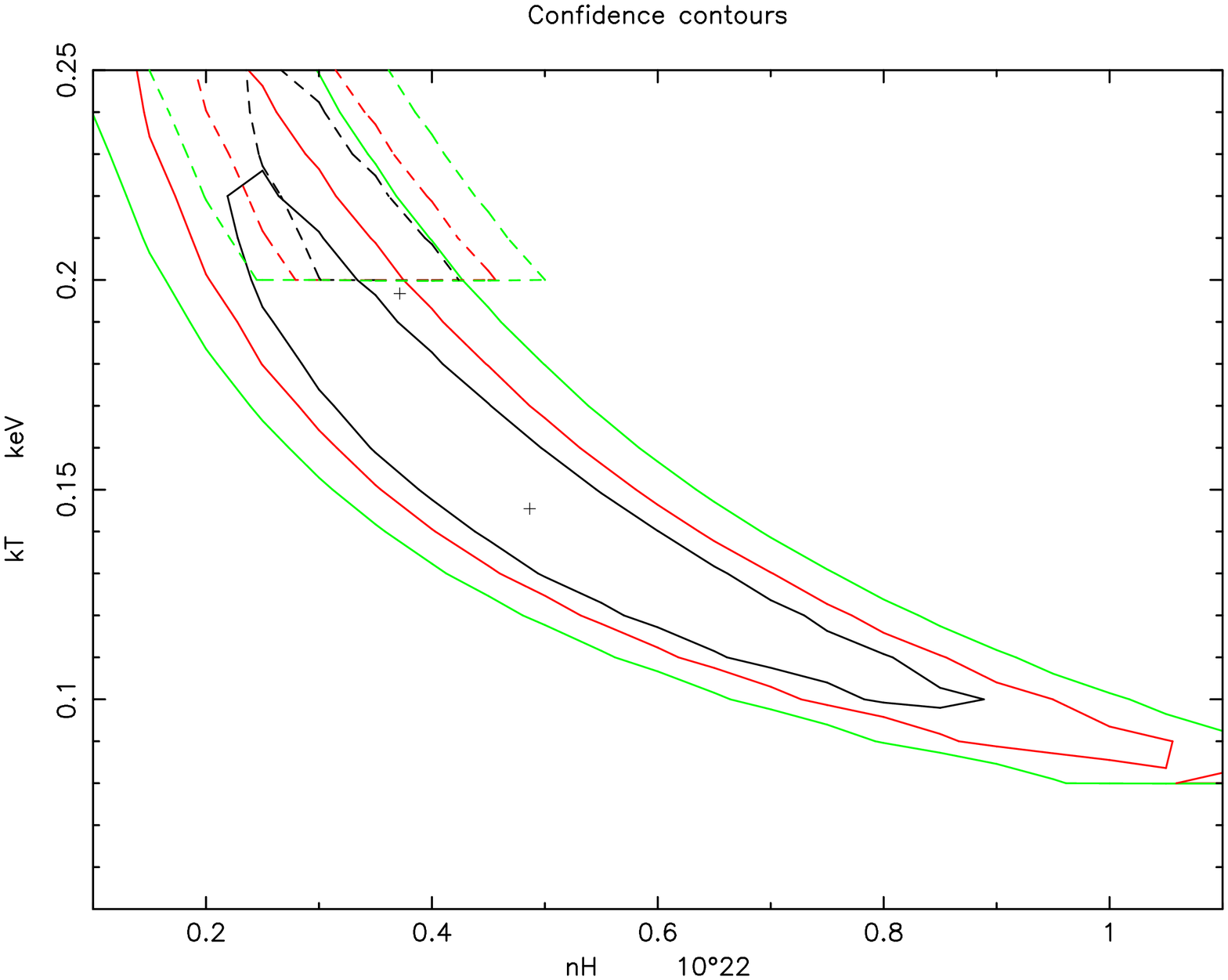}}
\end{minipage}

\begin{minipage}[h]{\linewidth}
\resizebox{.3\linewidth}{!}{\includegraphics[angle=-90]{f9_7a.eps}}
\resizebox{.3\linewidth}{!}{\includegraphics[angle=-90]{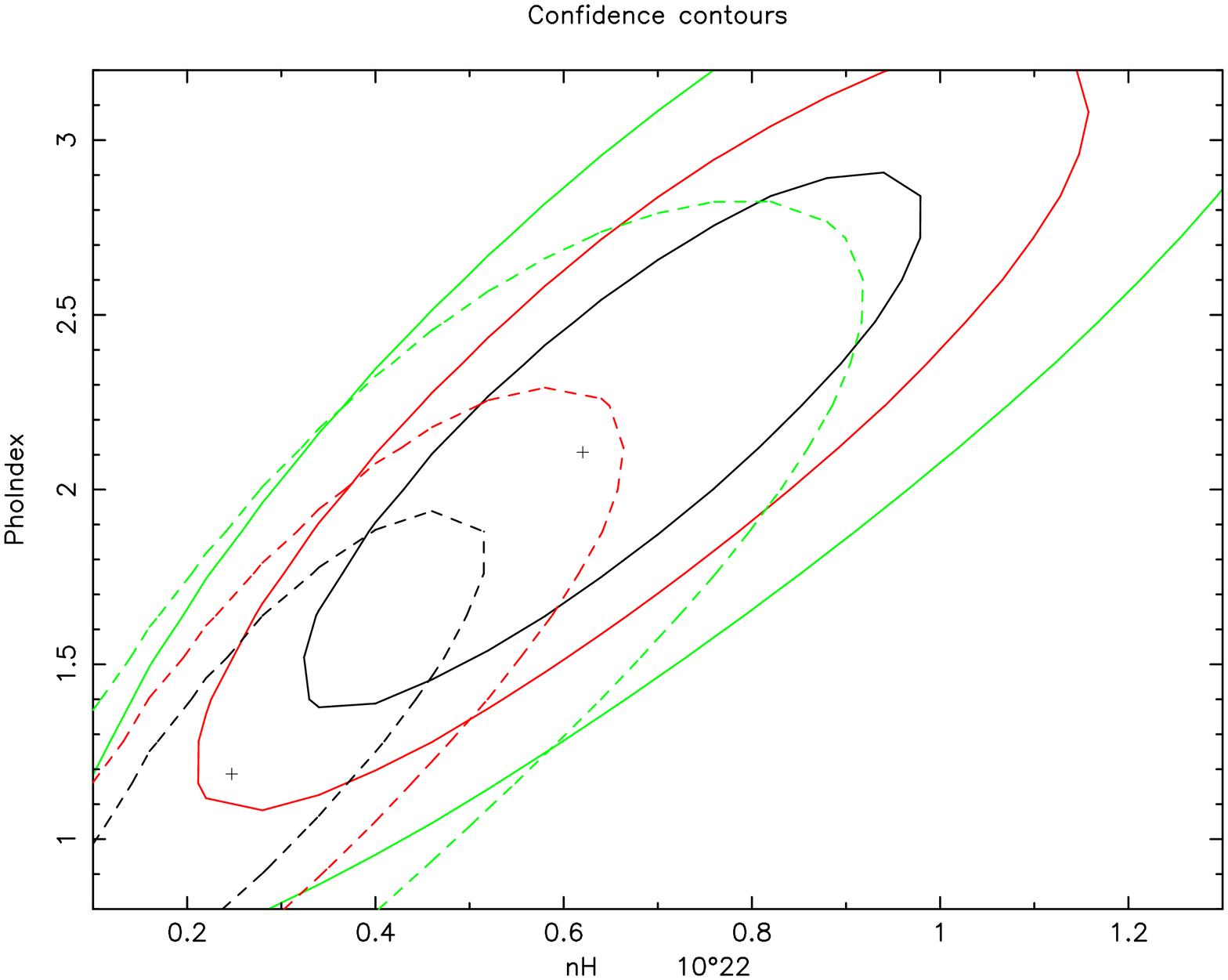}}
\end{minipage}

\begin{minipage}[h]{\linewidth}
\resizebox{.3\linewidth}{!}{\includegraphics[angle=-90]{f9_8a.eps}}
\resizebox{.3\linewidth}{!}{\includegraphics[angle=-90]{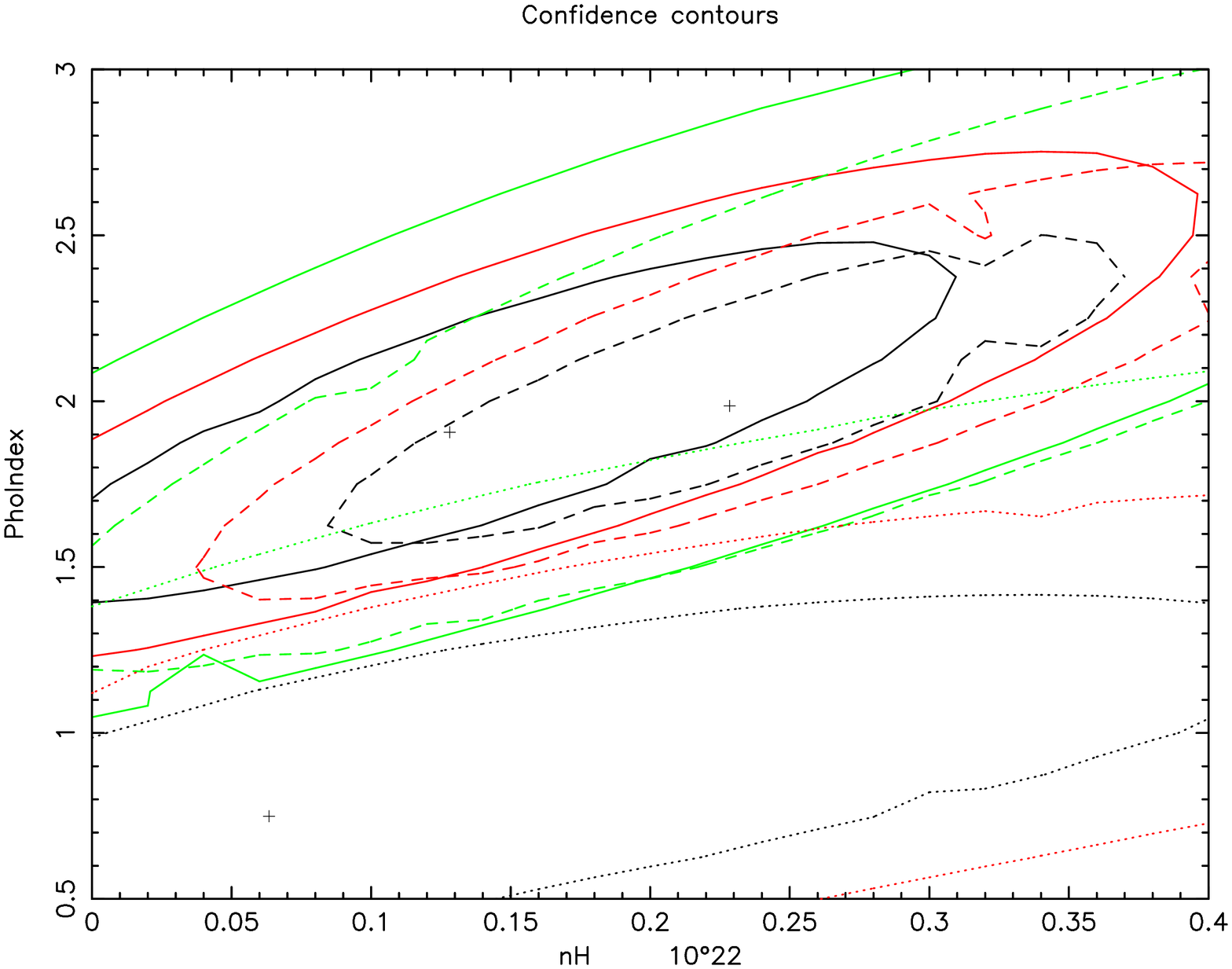}}
\end{minipage}

\begin{minipage}[h]{\linewidth}
\resizebox{.3\linewidth}{!}{\includegraphics[angle=-90]{f9_9a.eps}}
\resizebox{.3\linewidth}{!}{\includegraphics[angle=-90]{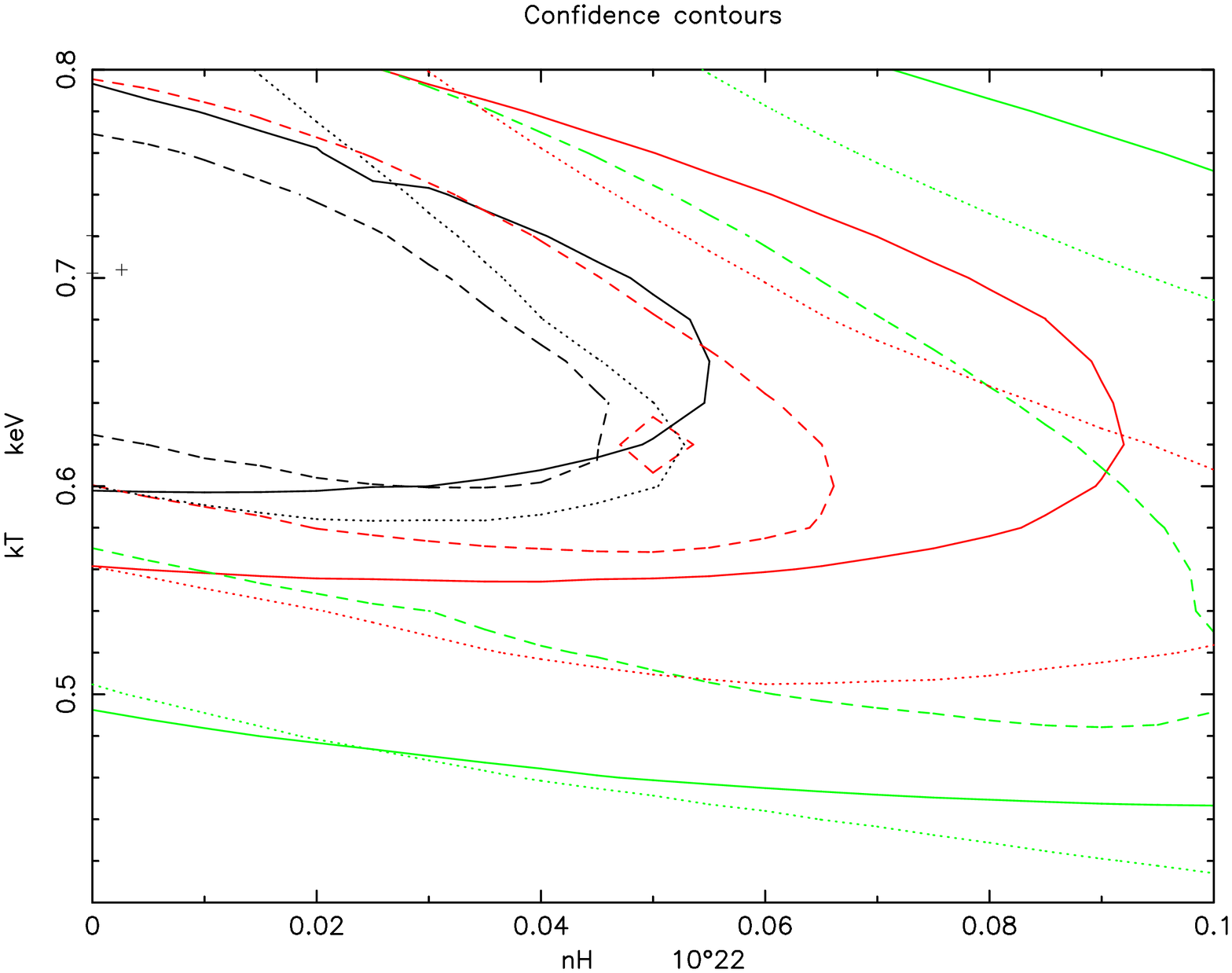}}
\resizebox{.3\linewidth}{!}{\includegraphics[angle=-90]{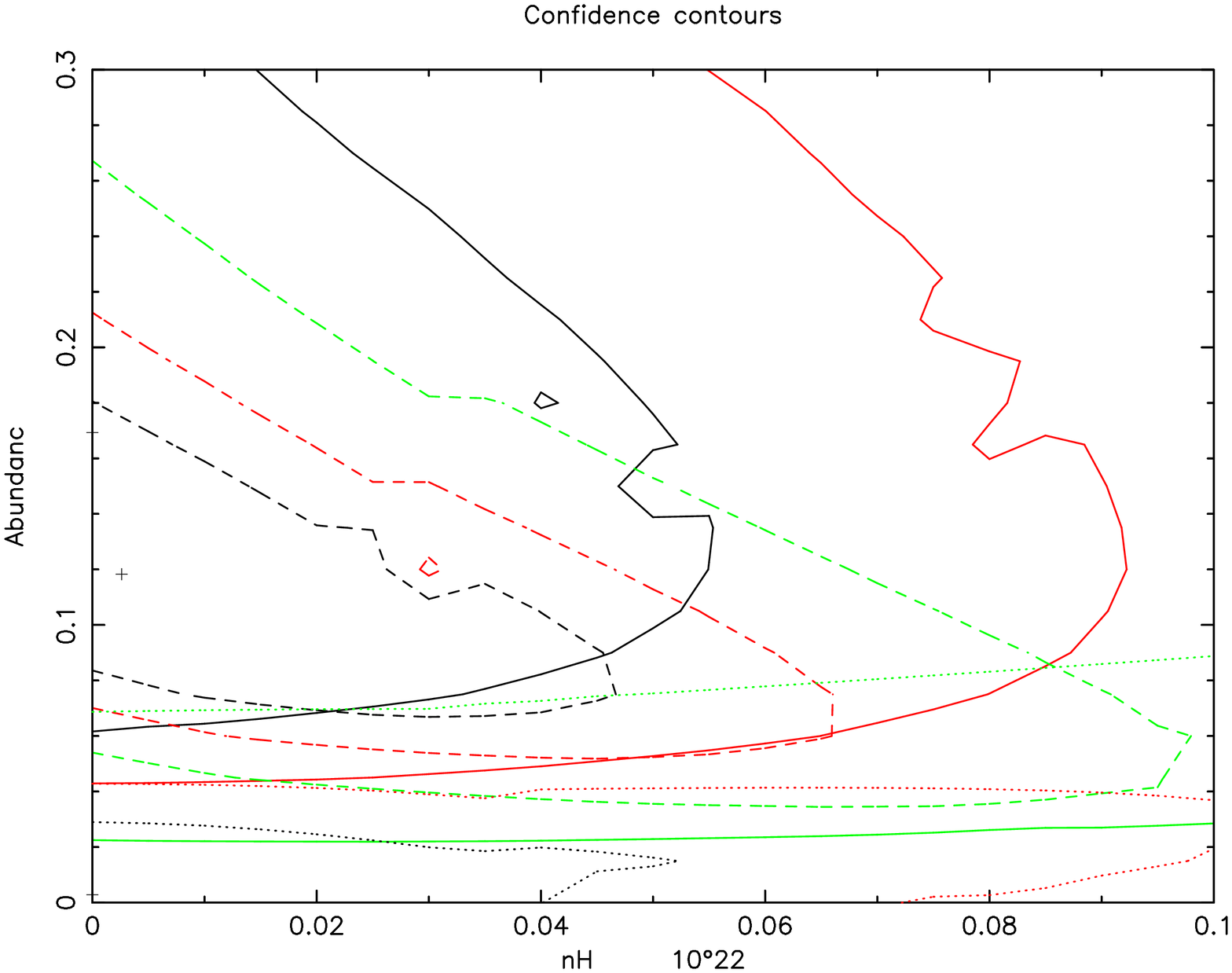}}
\end{minipage}

\begin{minipage}[h]{\linewidth}
\resizebox{.3\linewidth}{!}{\includegraphics[angle=-90]{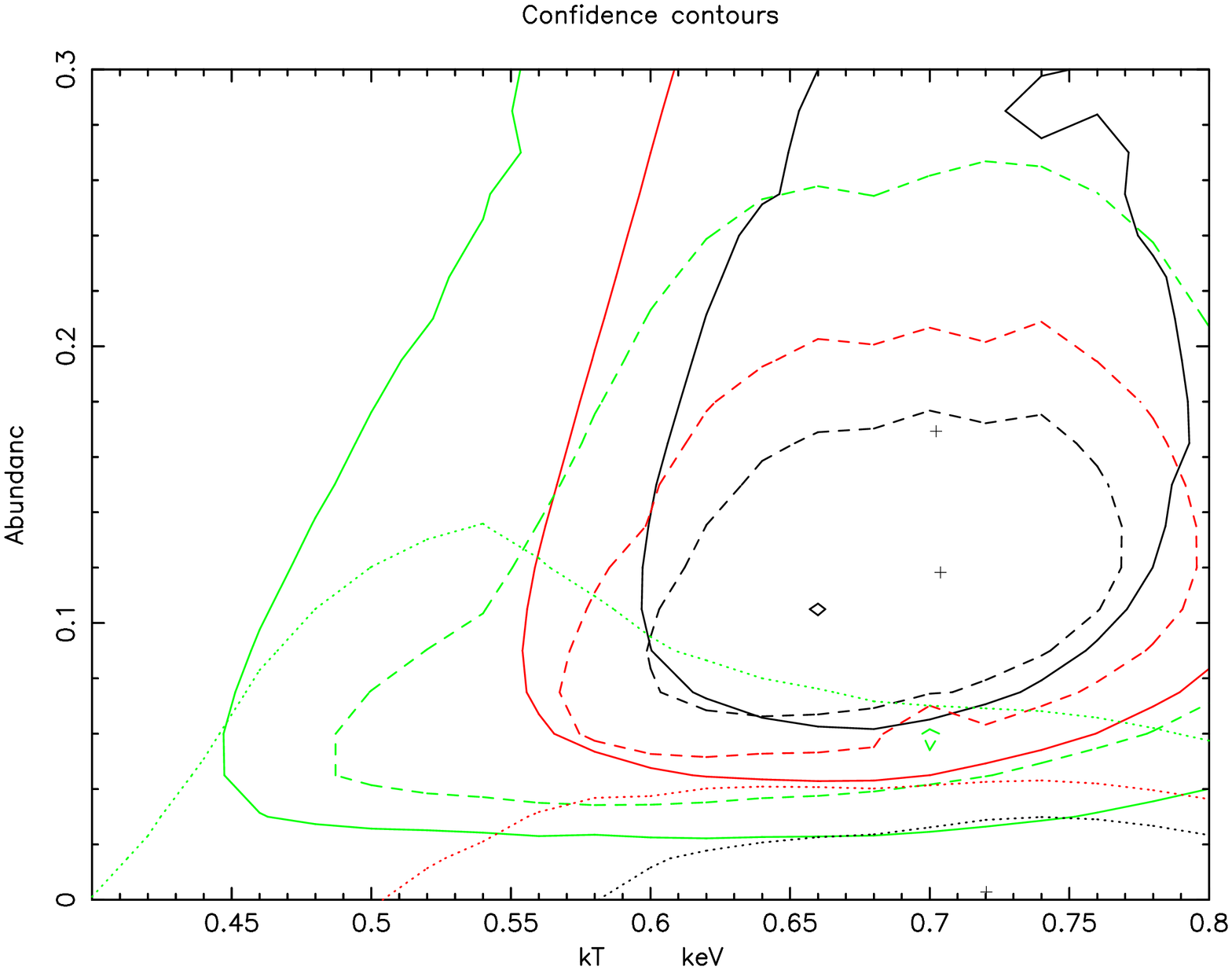}}
\end{minipage}

\end{figure*}
\newpage
\begin{figure*}

\begin{minipage}[h]{\linewidth}
\resizebox{.3\linewidth}{!}{\includegraphics[angle=-90]{f9_10a.eps}}
\resizebox{.3\linewidth}{!}{\includegraphics[angle=-90]{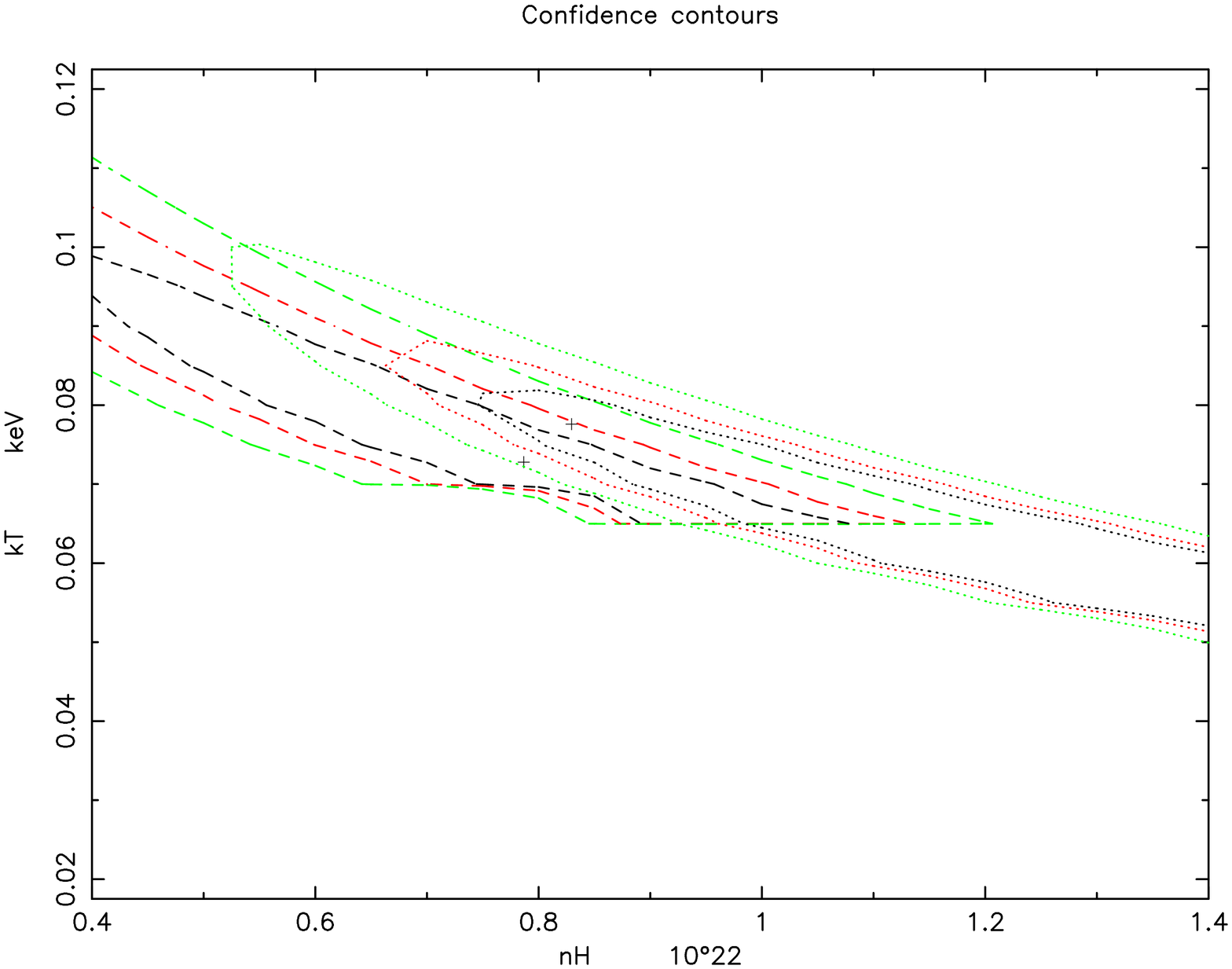}}
\end{minipage}

\begin{minipage}[h]{\linewidth}
\resizebox{.3\linewidth}{!}{\includegraphics[angle=-90]{f9_11a.eps}}
\resizebox{.3\linewidth}{!}{\includegraphics[angle=-90]{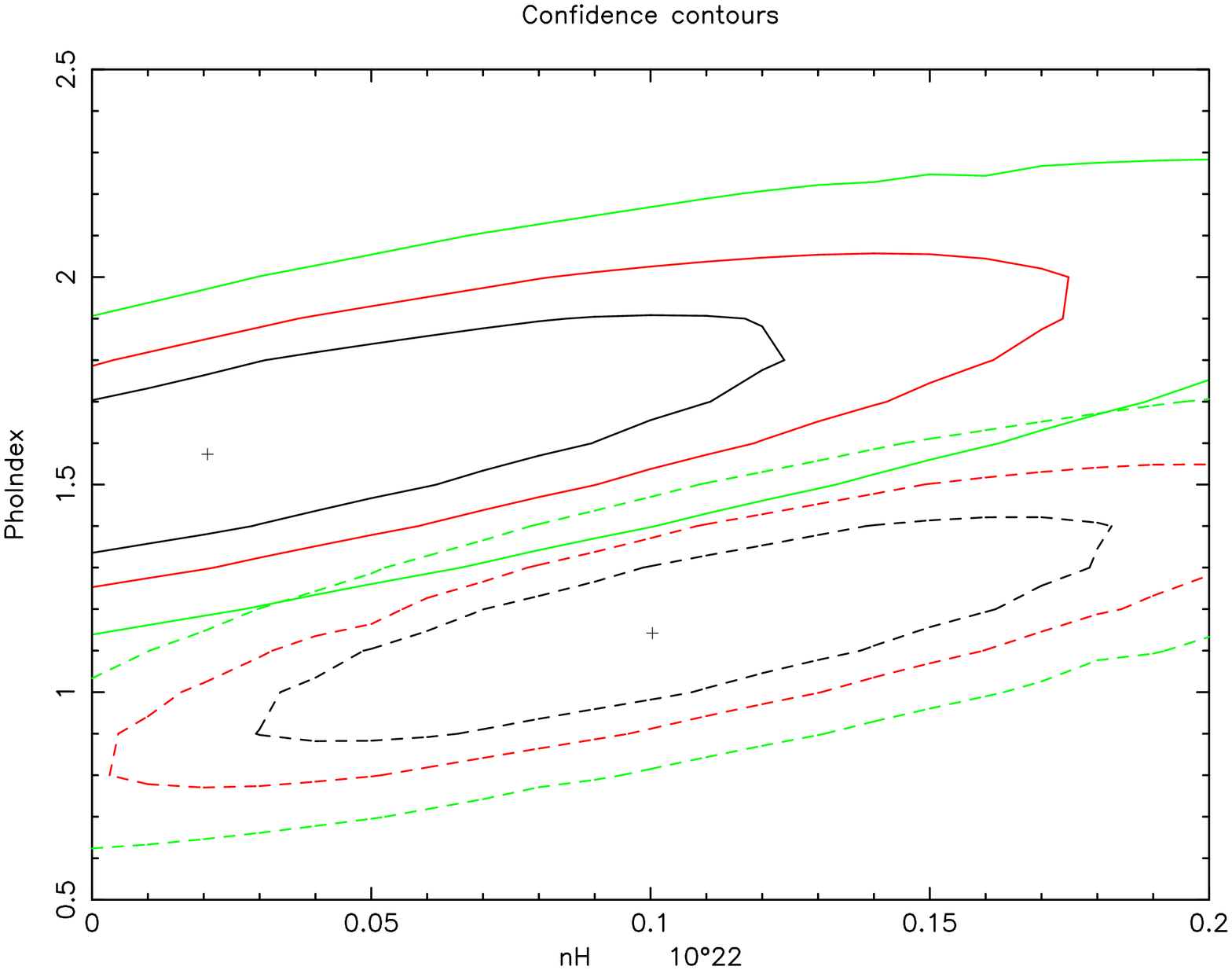}}
\end{minipage}

\begin{minipage}[h]{\linewidth}
\resizebox{.3\linewidth}{!}{\includegraphics[angle=-90]{f9_12a.eps}}
\resizebox{.3\linewidth}{!}{\includegraphics[angle=-90]{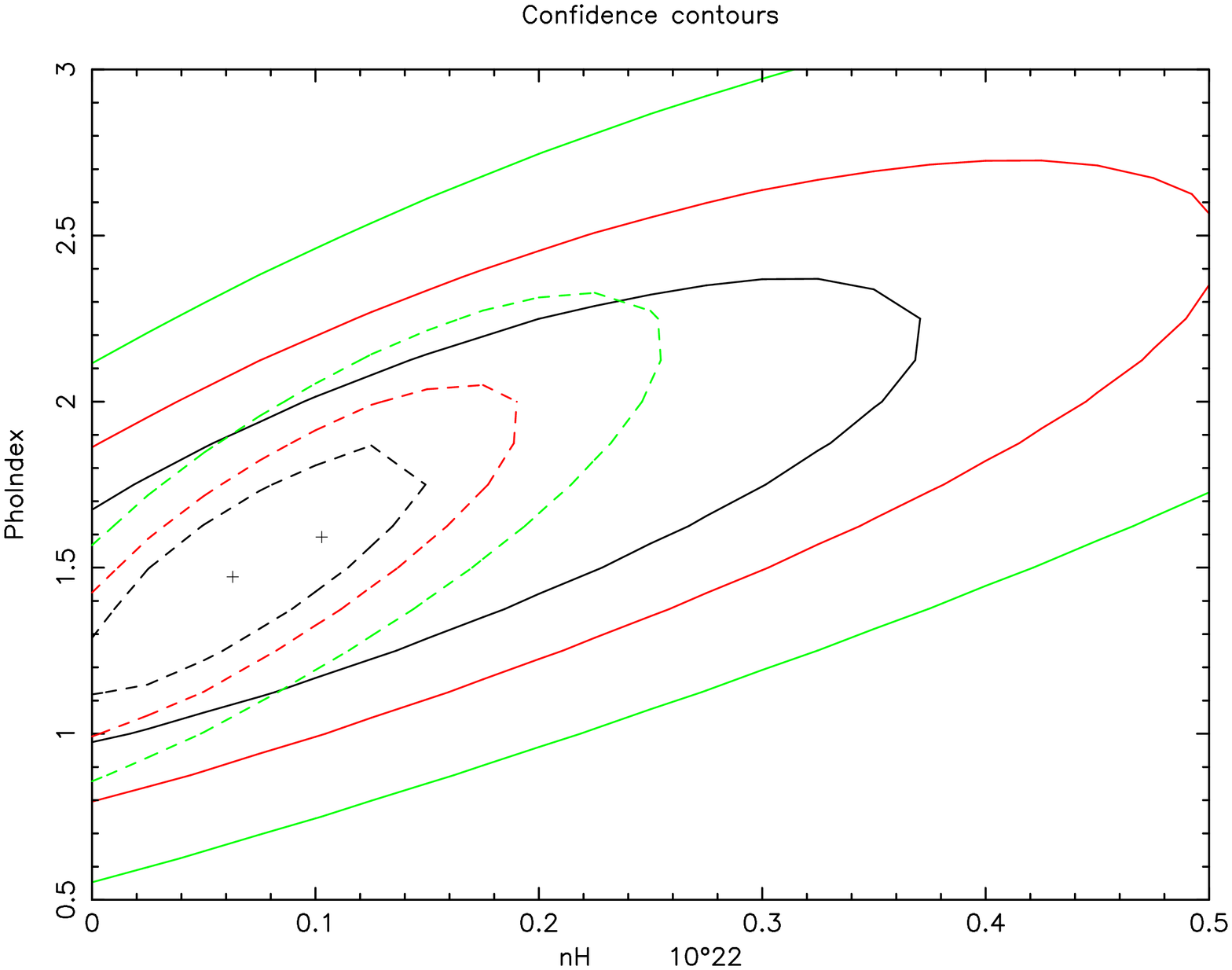}}
\end{minipage}

\begin{minipage}[h]{\linewidth}
\resizebox{.3\linewidth}{!}{\includegraphics[angle=-90]{f9_13a.eps}}
\resizebox{.3\linewidth}{!}{\includegraphics[angle=-90]{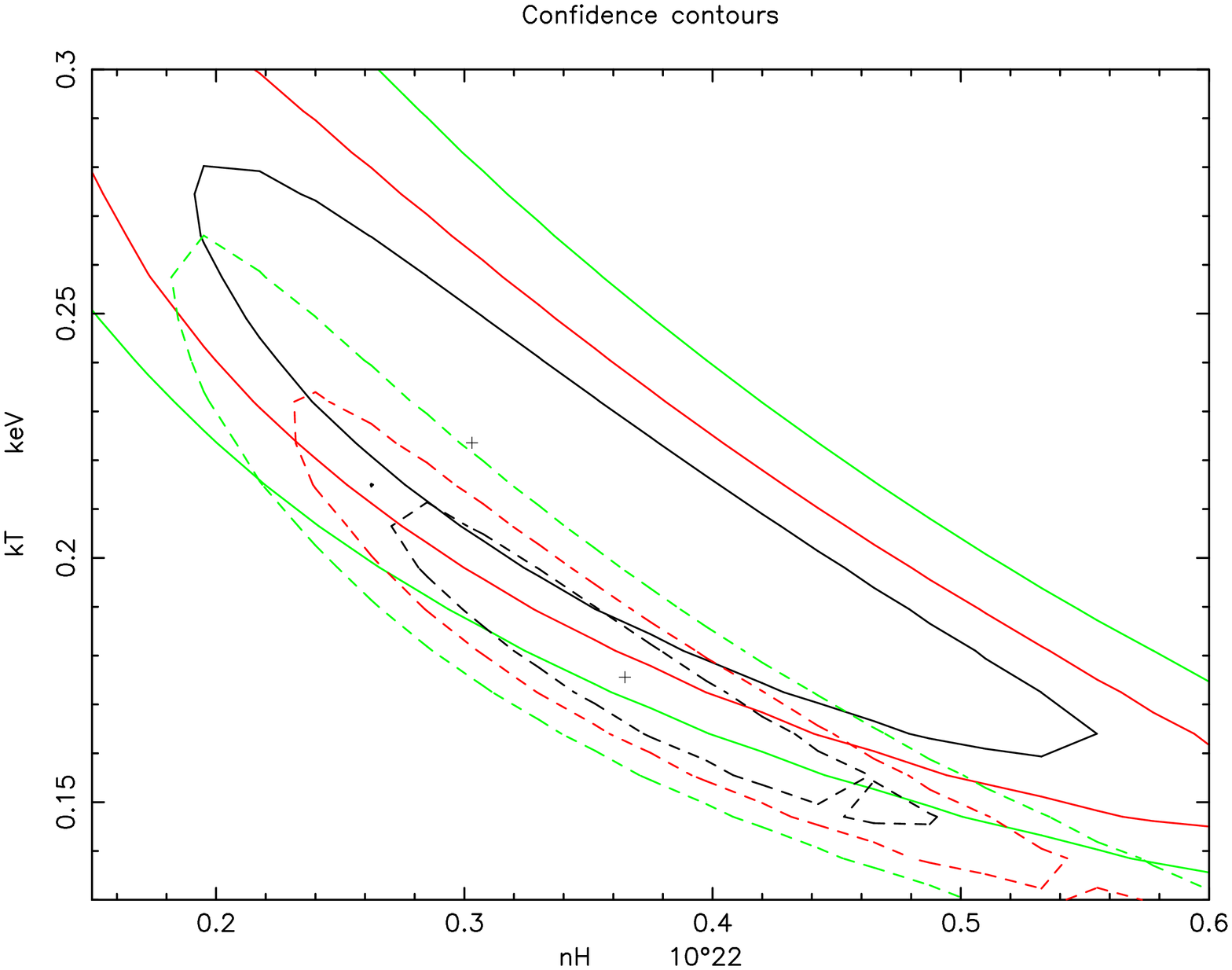}}
\end{minipage}

\begin{minipage}[h]{\linewidth}
\resizebox{.3\linewidth}{!}{\includegraphics{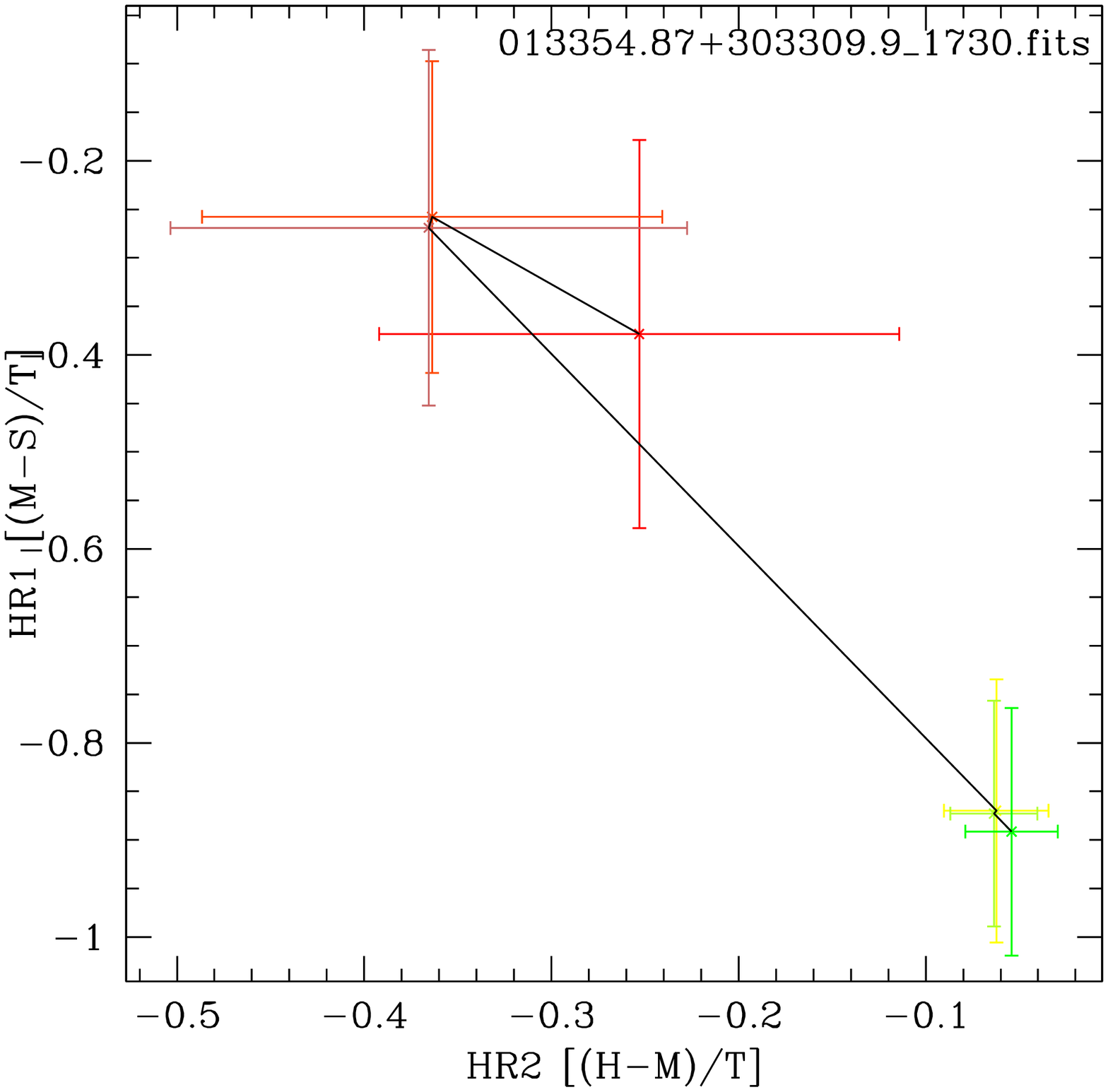}}
\resizebox{.3\linewidth}{!}{\includegraphics{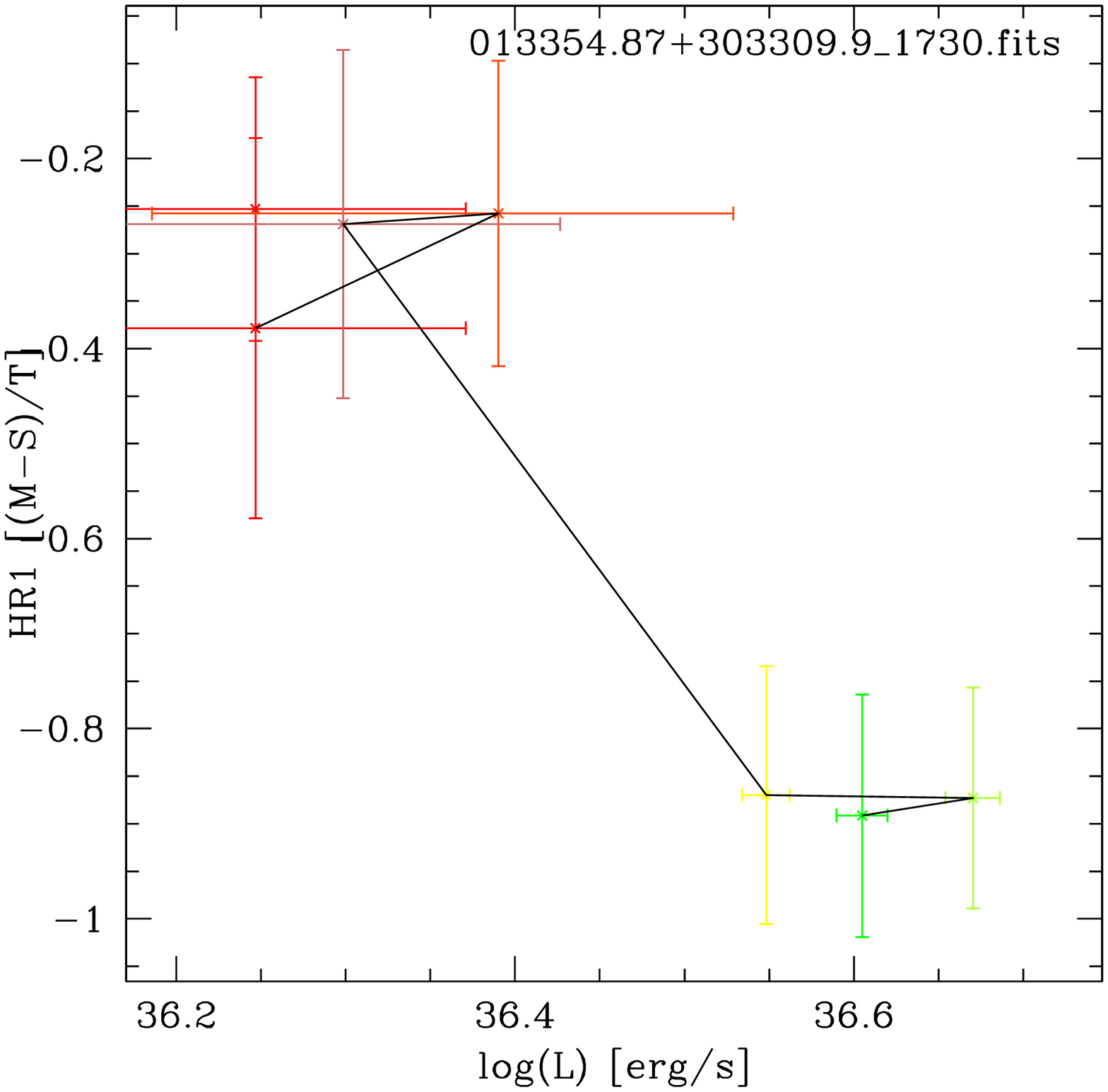}}
\resizebox{.3\linewidth}{!}{\includegraphics{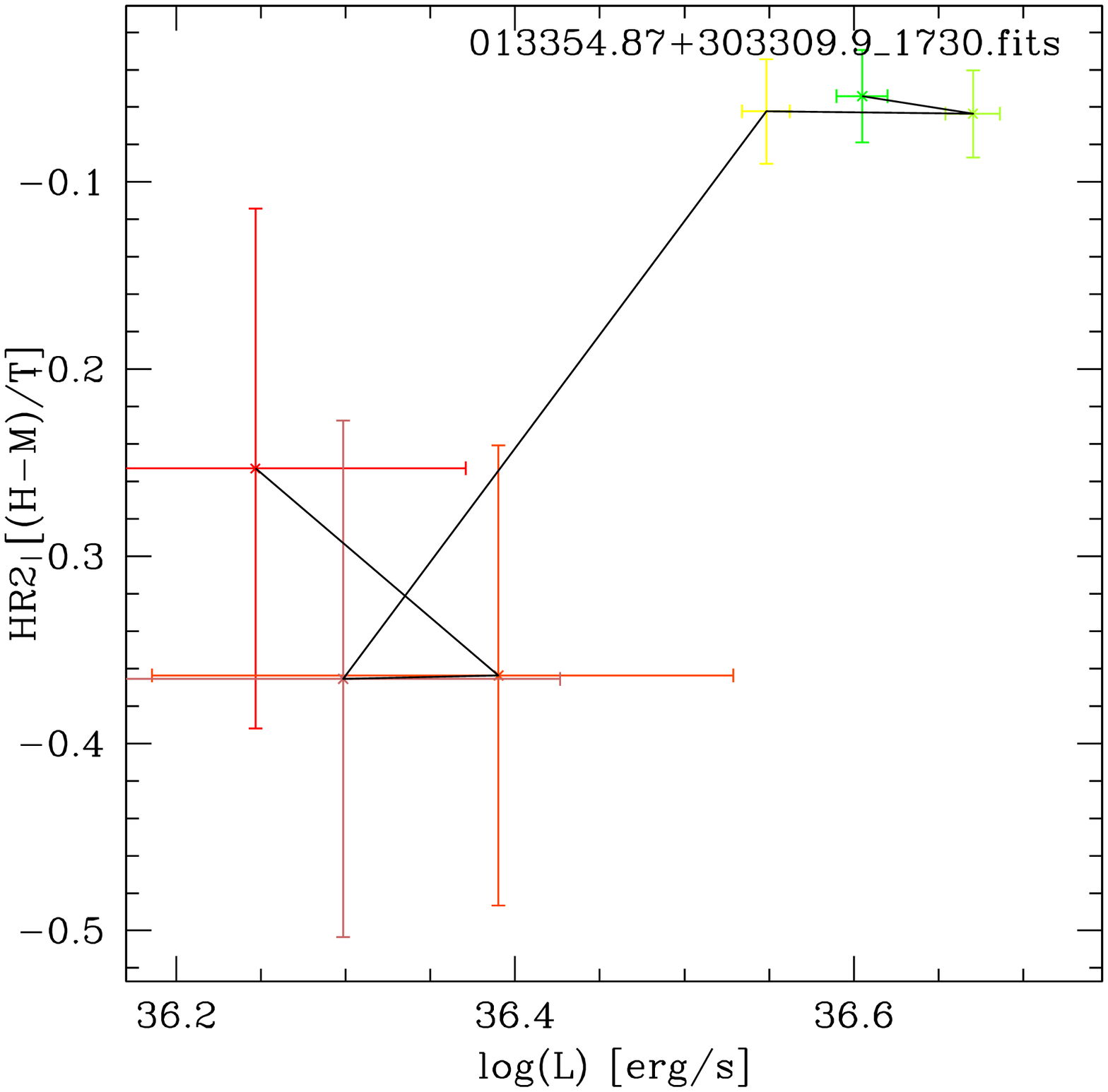}}
\end{minipage}

\end{figure*}
\newpage
\begin{figure*}

\begin{minipage}[h]{\linewidth}
\resizebox{.3\linewidth}{!}{\includegraphics[angle=-90]{f9_14a.eps}}
\resizebox{.3\linewidth}{!}{\includegraphics[angle=-90]{f9_14b.eps}}
\end{minipage}

\begin{minipage}[h]{\linewidth}
\resizebox{.3\linewidth}{!}{\includegraphics[angle=-90]{f9_15a.eps}}
\resizebox{.3\linewidth}{!}{\includegraphics[angle=-90]{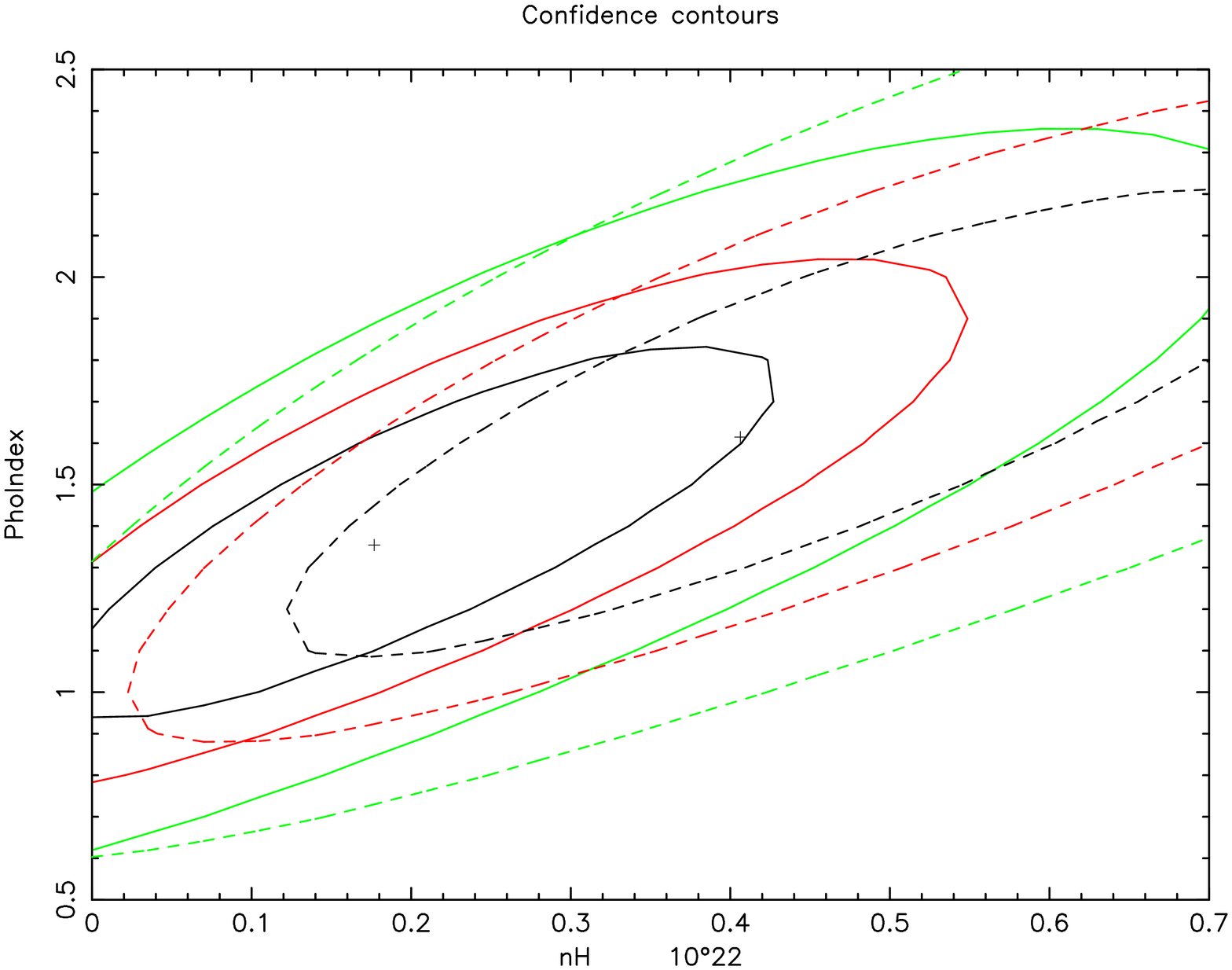}}
\end{minipage}

\begin{minipage}[h]{\linewidth}
\resizebox{.3\linewidth}{!}{\includegraphics[angle=-90]{f9_16a.eps}}
\end{minipage}

\begin{minipage}[h]{\linewidth}
\resizebox{.3\linewidth}{!}{\includegraphics{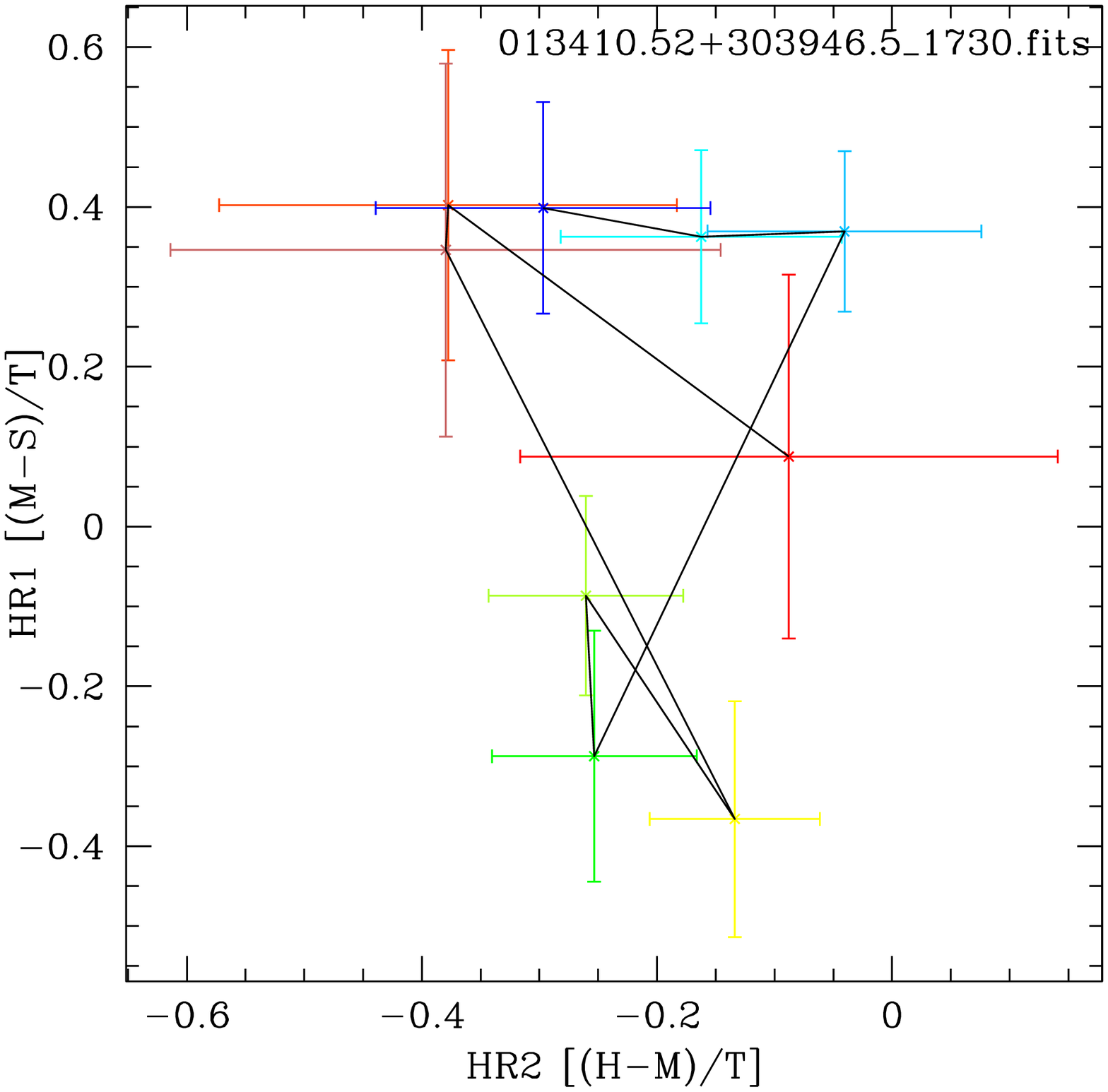}}
\resizebox{.3\linewidth}{!}{\includegraphics{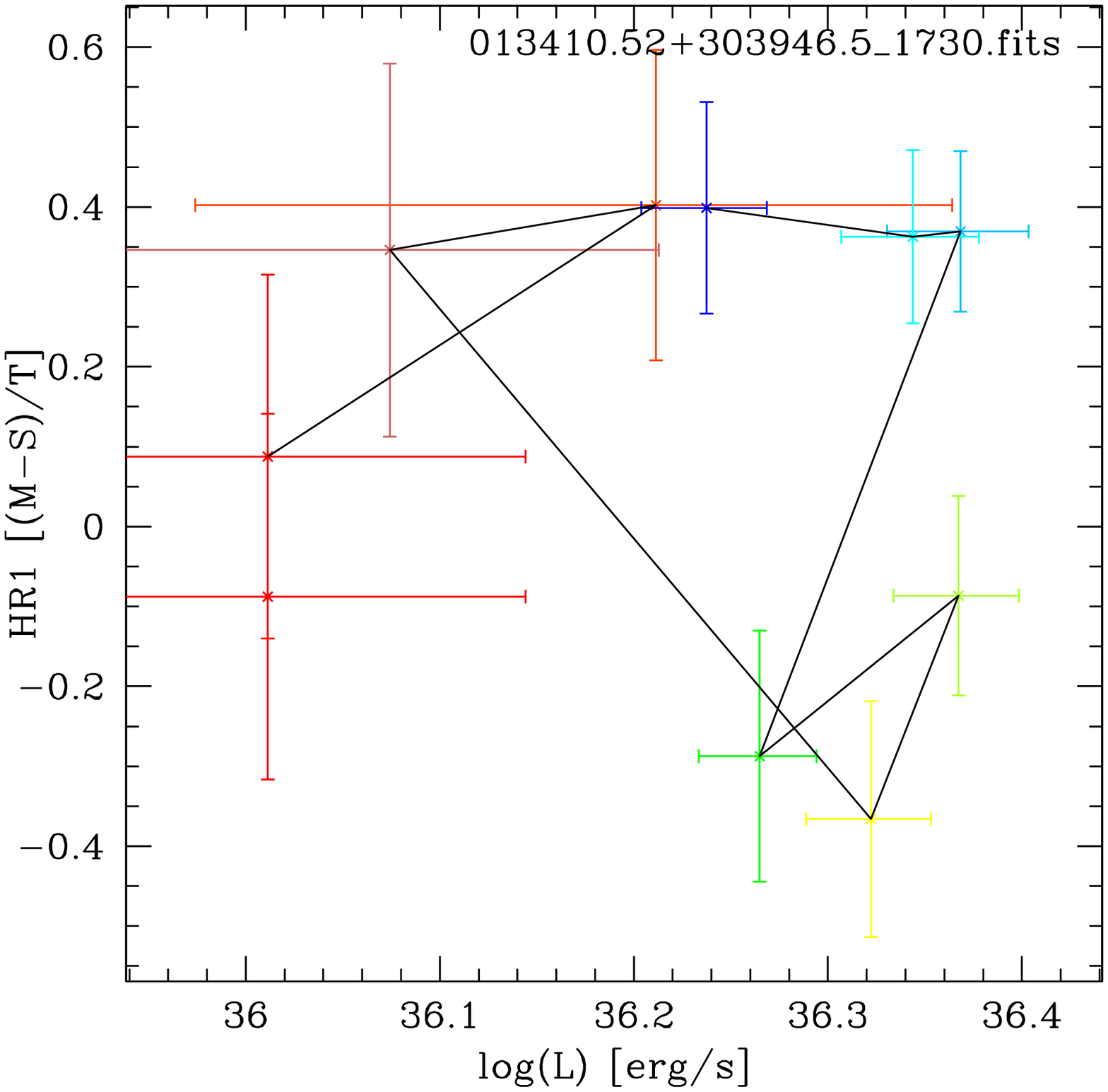}}
\resizebox{.3\linewidth}{!}{\includegraphics{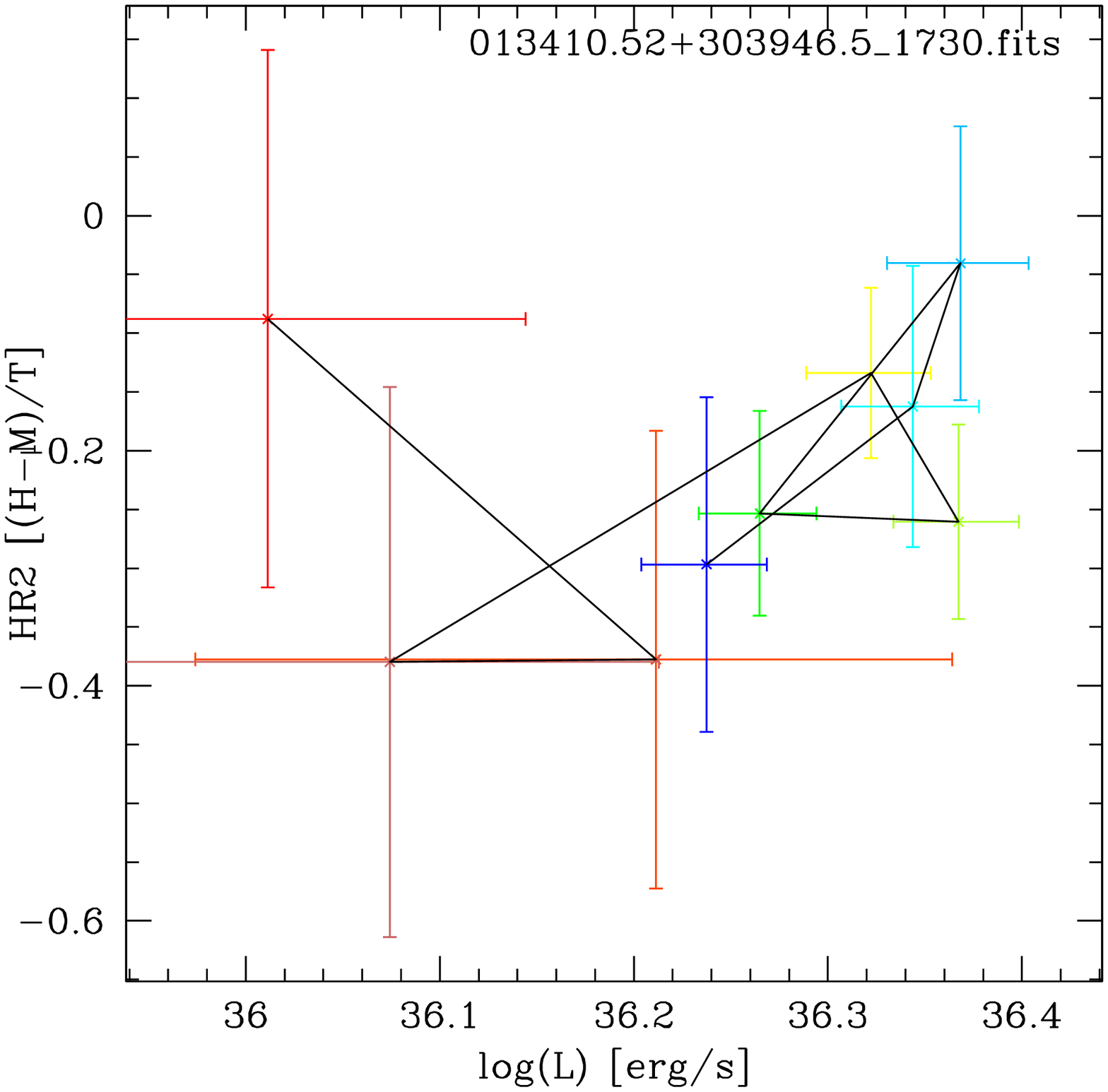}}
\end{minipage}

\begin{minipage}[h]{\linewidth}
\resizebox{.3\linewidth}{!}{\includegraphics[angle=-90]{f9_17a.eps}}
\resizebox{.3\linewidth}{!}{\includegraphics[angle=-90]{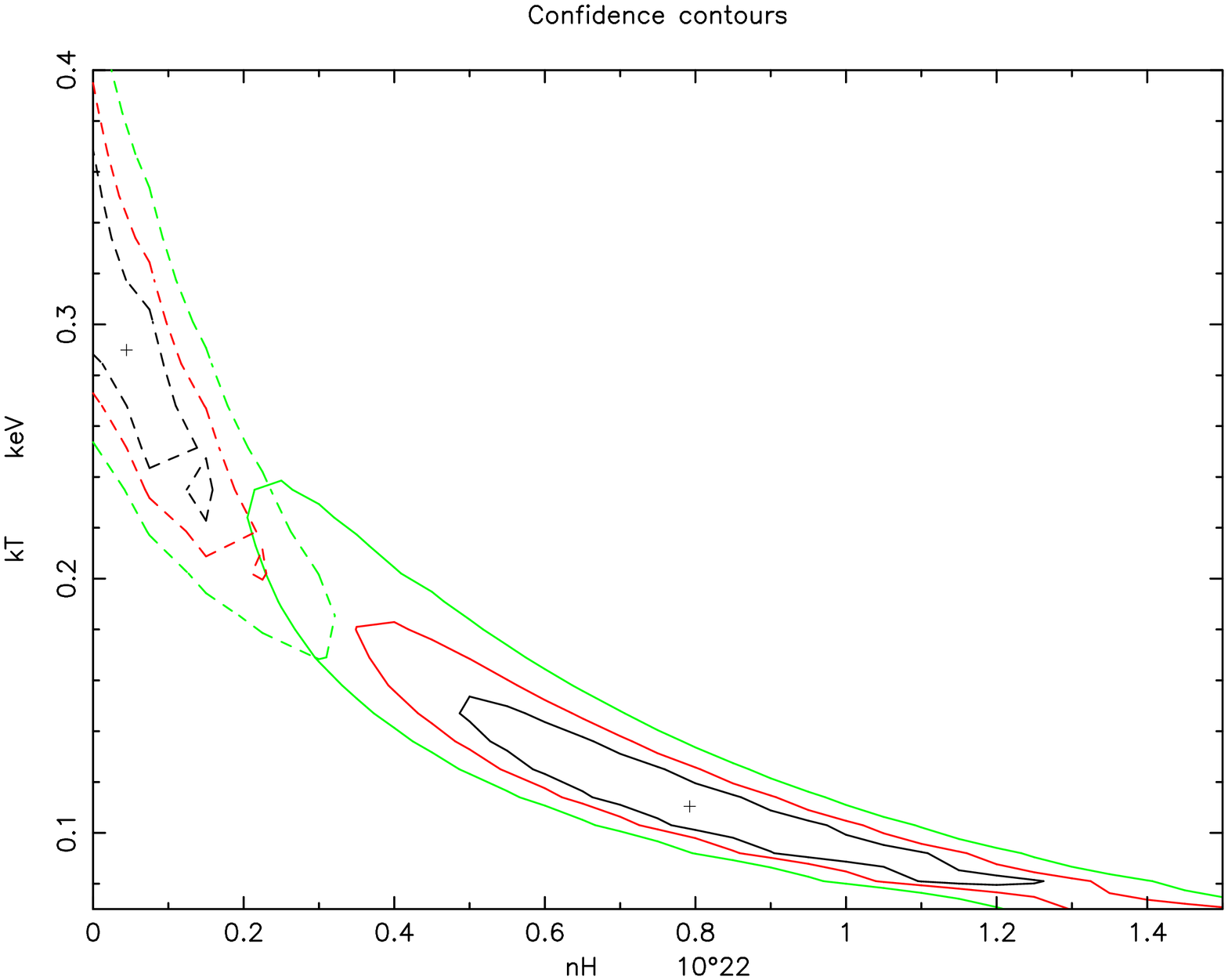}}
\end{minipage}

\end{figure*}
\newpage
\begin{figure*}

altern. APEC model:
\begin{minipage}[h]{\linewidth}
\resizebox{.3\linewidth}{!}{\includegraphics[angle=-90]{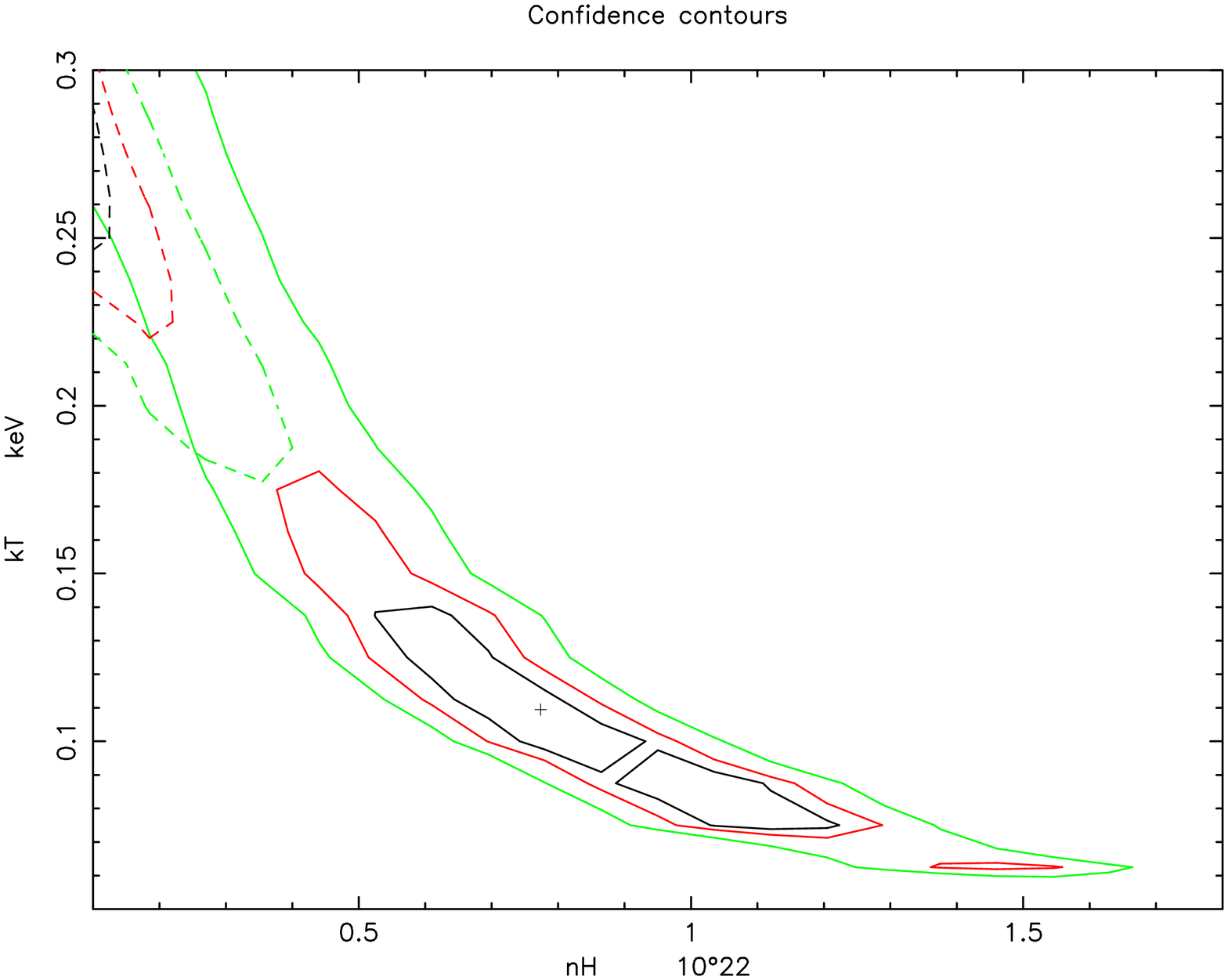}}
\end{minipage}

\begin{minipage}[h]{\linewidth}
\resizebox{.3\linewidth}{!}{\includegraphics[angle=-90]{f9_18a.eps}}
\resizebox{.3\linewidth}{!}{\includegraphics[angle=-90]{f9_18b.eps}}
\end{minipage}

\begin{minipage}[h]{\linewidth}
\resizebox{.3\linewidth}{!}{\includegraphics[angle=-90]{f9_19a.eps}}
\resizebox{.3\linewidth}{!}{\includegraphics[angle=-90]{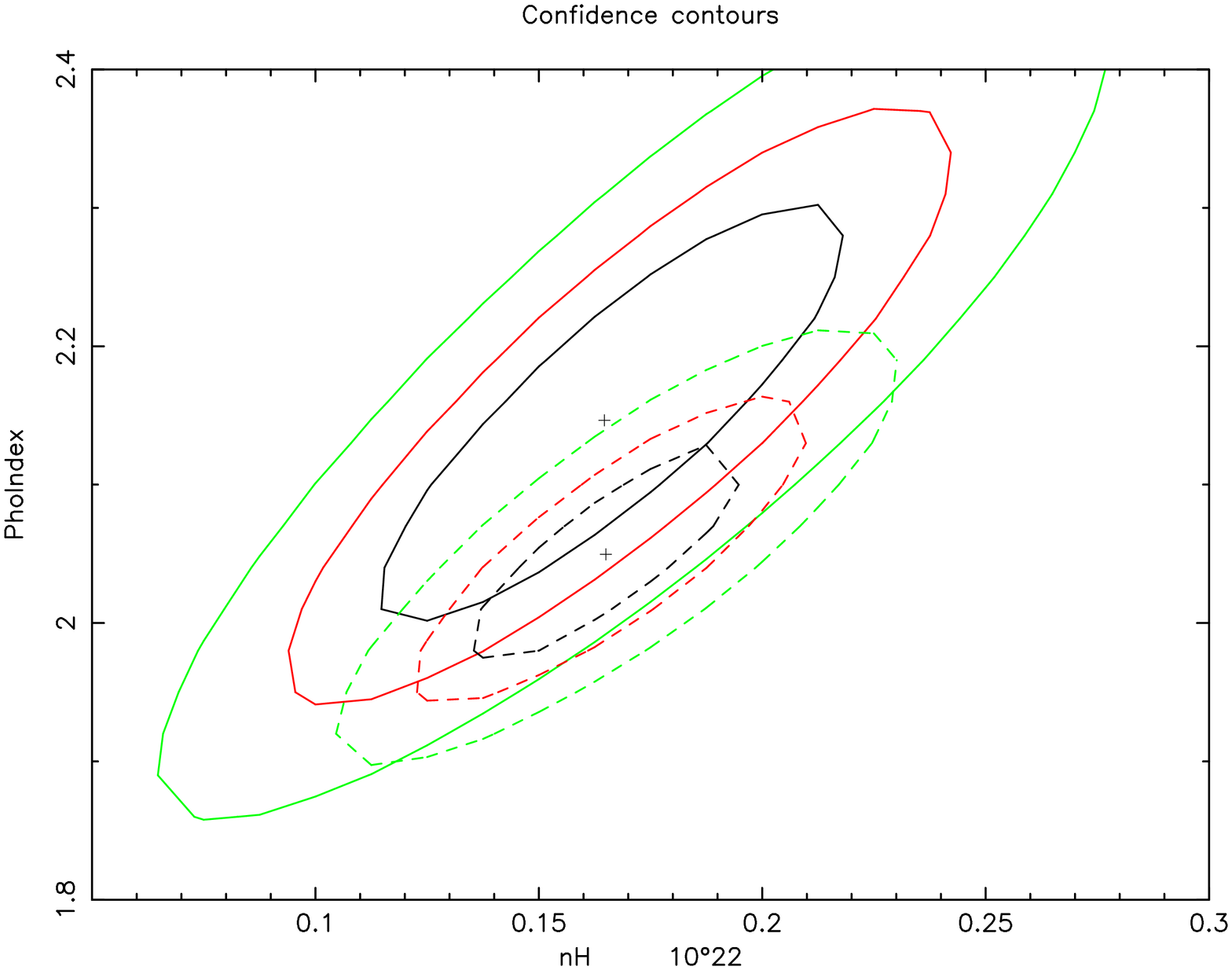}}
\end{minipage}

\begin{minipage}[h]{\linewidth}
\resizebox{.3\linewidth}{!}{\includegraphics[angle=-90]{f9_20a.eps}}
\resizebox{.3\linewidth}{!}{\includegraphics[angle=-90]{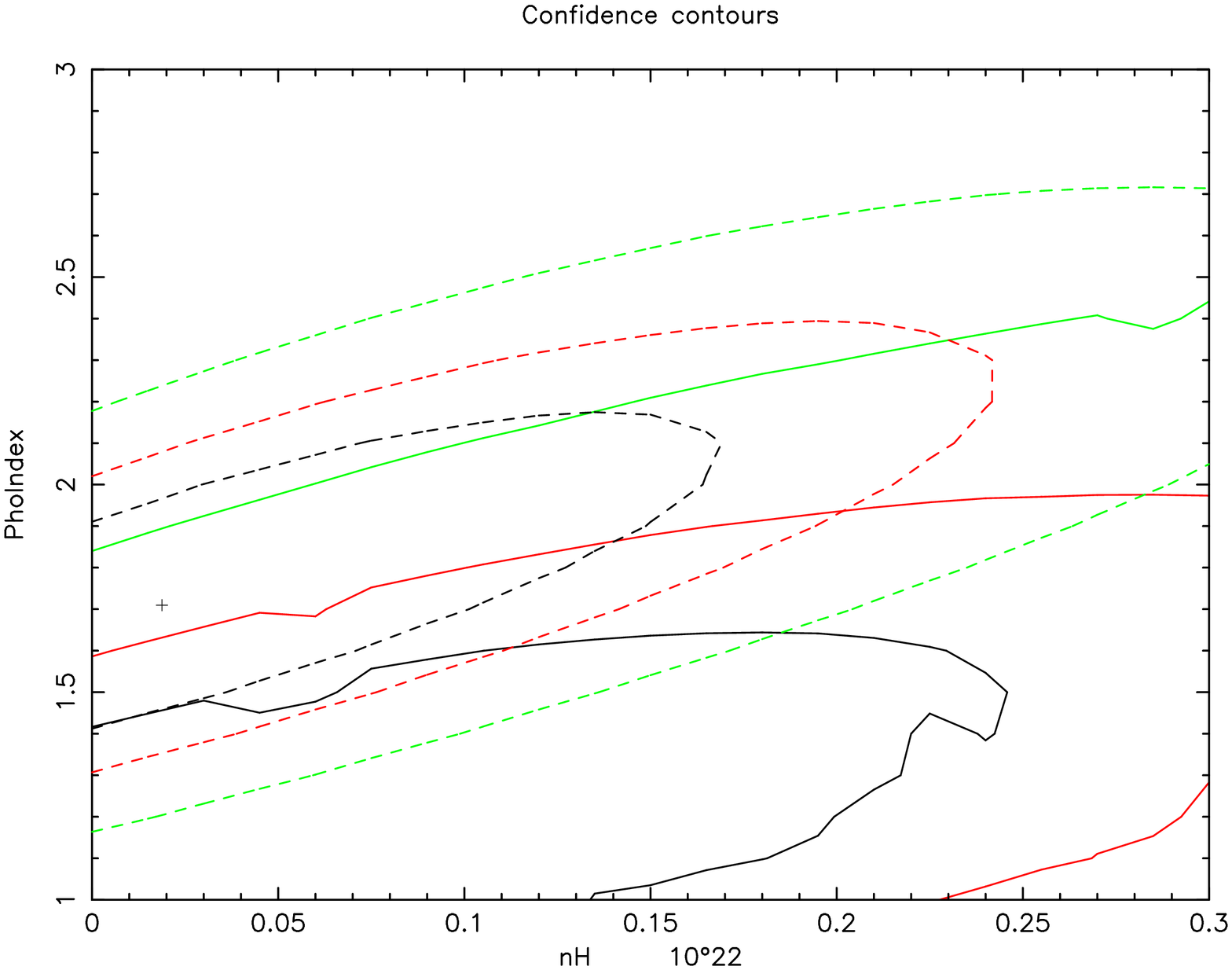}}
\end{minipage}

\begin{minipage}[h]{\linewidth}
\resizebox{.3\linewidth}{!}{\includegraphics[angle=-90]{f9_21a.eps}}
\resizebox{.3\linewidth}{!}{\includegraphics[angle=-90]{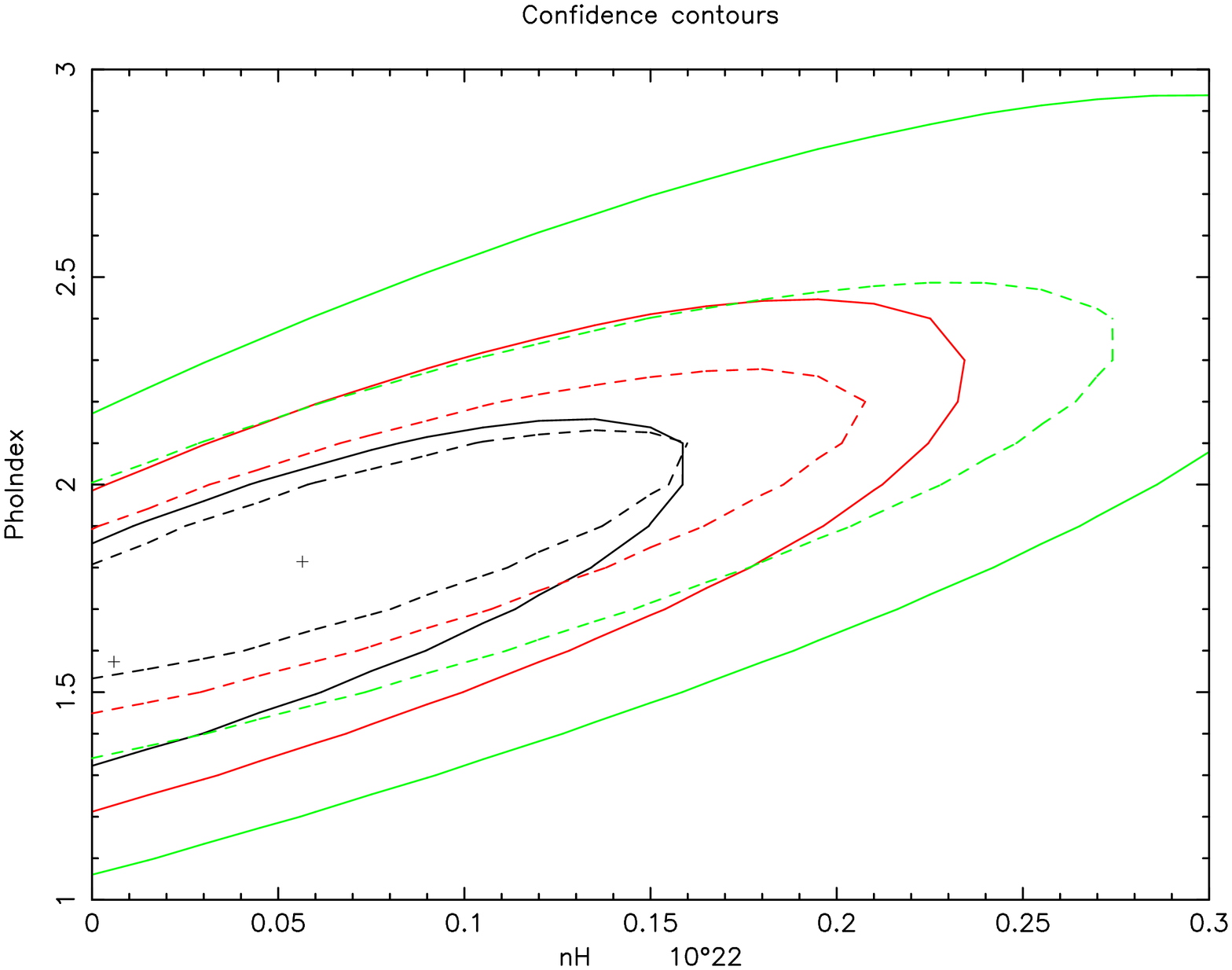}}
\end{minipage}

\end{figure*}
\newpage
\begin{figure*}

\begin{minipage}[h]{\linewidth}
\resizebox{.3\linewidth}{!}{\includegraphics[angle=-90]{f9_22a.eps}}
\resizebox{.3\linewidth}{!}{\includegraphics[angle=-90]{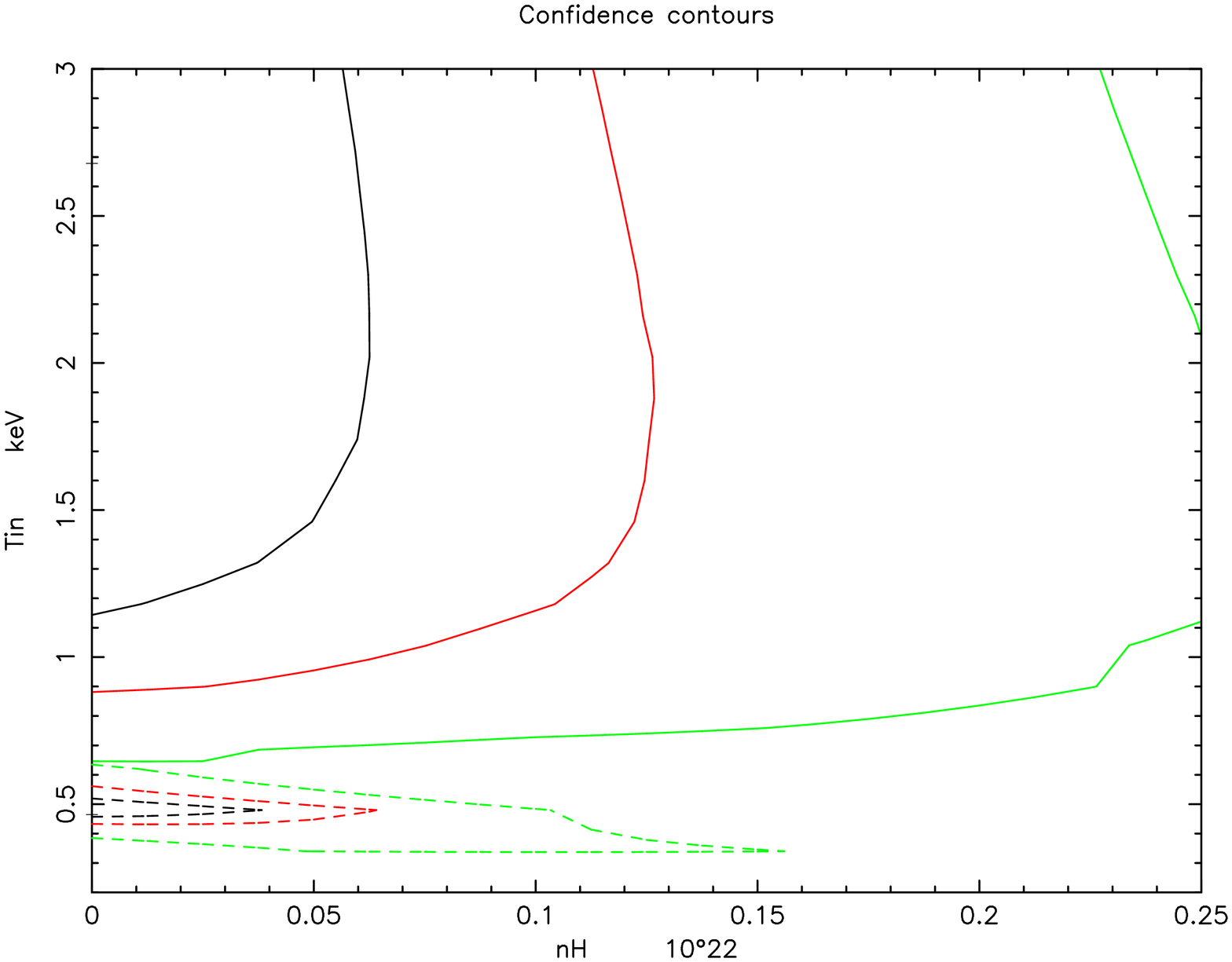}}
\end{minipage}

\begin{minipage}[h]{\linewidth}
\resizebox{.3\linewidth}{!}{\includegraphics[angle=-90]{f9_23a.eps}}
\resizebox{.3\linewidth}{!}{\includegraphics[angle=-90]{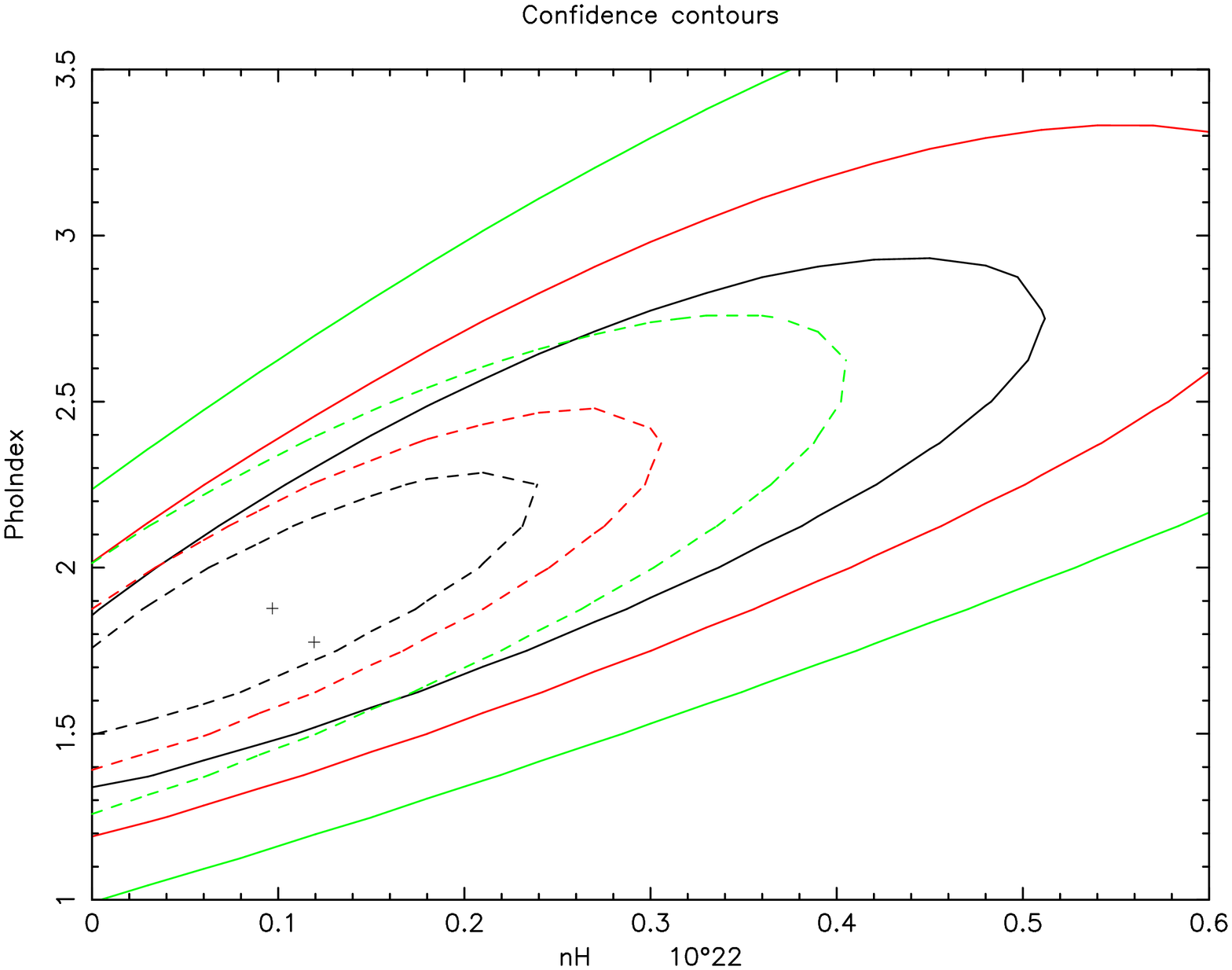}}
\end{minipage}

\begin{minipage}[h]{\linewidth}
\makebox[0.3\linewidth]{}
\end{minipage}

Spectra without well defined confidence contours and color variations:\\
\begin{minipage}[h]{\linewidth}
\resizebox{.3\linewidth}{!}{\includegraphics[angle=-90]{f9_24a.eps}}
\resizebox{.3\linewidth}{!}{\includegraphics[angle=-90]{f9_25a.eps}}
\resizebox{.3\linewidth}{!}{\includegraphics[angle=-90]{f9_26a.eps}}
\end{minipage}

\end{figure*}

\end{appendix}

\end{document}